\documentclass[nonacm]{acmart}
\onecolumn

\AtBeginDocument{%
  \providecommand\BibTeX{{%
    \normalfont B\kern-0.5em{\scshape i\kern-0.25em b}\kern-0.8em\TeX}}}

\copyrightyear{2020}
\acmYear{2020}
\acmDOI{10.1109/ACCESS.2020.2983652}

\usepackage{amsthm}
\usepackage{amsmath}
\usepackage{appendix}
\usepackage{arydshln}
\usepackage[english]{babel}
\usepackage{color}
\usepackage{comment}
\usepackage{flushend}
\usepackage{graphicx}
\usepackage{ifthen}
\usepackage{lineno}
\usepackage{mathrsfs}
\usepackage{mathtools}
\usepackage{multicol}
\usepackage{textcomp}
\usepackage[normalem]{ulem}
\usepackage{url}
\newboolean{REVISION}
\setboolean{REVISION}{true} 

\definecolor{violet}{rgb}{0.56,0.0,1.0}
\definecolor{brightmaroon}{rgb}{0.76, 0.13, 0.28}
\definecolor{amber}{rgb}{1.0, 0.75, 0.0}

\ifthenelse{\boolean{REVISION}}
{
    \newcommand{\vilc}   [1]{\textcolor{blue}   {[#1 -- VILC]}}
    \newcommand{\sadoc}  [1]{\textcolor{red}    {[#1 -- SADOC]}}
    \newcommand{\icc}    [1]{\textcolor{violet} {[#1 -- CUNHA]}}
    \newcommand{\cabral} [1]{\textcolor{cyan}   {[#1 -- CABRAL]}}
    \newcommand{\michael}[1]{\textcolor{magenta}{[#1 -- MG]}}

}{
    \newcommand{\vilc}   [1]{}
    \newcommand{\sadoc}  [1]{}
    \newcommand{\icc}    [1]{}
    \newcommand{\cabral} [1]{}
    \newcommand{\michael}[1]{}

}

\theoremstyle{plain}

\usepackage[english]{babel}
\usepackage{blindtext}

\usepackage{etoolbox}
\makeatletter
\patchcmd{\maketitle}{\@copyrightspace}{}{}{}
\makeatother

\makeatletter
\def\@copyrightspace{\relax}
\makeatother

\begin{document}

\title{Improving Predictability of User-Affecting Metrics to Support Anomaly Detection in Cloud  Services}

\author{Vilc Rufino}
\email{rufino@marinha.mil.br}
\orcid{0000-0001-5557-0519}
\affiliation{%
  \institution{Federal University of Rio de Janeiro(UFRJ) and Brazilian Navy}
  \city{Rio de Janeiro}
  \state{RJ}
  \postcode{21941-590}
  \country{Brazil}
}

\author{Mateus Nogueira}
\email{msznogueira@gmail.com}
\orcid{0000-0001-8851-8987}
\author{Daniel Menasch{\'e}}
\email{sadoc@dcc.ufrj.br}
\orcid{0000-0002-8953-4003}
\author{Cabral Lima}
\email{cabrallima@ufrj.br}
\orcid{0000-0003-1754-1666}
\affiliation{%
  \institution{UFRJ}
  \city{Rio de Janeiro}
  \state{RJ}
  \postcode{21941-590}
  \country{Brazil}
}
\author{Alberto Avritzer}
\email{beto@esulabsolutions.com}
\orcid{0000-0002-9401-9663}
\affiliation{%
  \institution{Esulab Solutions}
  \city{Plainsboro}
  \state{NJ}
  \country{USA}}

\author{Barbara Russo}
\email{brusso@unibz.it}
\orcid{0000-0003-3737-9264}
\author{Andrea Janes}
\email{ajanes@unibz.it}
\orcid{0000-0002-1423-6773}
\affiliation{%
  \institution{Free University of Bozen-Bolzano}
  \city{Bolzano-Bozen}
  \state{Bolzano}
  \country{Italy}}

\author{Vincenzo Ferme}
\email{vincenzo.ferme@usi.ch}
\orcid{0000-0002-6232-6554}
\affiliation{%
  \institution{Universita della Svizzera Italiana}
  \city{Lugano}
  \state{Ticino}
  \country{Italy}
}

\author{André van Hoorn}
\email{van.hoorn@informatik.uni-stuttgart.de}
\orcid{0000-0003-2567-6077}
\affiliation{%
  \institution{University of Stuttgart}
  \city{Stuttgart}
  \country{Germany}
}

\author{Henning~Schulz}
\email{henning.schulz@novatec-gmbh.de}
\orcid{0000-0001-9788-9982}
\affiliation{%
 \institution{Novatec Consulting GmbH}
 \city{Baden-Württemberg}
 \country{Germany}}

\renewcommand{\shortauthors}{Rufino, Nogueira, Avritzer, Menasch{\'e}, Russo, Lima, et al.}


\begin{abstract}
Anomaly detection systems aim to detect and report  attacks or unexpected behavior in networked systems. Previous work has shown that anomalies have an impact on system performance, and that performance signatures can be effectively used for implementing an IDS. In this paper, we present an analytical and an experimental study on the trade-off between anomaly detection based on performance signatures and system scalability. The proposed approach combines analytical modeling and load testing to find optimal configurations for the signature-based IDS.   We apply a heavy-tail bi-modal modeling approach, where ``long'' jobs represent large resource consuming transactions, e.g., generated by DDoS attacks; the model was parametrized using results obtained from controlled experiments.
For performance  purposes, mean response time is the key  metric to be minimized, whereas for security purposes, response time variance and classification accuracy  must be taken into account. The key insights from our analysis are: (i)~there is an optimal number of servers which minimizes the response time variance, (ii)~the sweet-spot number of servers that minimizes response time variance and maximizes classification accuracy is typically smaller than or equal to the one that minimizes  mean response time. Therefore, for security purposes, it may be worth slightly sacrificing performance to increase classification accuracy.
\end{abstract}



\maketitle

\section{Introduction} 
\label{sec:2_intro} 

\begin{figure}[htb]
        \centering
        \includegraphics[width=0.90\textwidth]{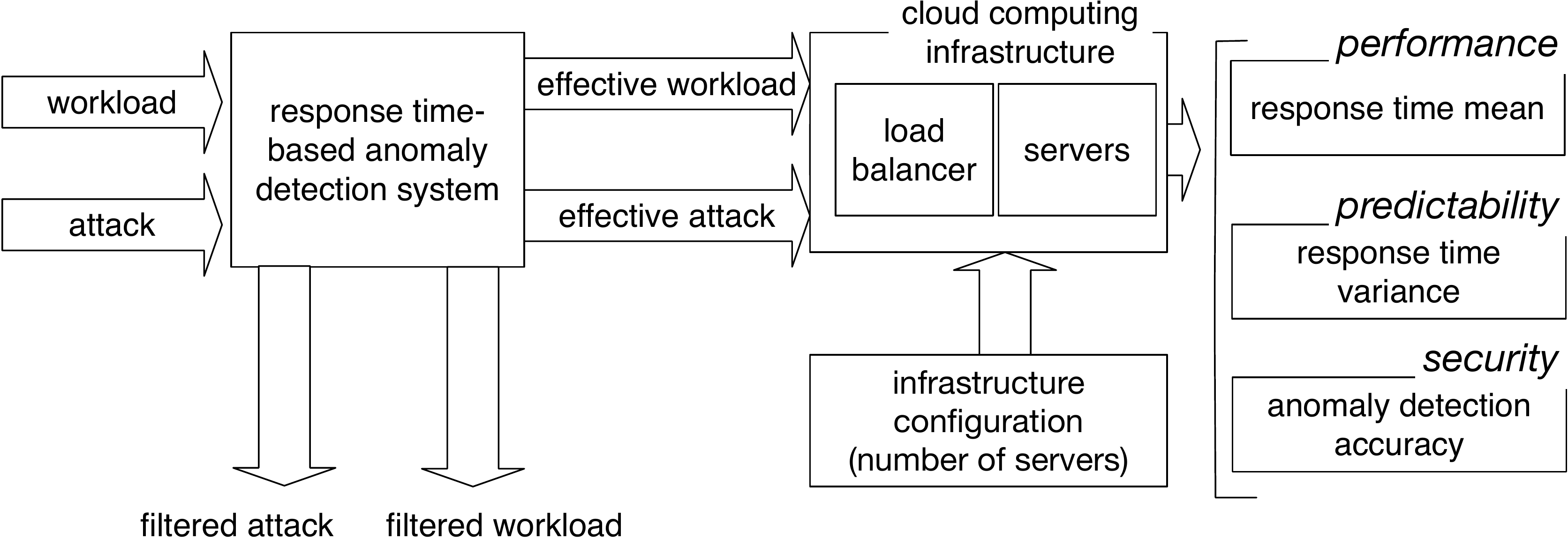}
        \vspace{-0.2in}
        \caption{Anomaly detection framework, accounting for infrastructure configuration actions at system deployment time.} 
        \label{fig:securityscheme}
    \end{figure}

    The multi-server paradigm with a central queueing system, e.g., encompassing  a load balancing front-end towards a cloud computing environment, has been used as the \emph{state of the practice}  for scalable server-side computation for the last several years~\cite{psounis2005systems, icac2018, tadakamalla2020autonomic}.  In spite of the  widespread adoption of the multi-server paradigm, the setup of  such systems still   poses a number of  challenges related to the shaping of the workload and the configuration of the servers. 
    Security threats against multi-server systems,  for instance, pose  a pressing need for a better understanding of how  the configuration of multi-server systems impacts  their performance, security and predictability (Figure~\ref{fig:securityscheme}).   

    In this paper, we  propose a framework for the analysis of the fundamental tradeoffs between performance, security and predictability in central-queue multi-server systems~\cite{harchol2013performance, menasce2002capacity}. In particular, we are interested in analyzing the effectiveness of anomaly detection approaches based on performance signatures, as for example, response time.  
    Such approaches rely on the response time being predictable.  

    Figure~\ref{fig:securityscheme} shows an example of how a multi-server system, comprising  a central queueing system, a load balancer and multiple servers, is integrated into an anomaly detection framework. It shows at the front-end the regular workload and attack streams being filtered by a response time based anomaly detection system. In addition, it shows at the back-end the analysis of  mean response time and variance  used to support response time predictability.

    Performance of cloud services is typically captured through its mean response time. Anomaly detection based on performance may rely on tracking mean and \emph{variance} of response times.   If response time variance is high, for instance, it may be difficult to distinguish between transient performance degradation and a denial of service attack.  It is well known that \emph{variance reduction techniques} can improve the task of anomaly detection~\cite{carter2008variance,chen2017outlier,timvcenko2017ensemble,rayana2016less,salehi2016smart}.
    One of the goals of this paper is to assess the impact of  the number of servers in a cloud computing environment   on the variance of response times and  accuracy of anomaly detectors.


    \begin{figure}[t]
        \centering
        \includegraphics[width=0.75\textwidth]{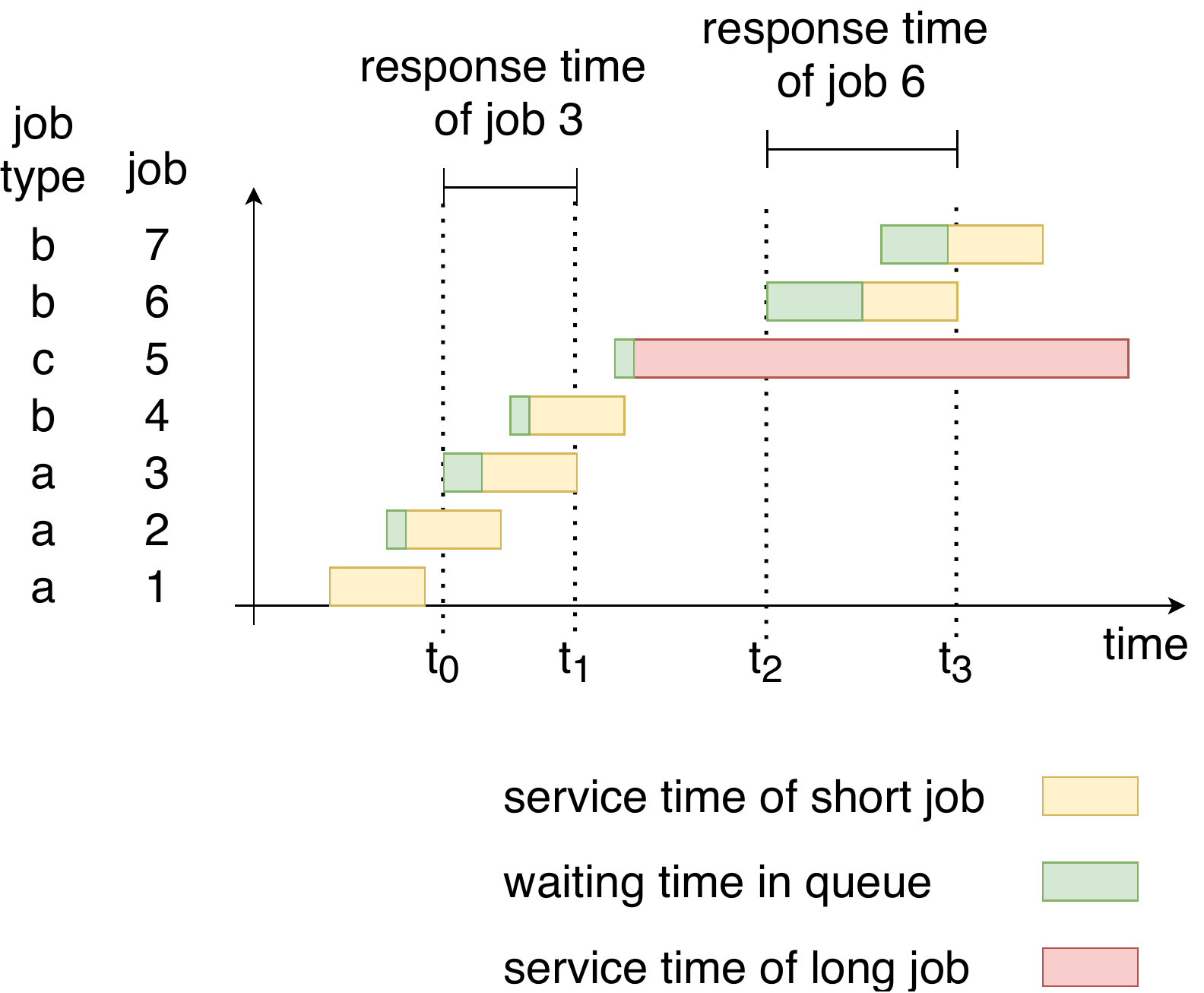}
        \caption{The considered anomaly detection mechanism aims to detect  anomalies solely based on job response times. When is it feasible to identify if the system is under attack solely based on response times?  Jobs of types $a$, $b$ and $c$ correspond to short jobs, short jobs impaired by attacks and long attack jobs, respectively. 
        Depending on the system setting, the waiting time of jobs in the waiting queue can be used to distinguish jobs of type $a$ and type $b$.  Under which system configurations does the waiting time contain enough information to allow the automatic distinction between jobs of types $a$ and $b$, implying the feasibility of locally detecting  anomalies solely based on response times?  } 
        \label{fig:anomalygeneral}
    \end{figure}

    \begin{figure*}[htb]
    \center
    \includegraphics[width=1.0\textwidth]{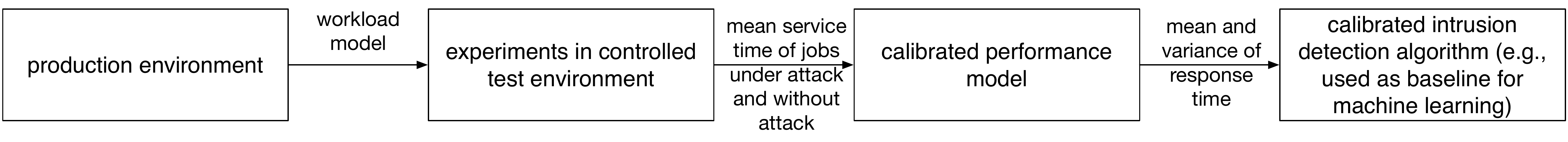}
    \vspace{-0.1in}
    \caption{Proposed framework for a performance-based anomaly detection system}
    \vspace{-0.1in}
     \label{fig2-label}
    \end{figure*}

    We consider  a fixed budget to be allocated in cloud premises, aiming at cost-optimal cloud deployments~\cite{li2013cost,tordsson2012cloud}.  
    The budget can be split into several servers  or few more powerful servers.
    We then pose the following questions:
     \begin{itemize}
         \item As the number of servers increases, and the capacity of each of them decreases, how does the mean and the variance of  response time vary?
         \item How does the accuracy of anomaly detection classifiers vary as a function of the number of servers?
     \end{itemize}
    While answering the first question, we discovered that both the mean and the variance of the response time first decrease and then increase as the number of servers grows. This, in turn, motivated the second question (Figure~\ref{fig:anomalygeneral}). 

    We focus on    metrics directly affecting customers without violating the user privacy, e.g., response times.  We indicate that those metrics are sensitive to  anomalies to the extent that they can then be used to support an anomaly detection framework.  As   customer-affecting metrics are intrinsically  publicly available, sharing them with third-parties responsible for anomaly detection has the benefit of not compromising user privacy. This should be contrasted against deep packet inspection, for instance, which is more computationally expensive and requires the sharing of sensible data.
    In addition, anomaly detection approaches based on performance signatures have the potential of detecting zero-day attacks~\cite{milenkoski2015evaluating}, as those  approaches  are based on detecting performance deviations from regular behavior and do  not require detailed knowledge of attack history.

    \emph{Prior art: } 
    Previous work has shown that performance engineering can be used for implementing attack detection methods. For instance, Avritzer et al.~\cite{avritzer2010monitoring} propose an approach that leverages performance signatures based on CPU, I/O, memory and network usage for the detection of security anomalies. The approach has been shown to be very accurate and to outperform, in some cases, anomaly detection systems~\cite{milenkoski2015evaluating} based on attack history. 
    The architecture of performance and security in cloud computing systems incurs several tradeoffs, as for example, the overhead of implementing security protocols~\cite{Wolter2010}, the configuration of  database partitioning systems for anomaly detection while optimizing performance~\cite{alomari2012autonomic}, and security-related cache invalidation analysis to ensure scalability of data-intensive applications~\cite{manjhi2006simultaneous}.
    In this work, we indicate that models can be instrumental to assess the impact of architecture design decisions in the tradeoff landscape. In particular, we evaluate the impact of the number of servers on the  mean and variance of response time at the application level, and on the accuracy of anomaly detectors. If response time is to be used as a metric to support anomaly detection, it is important to setup the system such that response time variability can be attributed to security anomalies and not to the variability in system performance. 

    \emph{Contributions: } In summary, our key contributions are twofold:

        \textbf{Analytical model to derive mean and variance of baseline response time: } We apply a heavy-tail bimodal modeling approach, where  \emph{short} jobs represent regular transactions and \emph{long} jobs represent  resource consuming transactions, e.g., due to  denial of service attacks. In our modeling approach, \emph{long} jobs are related to anomalies. Using the bi-modal modeling approach we are able to estimate the baseline system response time average and standard deviation,  for several system parameter values (e.g., different ratios of \emph{short} and \emph{long} jobs).  Although the bi-modal model is admittedly simple, it already serves  our purposes, namely to illustrate the tradeoffs between performance and predictability (Sections~\ref{sec:2_model}-\ref{sec:2_findings}) and accuracy of anomaly detection (Section~\ref{sec:2_anomaly}).
       
  
        \textbf{Optimal number of servers: } Using the proposed model, we consider the problem of determining the optimal number of servers accounting for scalability and security aspects, e.g., anomaly detection based on average response time.  
    
    Figure~\ref{fig2-label} illustrates the framework using the concepts studied in this paper and is further discussed in the upcoming section.
    
    The outline of the remainder of the paper is as follows. 
    In Section~\ref{sec:2_framework} we present a performance-based anomaly detection framework that is composed of 
        $(i)$  analytical  modeling, 
        $(ii)$ empirical experiments, 
        $(iii)$ calibrated performance models and 
        $(iv)$ anomaly detection algorithms based on tracking of customer-affecting metrics, as for example, response time.  
    Those aspects are treated, in that order,  in Sections~\ref{sec:2_model} to~\ref{sec:2_anomaly}.  
    Related work and     the limitations and broad  implications of our results are discussed in  Sections~\ref{sec:2_related} and~\ref{sec:2_limitations}.    Conclusions and directions for future research appear in Section~\ref{sec:2_conclusions}.

\section{Proposed Framework} 
\label{sec:2_framework}

In this section, we describe the proposed performance-based anomaly detection framework. Figure~\ref{fig2-label} shows the building blocks that compose  the framework and the inputs and outputs of each process block, as follows:

\begin{enumerate}
    \item The \textbf{ production environment} is monitored using application performance monitoring (APM) tools~\cite{HegerHMO17} to generate a workload model, e.g., user profile distribution~\cite{avritzer2018}. The obtained workload model is the basis of the proposed analytical model {(Section~\ref{sec:2_model})} and is used to support the specification of the load levels to be used in the performance testing experiments. 
    \item \textbf{ Experiments in a controlled test environment} are executed to estimate the  mean service time of jobs under attack and without attack. The load tests are  executed  using load levels specified by using the derived workload model.  Laboratory experiments are run using load testing tools~\cite{jiang2008automatic} to measure mean and standard deviation of service time, with and without anomalies. The submitted transaction rate is derived from APM tools. Experimental results to  calibrate the performance model are  presented in Section~\ref{sec:2_modelparam}.
    \item  \textbf{ Calibrate the analytic performance model} to assess the dependency of  the mean and standard deviation of response times with respect to  the number of servers. The calibration leverages experimental results reported in Section~\ref{sec:2_modelparam} to characterize a bi-modal distribution. Short jobs correspond to service times of typical users, and long jobs correspond to security attacks.  The analytical model results (reported in   Section~\ref{sec:2_findings}) are used to estimate the number of servers that minimizes   response time average and variance, as shown in Table~\ref{tab:optimal}.   
    \item \textbf{ Calibrate the anomaly detection algorithms} using the mean and standard deviation of response time for the optimal number of servers. Machine learning algorithms, such as SVM~\cite{jung2004fast}, can be used to perform   classification  between regular and anomalous service  times, so as to efficiently detect anomalies at an early stage, soon after they occur.  Such approach is illustrated in Section~\ref{sec:2_anomaly}, indicating how  the number of servers $K$ impacts classification  performance metrics such as accuracy.
\end{enumerate}

\section{Analytical Model} 
\label{sec:2_model}

In this section, we describe the proposed analytical model, which extends results by Psounis \emph{et al.}~\cite{psounis2005systems}.  The model aims at determining the impact of different system characteristics on the average and standard deviation of the response time.   
In particular, we account for the impact of 
the number of servers, the service time of regular jobs, and  the service time of jobs  during an attack period.

\begin{table}[t]
\center
\caption{Table of notation}
\begin{tabular}{p{0.1\textwidth}p{0.4\textwidth}p{0.2\textwidth}}
\hline
Variable & Description  & Comment \\
\hline
\hline
    $\alpha$ & \multicolumn{2}{l}{fraction of regular jobs (short jobs)} \\
    $\lambda_S$ & arrival rate of short jobs & $\alpha=\lambda_S/\lambda$  \\
    $\lambda_L$ & arrival rate of long jobs & $1-\alpha = \lambda_L/\lambda$ \\
    $\lambda$ & arrival rate of jobs & 
    $\lambda=\lambda_S + \lambda_L$  \\
\hline
\multicolumn{3}{c}{single server metrics} \\
\hline
    $E(X)$ &   service time in single server system & unit: millisecond (ms) \\
    $E(X_S)$ &   short job service time  &  $E(X)= \alpha E(X_S) +$ \\
    $E(X_L)$ &   long job service time  &  $\quad +(1-\alpha)E(X_L)$\\
    $\mu$ &  service rate  of single server system & $\mu=1/E(X)$ \\
    $\rho$ & utilization & $\rho = \rho_S + \rho_L = \lambda E(X)$ \\
\hline
\multicolumn{3}{c}{multi-server metrics} \\
\hline
    $\mu_K$ & service rate per server in multi-server system & $\mu_K= \mu/K$ \\
\hline
    $T$ & \multicolumn{2}{l}{metric of interest: response time} \\
\hline
    $E(T)$ & mean response time &  unit: ms \\
    $\sigma^2=V(T)$ & variance of response time  & unit: ms$^2$ \\
\hline
    $K^*_{\mu}$ & number of servers that minimizes $E(T)$ & $K^*_{\mu} = \textrm{argmin}_K E(T)$ \\ 
    $K^*_{\sigma}$ & number of servers that minimizes $\sigma$ & $K^*_{\sigma} = \textrm{argmin}_K \sqrt{V(T)}$ \\ 
    ${\mu}^*$ & minimum value of $E(T)$ & ${\mu}^* = \min_K E(T)$ \\ 
    ${\sigma}^*$ & minimum value of $\sigma$ & ${\sigma}^* = \min_K \sqrt{V(T)}$ \\
\hline
\end{tabular} \label{tab:2notation}
\end{table}

\subsection{Model description}

We consider arrivals that occur according to a Poisson process   to a queue  served by  $K$ servers.  When a service completes, the corresponding  server starts working on the next job at the head of the line.     Under the Kendall notation, such system is referred to as   $M/G/K$, where $M$ stands for arrivals that occur according to a  memoryless arrival process, and $G$ indicates that they are served by a server with  general service time distribution~\cite{harchol2013performance, menasce2002capacity}.
Although the $M/G/K$ is one of the simplest models to capture the essence of  a FIFO multi-server cloud service, most of its performance metrics are still not known in closed form,  and understanding their behavior remains an open problem (\cite[Figure 1.7]{harchol2013performance},~\cite{gupta2010inapproximability} and~\cite[Section 7.4]{grosof2018srpt}).

\begin{table*}
\center
\caption{Table contrasting the performance model from~\cite{psounis2005systems} with its security instantiation introduced in this paper} \label{tab:contrast}
\begin{tabular}{p{0.3\textwidth}p{0.3\textwidth}p{0.3\textwidth}}
\hline 
& performance model & security model \\
\hline
\hline
long jobs (highly demanding service)  & long tasks & attacks/anomalies \\
\hline
 utilization, $\rho$ & accounts for regular and long tasks & accounts for regular tasks and attacks \\
\hline
probability of anomalous service time, $1-\alpha$ & probability of long service time  & probability of arrival  being due to an attack \\ 
\hline
standard deviation of service time  & important for provisioning purposes  & important for attack detection purposes \\
\hline
optimal number of servers & reduces mean response time & must account for standard deviation of response time  \\   
\hline
\end{tabular}
\end{table*}

Let $\lambda$ be the request arrival rate.
Such requests may  correspond to typical user requests or attacks.  

\begin{figure}[b]
    \centering
    \includegraphics[width=0.7\columnwidth]{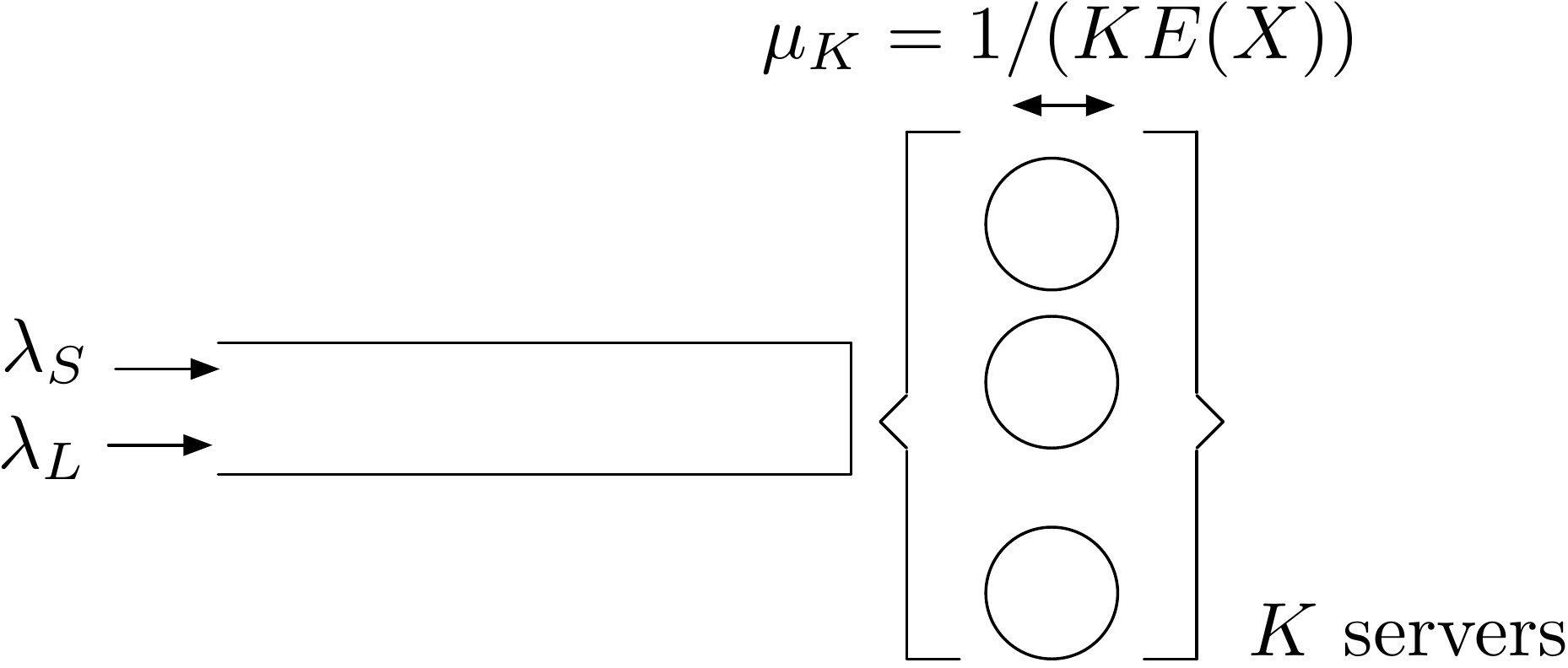}
    \caption{$M/G/K$ model: 
     the proposed analytical model is used to assess the impact of the number of servers $K$ on  standard deviation and mean response times.} 
    \label{fig:mgkmodel}
\end{figure}

To characterize the service times of jobs, we begin by considering a single server system.
We denote by $E(X)$ the mean service time of requests, assuming a single server system.  
The mean service times of  user requests and attacks are given by $E(X_S)$ and $E(X_L)$,  respectively.  The subscripts $S$ and $L$ follow from the assumption that user requests are  short and attacks are long, i.e., $E(X_S) \le E(X_L)$.  
Figure~\ref{fig:mgkmodel} summarizes the key elements of the model. Table~\ref{tab:2notation} summarizes the model parameters and notation. 

Let $\lambda_S$ and $\lambda_L$ correspond to the arrival rate of regular jobs (short jobs) and attacks (long jobs), respectively. Let $\alpha$ be the fraction of requests corresponding to regular traffic,
\begin{equation}
    \lambda = \lambda_S+ \lambda_L, \quad
    \alpha = {\lambda_S}/{\lambda}.
\end{equation}
Then,
\begin{equation}
E(X) = \alpha E(X_S) + (1-\alpha) E(X_L).
\end{equation}
We denote by $\rho$ the   utilization of the single server system, and by $\rho_S$ and $\rho_L$ the utilization due to short and long jobs. Then,
\begin{equation}
\rho = \rho_S + \rho_L, \quad
\rho_S = \lambda_S E(X_S), \quad  \rho_L = \lambda_L E(X_L). \label{eq:rhoxsxl}
\end{equation}

The $M/G/K$ queue is used to characterize response times, i.e., request sojourn times.  We denote by $\mu$ the mean service capacity (service rate) of a single server system, measured in jobs served per time unit,
\begin{equation}
\mu = 1/E(X).
\end{equation}
We denote by
$\mu_K$  the mean service capacity of each server when there are $K$ servers in the system.  
Note that whereas $\mu$, $E(X_S)$, $E(X_L)$ and $E(X)$ are functionally independent of $K$, $\mu_K$ is a s function of $K$.

\subsubsection{Scaling law}
In the remainder of this paper, we consider a   fixed budget to be allocated in the deployment of multiple servers.  As the number of servers grows, the constant system capacity $\mu$ is subdivided into $K$ servers of capacity $\mu_K$. 
In particular, we assume that the total system capacity is uniformly distributed among servers,
\begin{equation}
\mu_K = \frac{\mu}{K}= \frac{1}{K E(X)}.     
\end{equation}
As the number of servers $K$ grows,   the mean service time per server  linearly increases as $1/\mu_K=KE(X)$.

\subsubsection{Response time} \label{sec:2_responsetime}
We denote by $T$ the system response time, our key metric of interest. 
We  leverage results by Psounis \emph{et al.}~\cite{psounis2005systems} to approximate $E(T)$. In essence, $E(T)$ is comprised of two components: $(i)$ the time in the server, $KE(X)$, and $(ii)$ for \emph{blocked} jobs that have to wait in line, the  time in the waiting queue which is approximated as the waiting time of an $M/G/1$ queue, given by ${\rho} E(X_r) /({1-\rho})$, where $E(X_r)$ is the residual service time.     Then,
\begin{align}
E(T) & = K E(X) +  p_B(K,\rho_S,\rho_L) E(W|W>0),  \label{eq:et1}
\end{align}
where $E(W)$ is the mean time in the waiting queue, or simply \emph{waiting time},  and  $p_B$ is the \emph{blocking} probability.
 To obtain $E(W|W>0)$, Psounis \emph{et al.}~\cite{psounis2005systems} propose to consider the   $M/G/1$ queue as a proxy for the $M/G/K$, indicating the accuracy of the following approximation,
\begin{align}
E(W|W>0) & \approx \frac{\rho}{1-\rho} E(X_r) =  \frac{\rho}{1-\rho} \frac{E(X^2)}{2 E(X)},  \end{align}
where $E(X_r)$ is the residual life of the service time.  
We refer the reader to~\cite{psounis2005systems} for further details, including the derivation of the blocking probability. 
Similarly,
\begin{align}
V(T) & =   K^2 V(X) +  V(W|W>0)  p_B(K,\rho_S,\rho_L) \label{eq:et21a} \\ 
& \approx K^2 E(X^2) + \frac{\rho}{1-\rho} \frac{E(X^3)}{3 E(X)} p_B(K,\rho_S,\rho_L),   \label{eq:et21} \\
\sigma(T) &= \sqrt{V(T)} 
\end{align}
where~\eqref{eq:et21} follows from~\eqref{eq:et21a} when the service time of jobs under attack is very long compared {against} the service time of short jobs. In that case, as argued by Psounis \emph{et al.}~\cite[Section 3.4]{psounis2005systems}, 
we  have  
$V(T) \approx E(T^2)$.

Finally, 
throughout this paper, in our analytical results we also assume that service times of short and long jobs are roughly deterministic,
\begin{equation}
       E(X^i) \approx \alpha E(X_S)^i + (1-\alpha) E(X_L)^i.    
\end{equation}

The simplifying assumptions above serve to produce plots that are easy to visualize without noise that is inherent to simulations or more refined approximations (contrast  Figures~\ref{fig:response_time_factor_0_5686}  and~\ref{fig:response_time} produced with the approximations above against simulation results shown in Figure~\ref{fig:response_time_simulation_analitical_rho05}).
We conducted a simulation campaign to validate the adopted simplifications, and  verified that they  serve to extract the essence of our results (see Section~\ref{sec:2_validation}).

\subsubsection{Model summary}  We have introduced the  analytical model considered throughout this work:  an $M/G/K$ queue subject to short and long jobs.  Although the $M/G/K$ is very simple, there are no known closed form expressions for the moments of its response time~\cite{grosof2018srpt}, motivating simplifying approximations as suggested by~\cite{psounis2005systems, brumelle1971some}.    In the following sections, we consider the model in its simplest form,  and indicate the insights that it entails with respect to system predictability and security.  


The source code  to reproduce our results  is made available online~\cite{techrep}.
It is quite remarkable that a simple and classic model such as the $M/G/K$ still allows us to derive fundamentally novel insights on the tradeoff between predictability and performance (Section~\ref{sec:2_key}) and that the  insights derived from such simple model already serve to guide practitioners (Section~\ref{sec:2_discussion}).

\subsection{Optimal number of servers}

\label{sec:2_optimalserv}

The analytical model can be used to determine the impact of different system parameters on the mean and variance of response times. In particular, it can be used to assess how $K$, $\rho$, $\alpha$, $E(X_L)$ and $E(X_S)$ impact $V(T)$ and $E(T)$.

The tuning of the   number of servers $K$ must account for  $E(T)$, $V(T)$ and the accuracy of anomaly detectors.  
Table~\ref{tab:contrast} summarizes the key differences between a performance perspective towards finding the optimal value of $K$ and a joint performance-security perspective, wherein both $V(T)$ and $E(T)$ must be taken {into}  account. In essence, a performance perspective should focus on the average response time.  A security perspective  must also account for the response time standard deviation, as it is key for anomaly detection systems to distinguish between normal system performance and system performance under attacks (Figure~\ref{fig:securityscheme}).

\subsection{Key insights}
\label{sec:2_key}

{In this section we}   briefly report  the key insights derived  from the proposed model.
They relate to how the system behaves as the number of servers $K$ increases. 

\textbf{There is a sweet-spot to reduce mean response time}:  when the number of servers is small, all  servers may be
\emph{blocked} by  long jobs, preventing  other jobs to progress.  This favors an increase in the number of servers $K$, as far as the gain due to a reduction in blocking probability  $p_B$ (second term in~\eqref{eq:et1}) is greater than the corresponding increase in the service time $KE(X)$ (first term in~\eqref{eq:et1}). When the number of servers is  large, in contrast, a significant fraction of the servers will be idle, i.e., $p_B\approx 0$, favoring a reduction in the number of servers.   In that case, an increase in  $K$ will cause an increase in service time $KE(X)$ but  produce marginal gains with respect to the blocking probability, as the latter is low at first place (second term in~\eqref{eq:et1}).   Together, these two observations imply that there is an optimal number of servers that minimizes  mean response time.\footnote{It is well known that, under the considered scaling laws,  the mean response time of the  $M/M/1$ queueing system is smaller than $M/M/K$, for any $K \ge 2$, irrespectively of whether one considers a single queue or multiple queues~\cite{kleinrock1976queueing}.  Clearly, when considering more general service times the optimality of the single server system may not  hold. }

 \textbf{Reduction in number of servers favors increased response time predictability:}
 we found 
that for all the considered setups there is an optimal number of servers that minimizes the response time variance of a typical job.  
 For all the considered scenarios except one, the number of servers that minimizes response time variance is  smaller than or equal to  the one that minimizes the response time mean. Therefore,  it may be worth slightly sacrificing mean response time to reduce response time variance.

 \textbf{Reduction in number of servers favors early detection of attacks based on job response times:}
the variability in response times is usually 
a key component that impacts the detection of attacks---reducing variance builds security.  In summary, our model suggests that a system with more servers is typically more scalable, but less robust against attacks, presenting higher variability in response times (see Section~\ref{sec:2_anomaly}).

The formal analysis of the $M/G/K$ system turns out to be a daunting task.  For this reason, most of the previous works resorted to approximations and numerical analysis to get insights from the model~\cite{psounis2005systems,wierman2006many}. In this paper, we take a similar approach and leave the formal proof of the   conjectures implied by the insights above as a subject for future work.

\section{Model parameterization:  lab experiments}
\label{sec:2_modelparam}

The parameterization of the proposed model involves the determination of the parameters presented in Table~\ref{tab:2notation}. In this paper, we vary $K$, $\lambda$, $\alpha$, and $E(X_S)/E(X_L)$ to illustrate their impact on the metrics of interest, which are the mean and the standard deviation of the response times (Section~\ref{sec:2_findings}) as well as the accuracy of mechanisms of  anomaly detection (Section~\ref{sec:2_anomaly}). 

We run controlled experiments to assess the values of $E(X_S)$ and $E(X_L)$ under known attacks such as those produced by the Mirai botnet~\cite{kambourakis2017mirai,antonakakis2017understanding}. 
%
%
%
Different events may cause an increase in response time.  In this paper, we focus on security threats, i.e., anomalies due to Mirai botnets.  Nonetheless, the framework is general, and can encompass both endogenous and exogenous attack vectors~\cite{laszka2012survey}.  Future work consists in extending the  experiments to  account for a wider set of attack vectors. 

\subsection{Testbed setup}
{Next, we} describe  the experimental setup considered in our testbed to assess the values of $E(X_S)$ and $E(X_L)$ under a typical Mirai attack.
Our testbed comprised three machines, used as $(i)$~a web-server (running the Sock Shop sample application), referred to as \texttt{sockshop},\footnote{\label{ftn:demo}\url{https://github.com/microservices-demo/microservices-demo}} $(ii)$~a load generator (to generate load to the web-server)~\cite{avritzer2018}, and $(iii)$~a Mirai bot.  The three machines were connected to a Cisco switch, through a 100 Mb/s full duplex network.

We summarize the key aspects of our experimental setup as follows. 
  Three physical machines were connected to the same Ethernet link: 
    $(i)$ the first machine  was set as a web server, working as a virtual store (\texttt{sockshop}), over the \emph{Docker} platform;
    $(ii)$ the second machine  was set as a baseline load generator, using \emph{Apache JMeter} with 10 threads and issuing 10 requests per second;
    $(iii)$ the third machine was set as the attacker, running \emph{real malware} code from the \emph{Mirai} botnet (Section~\ref{sec:2_mirai}).  Its goal is to increase the system load.  
    From the
publicly available Mirai source code, we 
extracted a software module to conduct Mirai  attacks to a given target, for a configurable time.  Using that module, the machine launched a Mirai DDoS attack,  executing 256 threads  and using the GET method to  overload  the web  server with small HTTP requests. 

\begin{figure*}[t!]
    \begin{center}
    \begin{tabular}{@{\hspace{-1ex}}c @{\hspace{-3ex}}c@{\hspace{-3ex}}c}
        \includegraphics[width=0.37\textwidth]{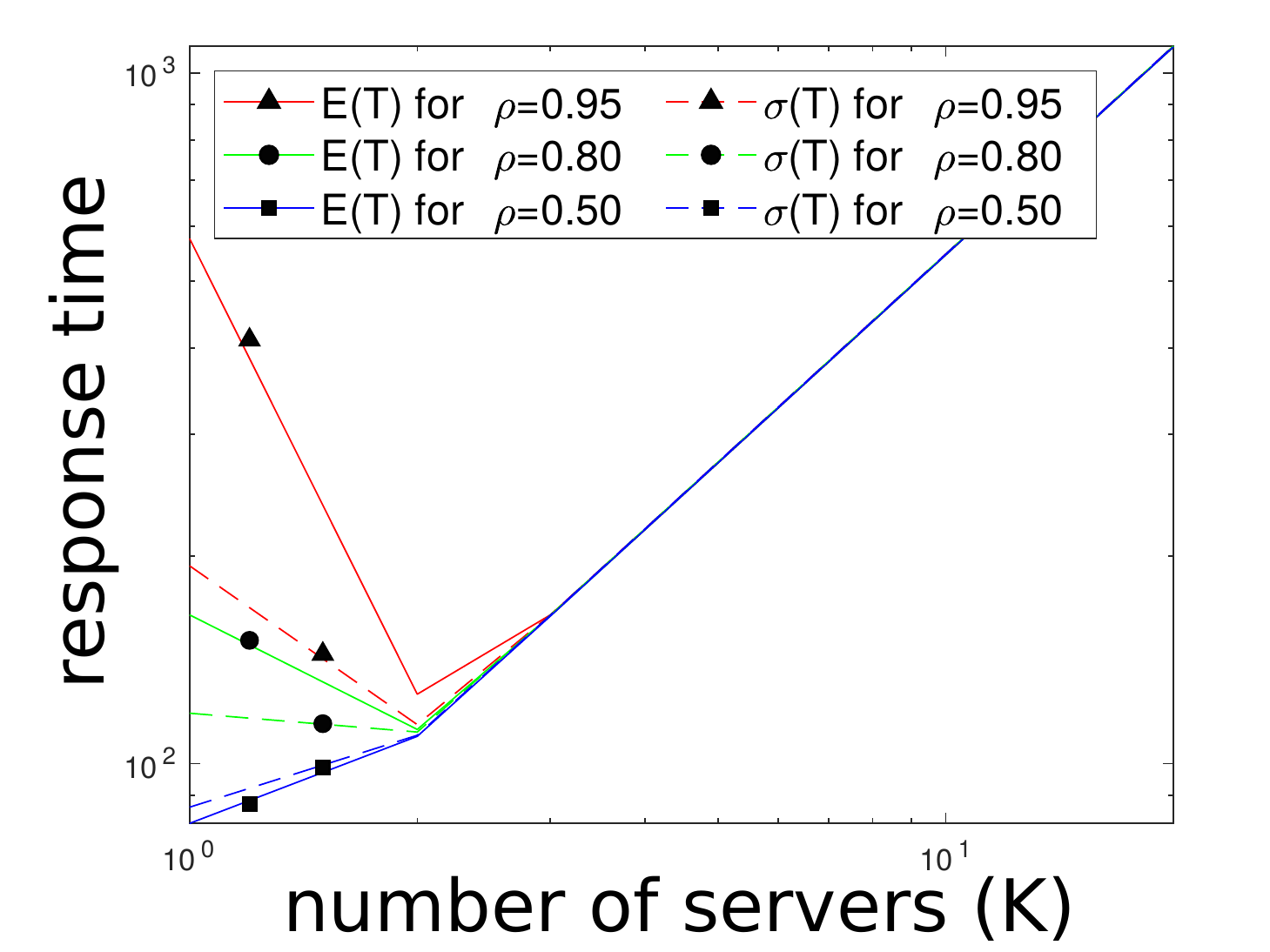}&
        \includegraphics[width=0.37\textwidth]{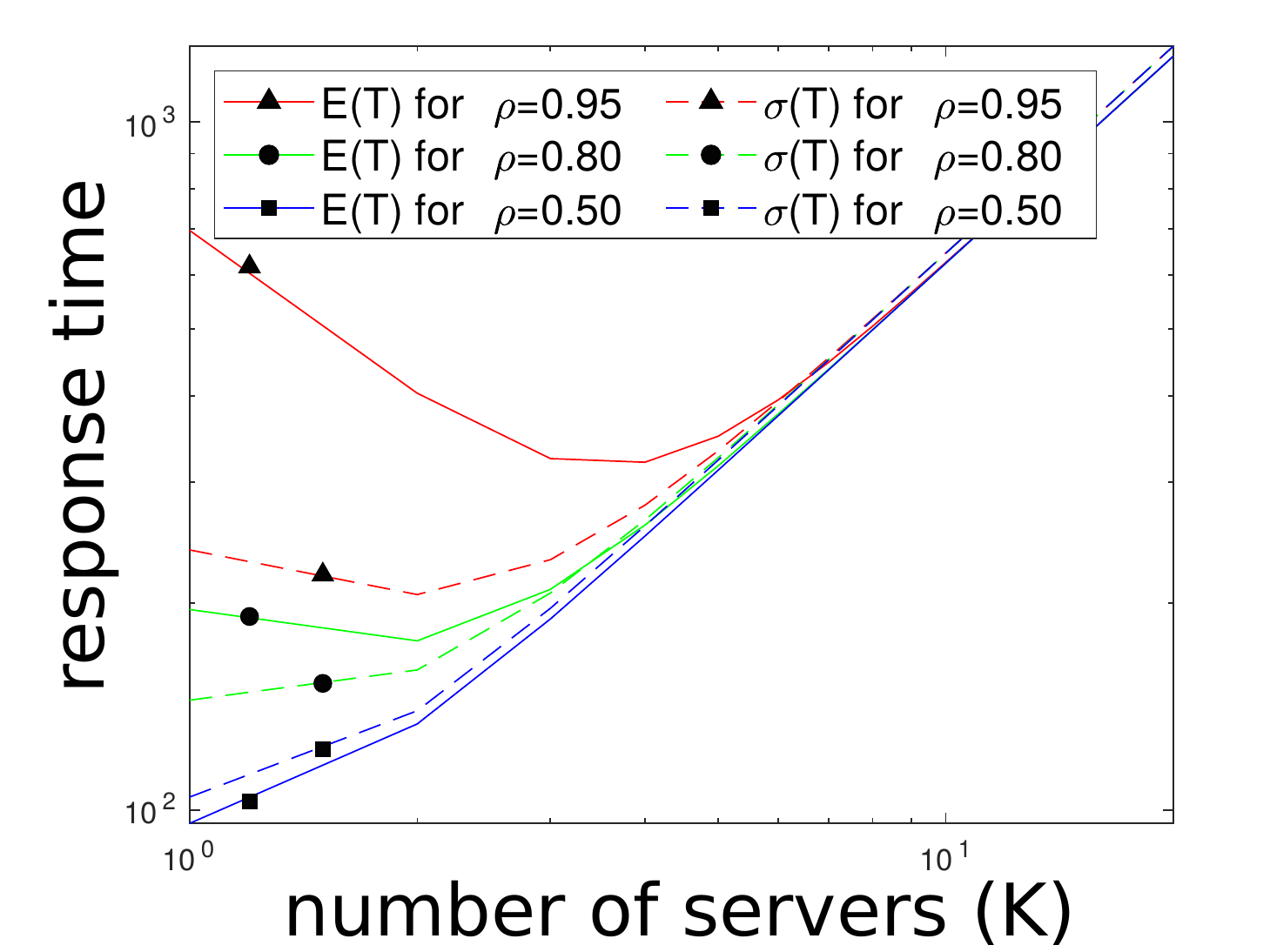}&
        \includegraphics[width=0.37\textwidth]{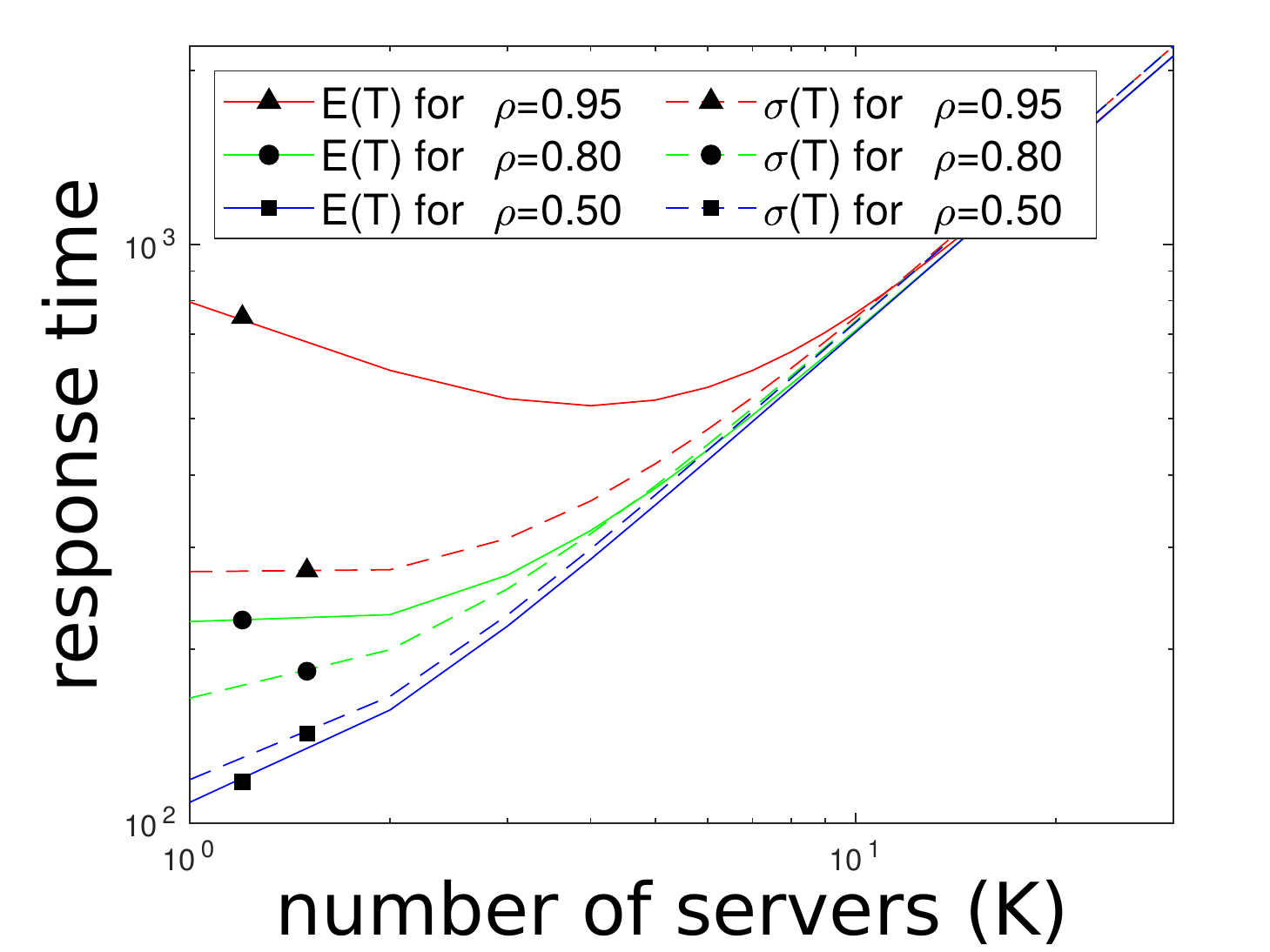}\\
        (a) $\alpha=0.99$ & (b) $\alpha=0.80$ & (c) $\alpha=0.60$
    \end{tabular} 
    \end{center} \vspace{-0.1in}
    \caption{Assessing the impact of $\alpha$ on response times, given  ratio $E(X_S)/E(X_L)=0.5686$ as obtained from  experiments (Section~\ref{sec:2_modelparam}). }
    \label{fig:response_time_factor_0_5686}
    
\end{figure*}

\begin{table}
\center
\caption{Empirical results on response times } 
\label{table:empirical-results}
\begin{tabular}{|c|c|} \hline
\multicolumn{2}{|c|}{Without anomaly}  \\
 Avg. response time  & Std. deviation of response time \\
54.13 ms & 39.23 ms \\ \hline \hline
 \multicolumn{2}{|c|}{With anomaly} \\
  Avg. response time  & Std. deviation of response time \\ \hline 
 95.20 ms & 83.72 ms \\
 \hline
\end{tabular} \label{tab:experiment_ufrj}
\end{table}

\subsection{Testbed results}
\label{sec:2_testbedresults}

We repeated  each of our experiments for five runs.   
When considering a setup without attacks, standard workload starts to be generated at time zero.  The mean and standard deviation of response times are continuously updated throughout the experiment.  Then, we recorded the   accumulated mean and standard deviation of  response times as observed at the end of the experiment, i.e., after 10 minutes.
When considering attacks, standard workload also starts to be generated at time zero. We wait for a warm-up of 3 minutes, before starting a 5-minute attack. 
The mean and standard deviation of response times are continuously updated throughout the experiment, and we report the accumulated measurements   collected  7~minutes after the experiment started.

In our baseline lab experiments without anomalies, under the setup described in the previous section, we found that the  response time mean  and the standard deviation  equal to $54.13$~ms and $39.23$~ms, respectively. With anomalies, those values increased to $95.20$~ms and $83.72$~ms, respectively.   



{In} our baseline setup the system utilization  was low,  implying a  negligible  idle  waiting time and blocking probability.  Hence, we assume that the response times measured in our experiments without attacks  correspond to regular service times (see~\eqref{eq:et1}).  We further assume that the service time of jobs corresponding to attacks is given by the response time  of  jobs under attacks, as measured in our experiments.   Then,  it  follows 
that 
\begin{align}
    E(X_S) = 54.13,   \quad  E(X_L) = 95.20. 
\end{align}    
All values above are measured in milliseconds (Table~\ref{tab:experiment_ufrj}).  
In what follows, such values are used   to estimate the average and the standard deviation of the response times, $E(T)$ and $\sigma(T)=\sqrt{V(T)}$, under various scenarios of interest. 

\begin{figure*}[ht!]
    \begin{center} \vspace{-0.1in}
    \begin{tabular}{@{\hspace{-1ex}}c @{\hspace{-3ex}}c@{\hspace{-3ex}}c}
        \includegraphics[width=0.37\textwidth]{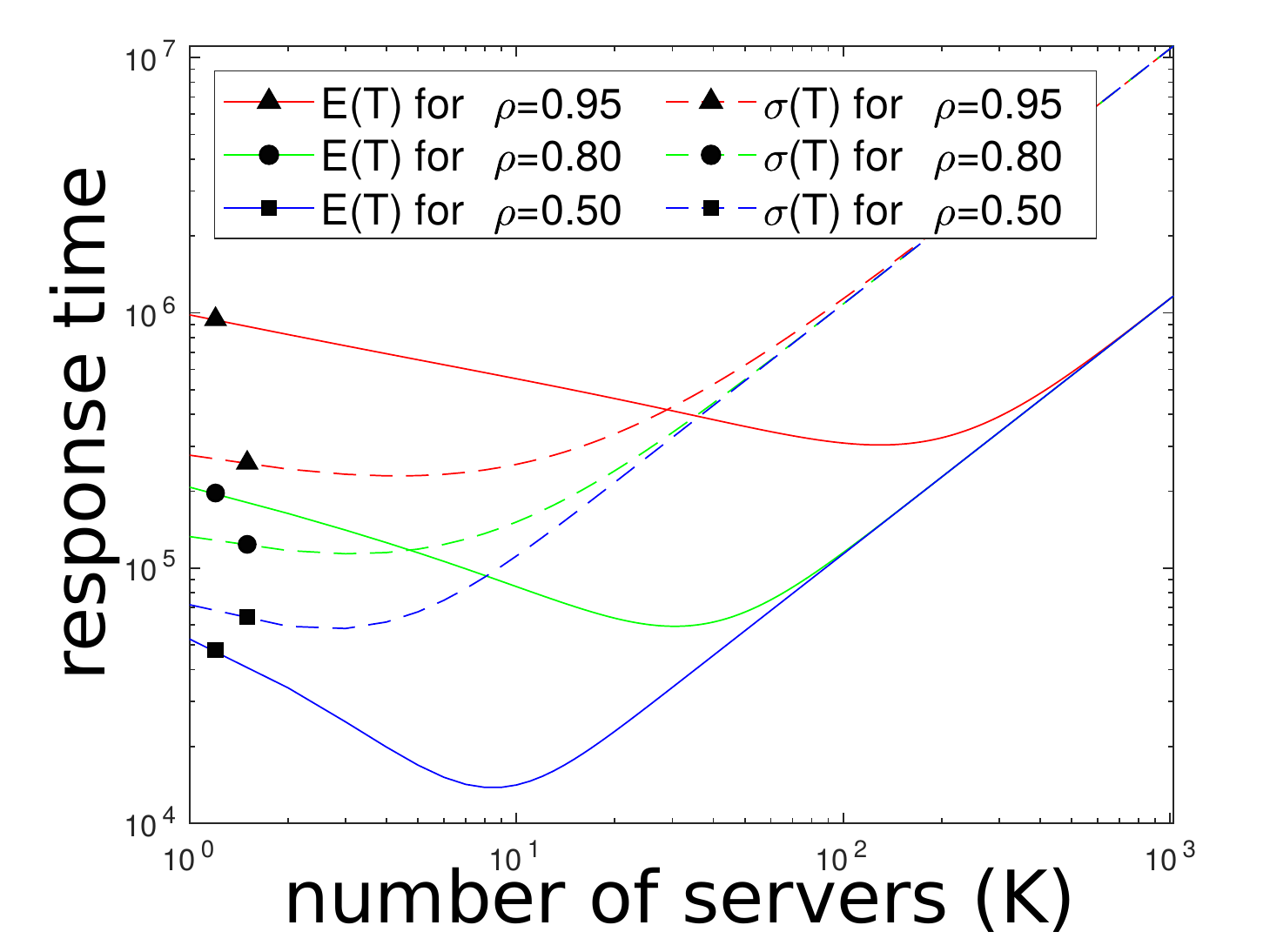}&
        \includegraphics[width=0.37\textwidth]{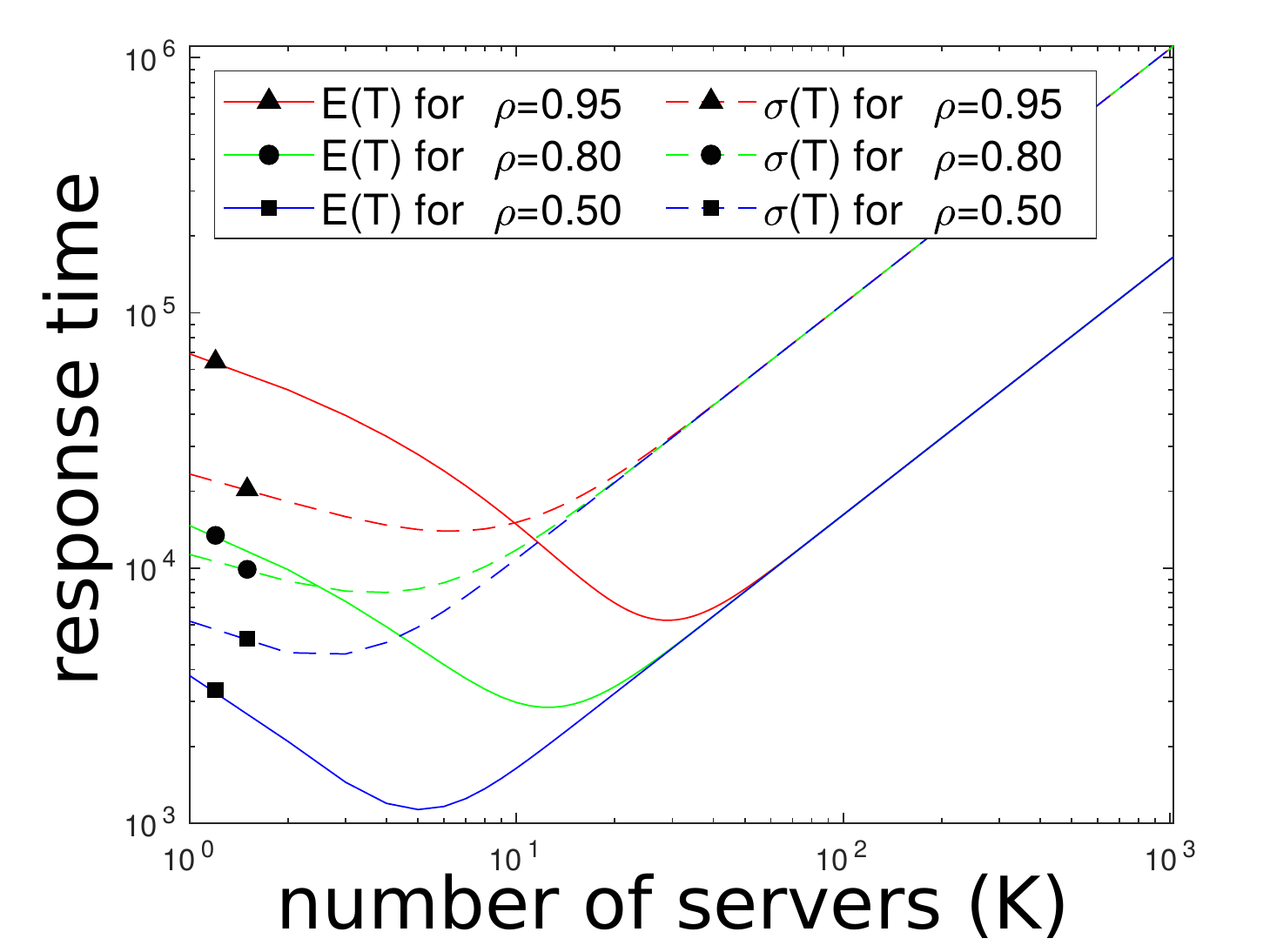} &
        \includegraphics[width=0.37\textwidth]{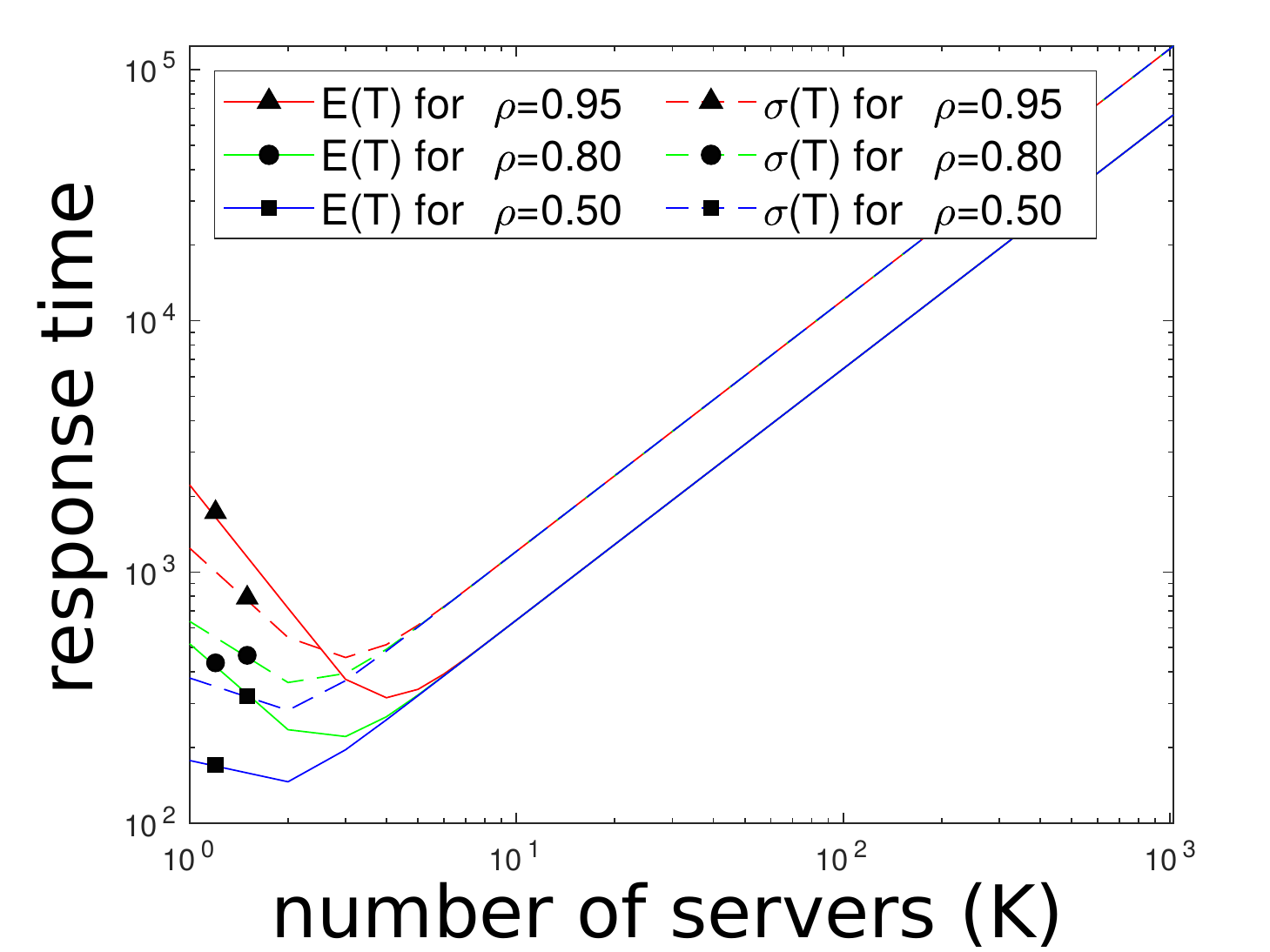} \\
        (a) $E(X_S)/E(X_L)=0.0005$ & (b) $ E(X_S)/E(X_L)=0.005$ & (c) $ E(X_S)/E(X_L)=0.05$
    \end{tabular}
    \end{center} \vspace{-0.1in}
    \caption{ Assessing the impact of $E(X_S)/E(X_L)$ on response times, given $\alpha=0.99$. 
    The number of servers that reduces the standard deviation of response times is typically smaller than or equal to the number of servers that reduces mean response times, indicating that to simplify anomaly detection one must trade between security and performance. }
    \label{fig:response_time} \vspace{-0.1in}
\end{figure*}

In summary, 
\begin{enumerate}
    \item given $E(X_L)$ and $E(X_S)$ obtained from experiments, with and without Mirai, we parametrize $\rho$ (equation~\eqref{eq:rhoxsxl});
    \item the resulting value of $\rho$ is used to compute $E(T)$ and $V(T)$ as a function of $\alpha$ and $K$ (equations~\eqref{eq:et1} and~\eqref{eq:et21a});
    \item varying $\alpha$ and  the number of servers $K$, we study its impact on the metrics of interest (Figure~\ref{fig:response_time_factor_0_5686});
    \item in addition, we also vary $E(X_L)$ around its  experimental estimate  to conduct what-if counterfactual model-based analysis accounting for  different attack vectors that may have different impact on $E(X_L)$ (Figure~\ref{fig:response_time}).
\end{enumerate}
The steps above are illustrated in the following section.

\section{Findings from analytical model}
\label{sec:2_findings}

Next, we assess the impact of the infrastructure setup on the performance and security of the system. In particular, we focus on the impact of the number of servers $K$ on the mean and standard deviation of the response time: 
 what are the recommended values of $K$ when the utilization, $\rho$, and  the fraction of load that is due to attacks, $1-\alpha$, are taken into account?

\subsection{Numerical results}
Figures~\ref{fig:response_time_factor_0_5686} and~\ref{fig:response_time}, together with  Tables~\ref{tab:optimal} and~\ref{tab:optimalnumbig}, respectively, 
present the numerical results obtained with the analytical model.  
Solid (resp., dashed) lines in Figures~\ref{fig:response_time_factor_0_5686} and~\ref{fig:response_time}  show the average (resp., standard deviation) of the response time as a function of the number of servers in the system, $K$ (see~\eqref{eq:et1} and~\eqref{eq:et21a}).

Figure~\ref{fig:response_time_factor_0_5686} {shows} results assuming $E(X_S)/E(X_L)=0.5686$ as estimated by our experimental results (see Section~\ref{sec:2_testbedresults}).  
Each subplot, from left to right, corresponds to increasing fraction of jobs due to anomalies, with $1-\alpha$ varying between  $0.01$, $0.2$ and $0.4$. 
In  Figure~\ref{fig:response_time}, we maintain $\alpha=0.99$ and  $E(X_S)=54.13$, and vary $E(X_L)$ so as to reach a target $E(X_S)/E(X_L)$.

\subsection{Reducing number of servers favors variance reduction}

In most of the scenarios considered, the value of $K$ that minimizes the standard deviation of the response time is  smaller than or equal to the value of $K$ that minimizes  mean response times (see Figures~\ref{fig:response_time_factor_0_5686} and~\ref{fig:response_time}, and Tables~\ref{tab:optimal} and~\ref{tab:optimalnumbig}).
This suggests that  one must trade between  performance (response time mean) and predictability (response time variance).   
A notable exception consists of the setup 
with $\rho=0.5$, for which the value of $K$ that minimizes response time variance (resp., mean)  is 2 (resp., 1), as shown in Table~\ref{tab:optimalnumbig}. 

Smaller values of $K$ make the system more predictable, at the expense of an increase in mean response times. As the fraction of anomalous jobs increases, i.e., as $\alpha$ decreases, we observe that to minimize the standard deviation of the response time one needs to consider a smaller number of servers.  In particular, for $\alpha=0.6$ the standard deviation of response times is minimal when  $K=1$ for all the considered scenarios (see Tables~\ref{tab:optimal} and~\ref{tab:optimalnumbig}).

\subsection{Reducing number of  servers may be  beneficial under attacks}

\subsubsection{Impact of fraction of anomalous jobs}
Next, we consider the impact of the fraction of anomalous jobs, $1-\alpha$, on response times. 
Under the model parameterized by our experiments (Figure~\ref{fig:response_time_factor_0_5686} and Table~\ref{tab:optimal}), the fraction of anomalous jobs $1-\alpha$ did not have a significant impact on the optimal number of servers. 
In the additional  scenarios considered in 
Table~\ref{tab:optimalnumbig},   we observe that   as  $1-\alpha$ increases the optimal number of servers decreases, irrespectively of whether we account for the mean or the variance as our decision criterion.  This is because as $1-\alpha$ increases, the potential benefits of increasing the number of servers to reduce blocking probability diminish, as the additional servers will more likely be  occupied by long jobs.

\subsubsection{Impact of utilization} As the utilization increases, the optimal number of servers consistently increases, accounting either for the minimization of the mean or the standard deviation of the response time (see   
Figures~\ref{fig:response_time_factor_0_5686} and~\ref{fig:response_time}, and Tables~\ref{tab:optimal} and~\ref{tab:optimalnumbig}).
In particular, for low utilization ($\rho=0.5$), when $\alpha \in \{0.8, 0.6\}$ the optimal number of servers equals 1 or 2 under all the considered scenarios (Figures~\ref{fig:response_time_factor_0_5686} and~\ref{fig:response_time} and Table~\ref{tab:optimalnumbig}).

\begin{table}[b]
    \centering
        \caption{Optimal number of servers minimizing average and standard deviation of response time  
        (setup of  Fig.~\ref{fig:response_time_factor_0_5686},   $E(X_S)/E(X_L)=0.5686$).}
\begin{tabular}{|c||c|c||c|c||c|c|}
\hline 
{Setup} & \multicolumn{2}{c||}{$\rho=0.5$}  & \multicolumn{2}{c||}{$\rho=0.8$} &  \multicolumn{2}{c|}{$\rho=0.95$} 
\\ 
 Fig.~\ref{fig:response_time_factor_0_5686} 	&  $K^{\star}_\mu$  	& $K^{\star}_\sigma$  &    $K^{\star}_\mu$  	& $K^{\star}_\sigma$  &   $K^{\star}_\mu$  	& $K^{\star}_\sigma$  \\
\hline 
 \ref{fig:response_time_factor_0_5686}a) $\alpha=0.99$ 	&  $  1$ 	& $  1$ & 2 & 2 & 2 &2 \\
\hline 
\ref{fig:response_time_factor_0_5686}b) $\alpha=0.80$	&  $  1$ 	& $  1$ & $2$  & $1$ & $4$ & $2$ 	\\
\hline 
\ref{fig:response_time_factor_0_5686}c) $\alpha=0.60$	&  $  1$ 	& $ 1$ &  $1$&$1$ &$4$ &$1$ \\
\hline
\end{tabular}
\label{tab:optimal}
\end{table}

\begin{table*}[!htb]
        \caption{Optimal number of servers minimizing average and standard deviation of response time   $ (K^{\star}_\mu$ and $K^{\star}_\sigma$, resp.).  The table also reports  the minimum value of the average and standard deviation of response time   $ (\mu^{\star}$ and $\sigma^{\star}$, resp.), and  the metrics for the corresponding M/G/1 system. } 
    \centering
\setlength\tabcolsep{1.5pt}
\begin{tabular}{|c|c|c|c|c|rrr|rrr|r|r|}
\hline
$E_{S}/$ &          &        & \multicolumn{8}{c|}{analytical model given by~\eqref{eq:et1}-\eqref{eq:et21} (simulation)} & \multicolumn{2}{c|}{M/G/1 $(K=1)$}\\
    \cline{4-13}
$E_{L}$  & $\alpha$ & $\rho$ & $K^{\star}_{\mu} = $ & $K^{\star}_{\sigma} =$ & \multicolumn{3}{c|}{$\mu^{\star} = \min_K E(T)$} & \multicolumn{3}{c|}{$\sigma^{\star} = \min_K \sqrt{V(T)}$} & \multicolumn{1}{c|}{$E(T)$} & \multicolumn{1}{c|}{$\sqrt{V(T)}$} \\
&&&  argmin &  argmin & &&&&&&&\\
&&&  $E(T)$ &  $\sqrt{V(T)}$ & &&&&&&&\\
		\hline
		 $0.0005$ 	& $0.99$ 	& $0.95$ 	& $110(100)$ 	& $4(6)$ 	& $305779.58$  ( 	& $673511.58\pm$ & $129216.26)$ 	& $229709.60$  (	& $1057153.18\pm$ & $258041.82)$ 	& 981122.33 	& 276771.84  \\
		\hline
		 $0.0005$ 	& $0.99$ 	& $0.80$ 	& $30(41)$ 	& $3(6)$ 	& $59155.35 $ ( 	& $85381.27\pm$ & $6111.62)$ 	& $113757.70 $ ( 	& $248643.70\pm$ & $11340.91)$ 	& 207449.06 	& 132850.39  \\
		\hline
		 $0.0005$ 	& $0.99$ 	& $0.50$ 	& $8(9)$ 	& $3(3)$ 	& $13832.47 $ ( 	& $15383.40\pm$ & $274.68)$ 	& $57995.10 $ ( 	& $73628.48\pm$ & $1257.95)$ 	& 52714.41 	& 71838.26  \\
		\hline
		\hline
		 $0.0005$ 	& $0.80$ 	& $0.95$ 	& $6(2)$ 	& $1(1)$ 	& $819009.29 $ ( 	& $1076053.40\pm$ & $61577.73)$ 	& $320591.69 $ ( 	& $1171635.82\pm$ & $139011.37)$ 	& 1048113.50 	& 320591.69  \\
		\hline
		 $0.0005$ 	& $0.80$ 	& $0.80$ 	& $3(1)$ 	& $1(1)$ 	& $214297.84 $ ( 	& $235356.55\pm$ & $4113.42)$ 	& $173298.43 $ ( 	& $251482.63\pm$ & $6119.26)$ 	& 237783.34 	& 173298.43  \\
		\hline
		 $0.0005$ 	& $0.80$ 	& $0.50$ 	& $1(1)$ 	& $1(1)$ 	& $75717.31 $ ( 	& $75562.99\pm$ & $505.50)$ 	& $110856.90 $ ( 	& $93023.28\pm$ & $588.99)$ 	& 75717.31 	& 110856.90  \\
		\hline
				\hline
		 $0.0005$ 	& $0.60$ 	& $0.95$ 	& $4(1)$ 	& $1(1)$ 	& $924227.47 $ ( 	& $1051415.46\pm$ & $16675.28)$ 	& $340815.90 $ ( 	& $1033144.64\pm$ & $46522.04)$ 	& 1071036.09 	& 340815.90  \\
		\hline
		 $0.0005$ 	& $0.60$ 	& $0.80$ 	& $2(1)$ 	& $1(1)$ 	& $259296.47 $ ( 	& $261490.71\pm$ & $3549.80)$ 	& $193430.68 $ ( 	& $257243.25\pm$ & $5317.19)$ 	& 259694.29 	& 193430.68  \\
		\hline
		 $0.0005$ 	& $0.60$ 	& $0.50$ 	& $1(1)$ 	& $1(1)$ 	& $97425.93 $ ( 	& $97330.07\pm$ & $134.80)$ 	& $130950.16 $ ( 	& $98112.69\pm$ & $270.21)$ 	& 97425.93 	& 130950.16  \\
		\hline
				\hline
		 $0.0050$ 	& $0.99$ 	& $0.95$ 	& $29(90)$ 	& $6(7)$ 	& $6246.33 $ ( 	& $44824.63\pm$ & $8566.91)$ 	& $13960.01 $ ( 	& $65835.37\pm$ & $13221.24)$ 	& 69126.09 	& 23366.56  \\
		\hline
		 $0.0050$ 	& $0.99$ 	& $0.80$ 	& $12(23)$ 	& $4(6)$ 	& $2850.30 $ ( 	& $6838.74\pm$ & $120.64)$ 	& $8029.51 $ ( 	& $17074.55\pm$ & $251.18)$ 	& 14680.64 	& 11307.91  \\
		\hline
		 $0.0050$ 	& $0.99$ 	& $0.50$ 	& $5(6)$ 	& $3(3)$ 	& $1132.58 $ ( 	& $1496.67\pm$ & $29.86)$ 	& $4608.90 $ ( 	& $5863.73\pm$ & $52.78)$ 	& 3791.55 	& 6195.92  \\
		\hline
				\hline
		 $0.0050$ 	& $0.80$ 	& $0.95$ 	& $7(2)$ 	& $1(1)$ 	& $79372.55 $ ( 	& $99450.78\pm$ & $3766.02)$ 	& $31818.19 $ ( 	& $100033.26\pm$ & $6521.51)$ 	& 103048.98 	& 31818.19  \\
		\hline
		 $0.0050$ 	& $0.80$ 	& $0.80$ 	& $3(1)$ 	& $1(1)$ 	& $21086.50 $ ( 	& $23491.16\pm$ & $304.30)$ 	& $17219.40 $ ( 	& $24731.14\pm$ & $601.46)$ 	& 23438.08 	& 17219.40  \\
		\hline
		 $0.0050$ 	& $0.80$ 	& $0.50$ 	& $1(1)$ 	& $1(1)$ 	& $7515.90 $ ( 	& $7505.94\pm$ & $23.16)$ 	& $11030.59 $ ( 	& $9222.81\pm$ & $26.31)$ 	& 7515.90 	& 11030.59  \\
		\hline
				\hline
		 $0.0050$ 	& $0.60$ 	& $0.95$ 	& $4(1)$ 	& $1(1)$ 	& $91616.80 $ ( 	& $106977.06\pm$ & $4034.39)$ 	& $33990.33 $ ( 	& $107257.79\pm$ & $4833.27)$ 	& 106448.10 	& 33990.33  \\
		\hline
		 $0.0050$ 	& $0.60$ 	& $0.80$ 	& $2(1)$ 	& $1(1)$ 	& $25826.31 $ ( 	& $25898.44\pm$ & $129.70)$ 	& $19301.27 $ ( 	& $25242.58\pm$ & $322.41)$ 	& 25854.50 	& 19301.27  \\
		\hline
		 $0.0050$ 	& $0.60$ 	& $0.50$ 	& $1(1)$ 	& $1(1)$ 	& $9735.78 $ ( 	& $9741.73\pm$ & $25.59)$ 	& $13074.18 $ ( 	& $9760.89\pm$ & $31.85)$ 	& 9735.78 	& 13074.18  \\
		\hline
				\hline
		 $0.0500$ 	& $0.99$ 	& $0.95$ 	& $4(3)$ 	& $3(3)$ 	& $ 316.11 $ ( 	& $2234.81\pm$ & $80.60)$ 	& $ 457.04 $ ( 	& $2472.44\pm$ & $102.55)$ 	& 2220.74 	& 1244.74  \\
		\hline
		 $0.0500$ 	& $0.99$ 	& $0.80$ 	& $3(2)$ 	& $2(3)$ 	& $ 221.60 $ ( 	& $ 503.28\pm$ & $7.76)$ 	& $ 363.42 $ ( 	& $ 662.72\pm$ & $14.19)$ 	&  518.38 	&  636.56  \\
		\hline
		 $0.0500$ 	& $0.99$ 	& $0.50$ 	& $2(1)$ 	& $2(2)$ 	& $ 146.39 $ ( 	& $ 176.64\pm$ & $0.53)$ 	& $ 281.72 $ ( 	& $ 277.23\pm$ & $1.09)$ 	&  177.91 	&  378.74  \\
		\hline
				\hline
		 $0.0500$ 	& $0.80$ 	& $0.95$ 	& $8(1)$ 	& $2(1)$ 	& $5948.63 $ ( 	& $8698.78\pm$ & $367.26)$ 	& $2950.16 $ ( 	& $8540.61\pm$ & $617.28)$ 	& 8916.11 	& 2974.29  \\
		\hline
		 $0.0500$ 	& $0.80$ 	& $0.80$ 	& $2(1)$ 	& $1(1)$ 	& $1861.65 $ ( 	& $2071.57\pm$ & $6.73)$ 	& $1628.01 $ ( 	& $2175.33\pm$ & $31.03)$ 	& 2082.20 	& 1628.01  \\
		\hline
		 $0.0500$ 	& $0.80$ 	& $0.50$ 	& $1(1)$ 	& $1(1)$ 	& $ 715.42 $ ( 	& $ 712.31\pm$ & $1.66)$ 	& $1057.29 $ ( 	& $ 834.95\pm$ & $1.27)$ 	&  715.42 	& 1057.29  \\
		\hline
				\hline
		 $0.0500$ 	& $0.60$ 	& $0.95$ 	& $4(1)$ 	& $1(1)$ 	& $8445.50 $ ( 	& $10174.89\pm$ & $385.45)$ 	& $3313.95 $ ( 	& $10272.73\pm$ & $670.96)$ 	& 10068.56 	& 3313.95  \\
		\hline
		 $0.0500$ 	& $0.60$ 	& $0.80$ 	& $1(1)$ 	& $1(1)$ 	& $2487.21 $ ( 	& $2478.14\pm$ & $16.69)$ 	& $1891.78 $ ( 	& $2403.31\pm$ & $48.88)$ 	& 2487.21 	& 1891.78  \\
		\hline
		 $0.0500$ 	& $0.60$ 	& $0.50$ 	& $1(1)$ 	& $1(1)$ 	& $ 970.94 $ ( 	& $ 970.90\pm$ & $3.27)$ 	& $1288.88 $ ( 	& $ 934.07\pm$ & $3.59)$ 	&  970.94 	& 1288.88  \\
\hline 
\end{tabular}
    \label{tab:optimalnumbig}
\end{table*}

\begin{figure*}[ht!]
\begin{center}
\begin{tabular}{@{\hspace{0.2cm}}c @{\hspace{-3ex}}c@{\hspace{-3ex}}c}
		 \includegraphics[width=0.35\textwidth]{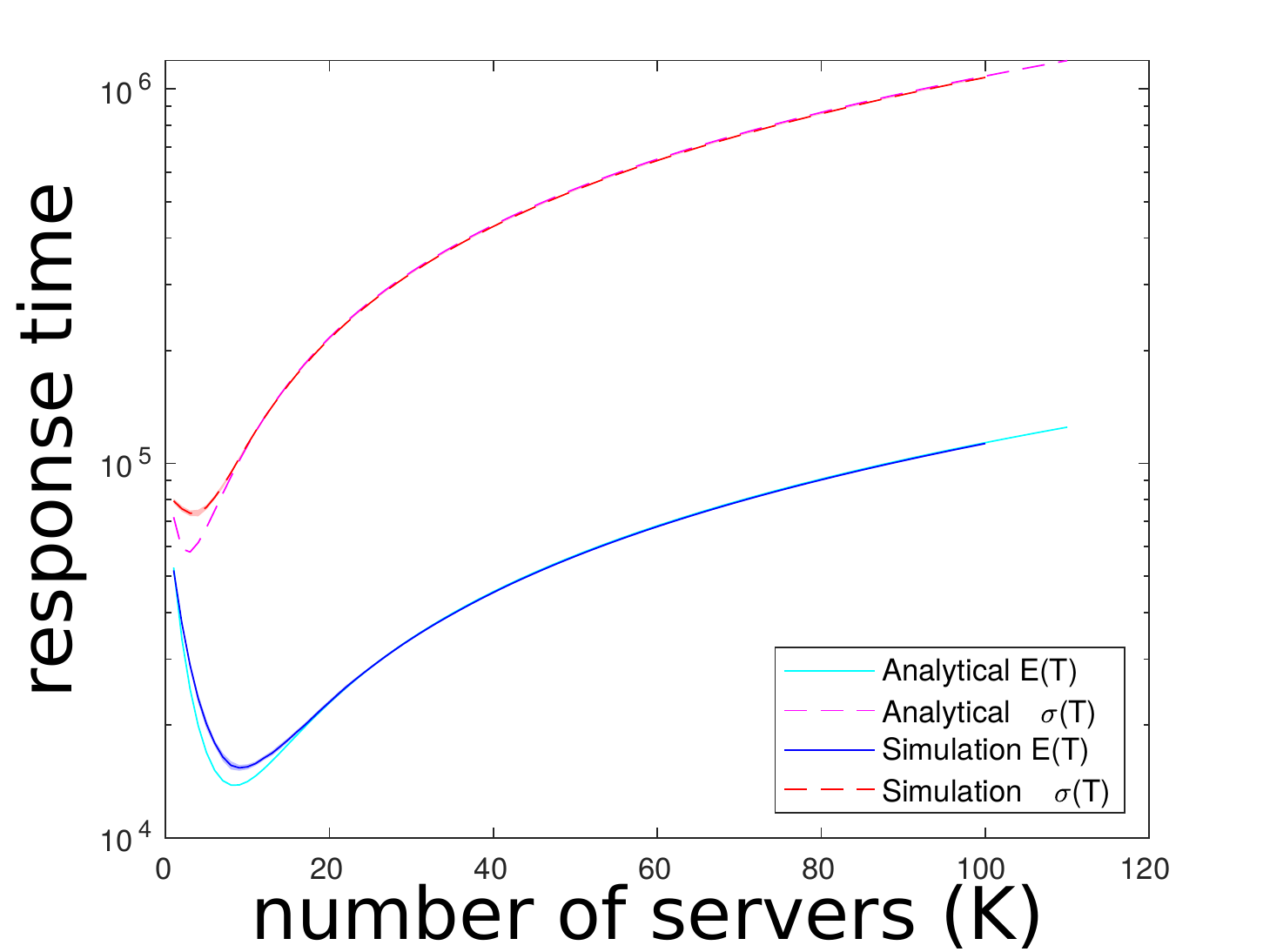}&
		 \includegraphics[width=0.35\textwidth]{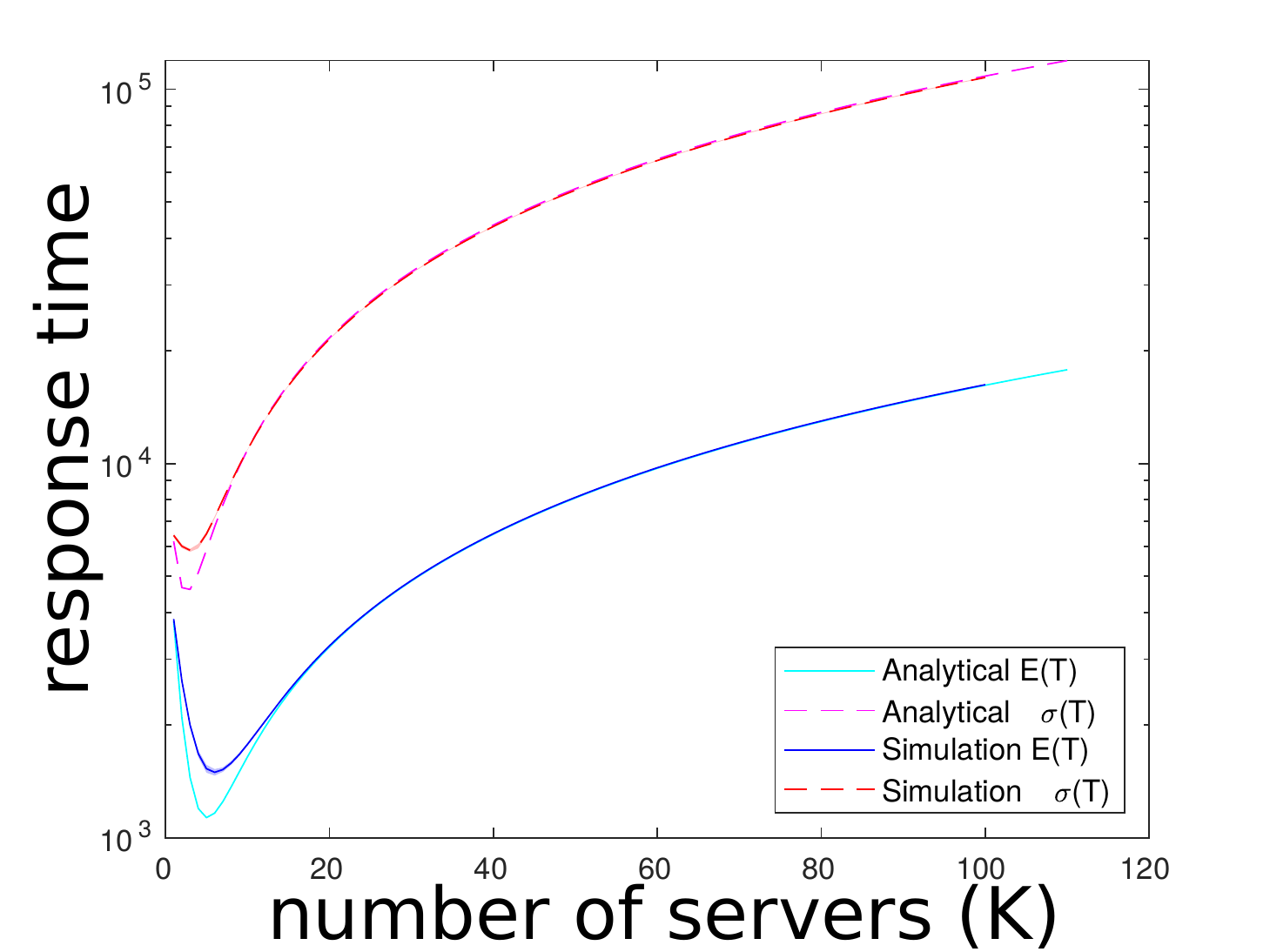}&	\includegraphics[width=0.35\textwidth]{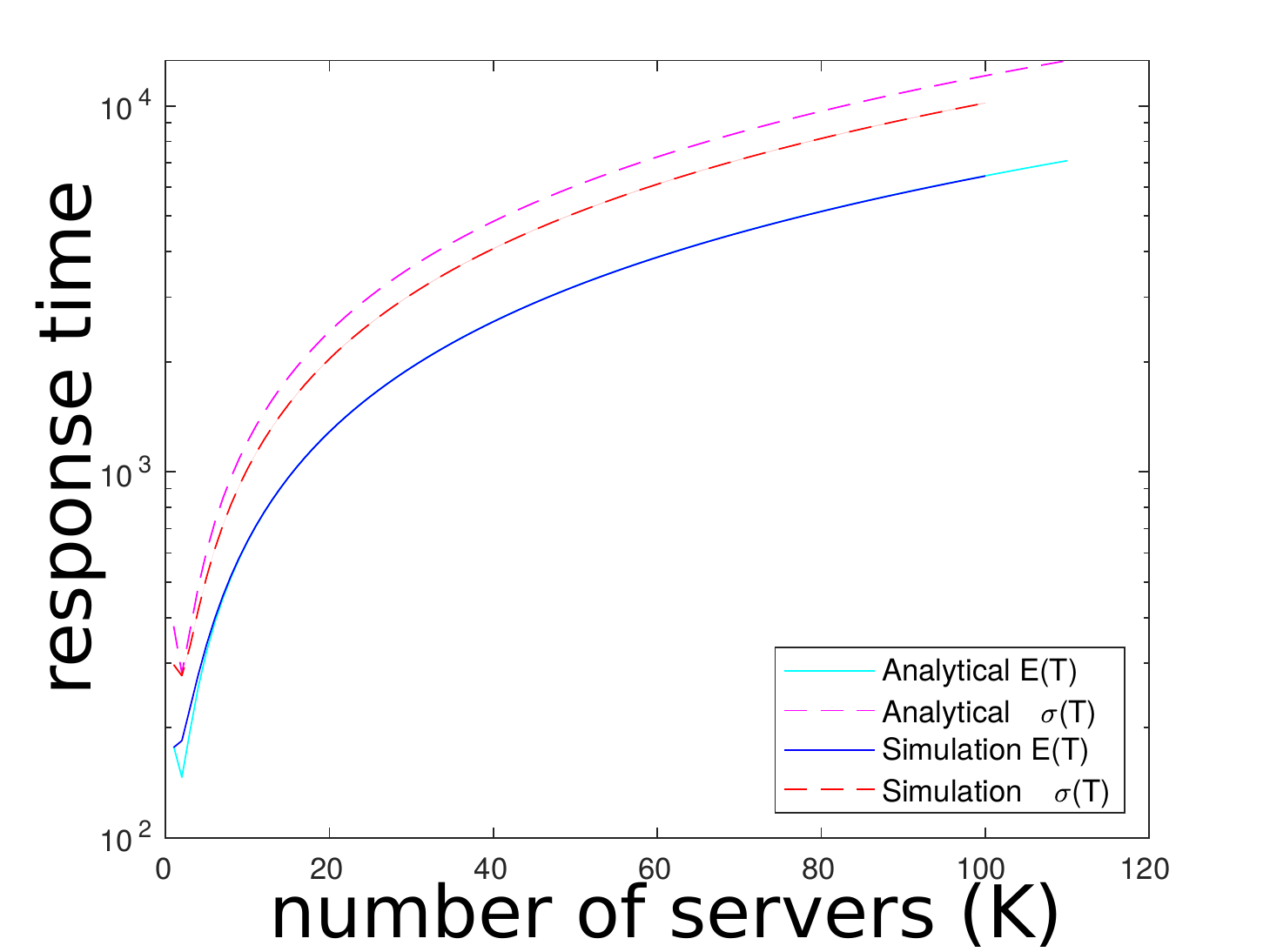}\\
		 (a) $\alpha=0.99$, $\frac{E(X_S)}{E(X_L)}=0.0005$ &		 (b) $\alpha=0.99$, $\frac{E(X_S)}{E(X_L)}=0.0050$&		 
		 (c) $\alpha=0.99$, $\frac{E(X_S)}{E(X_L)}=0.0500$\\
		 \includegraphics[width=0.35\textwidth]{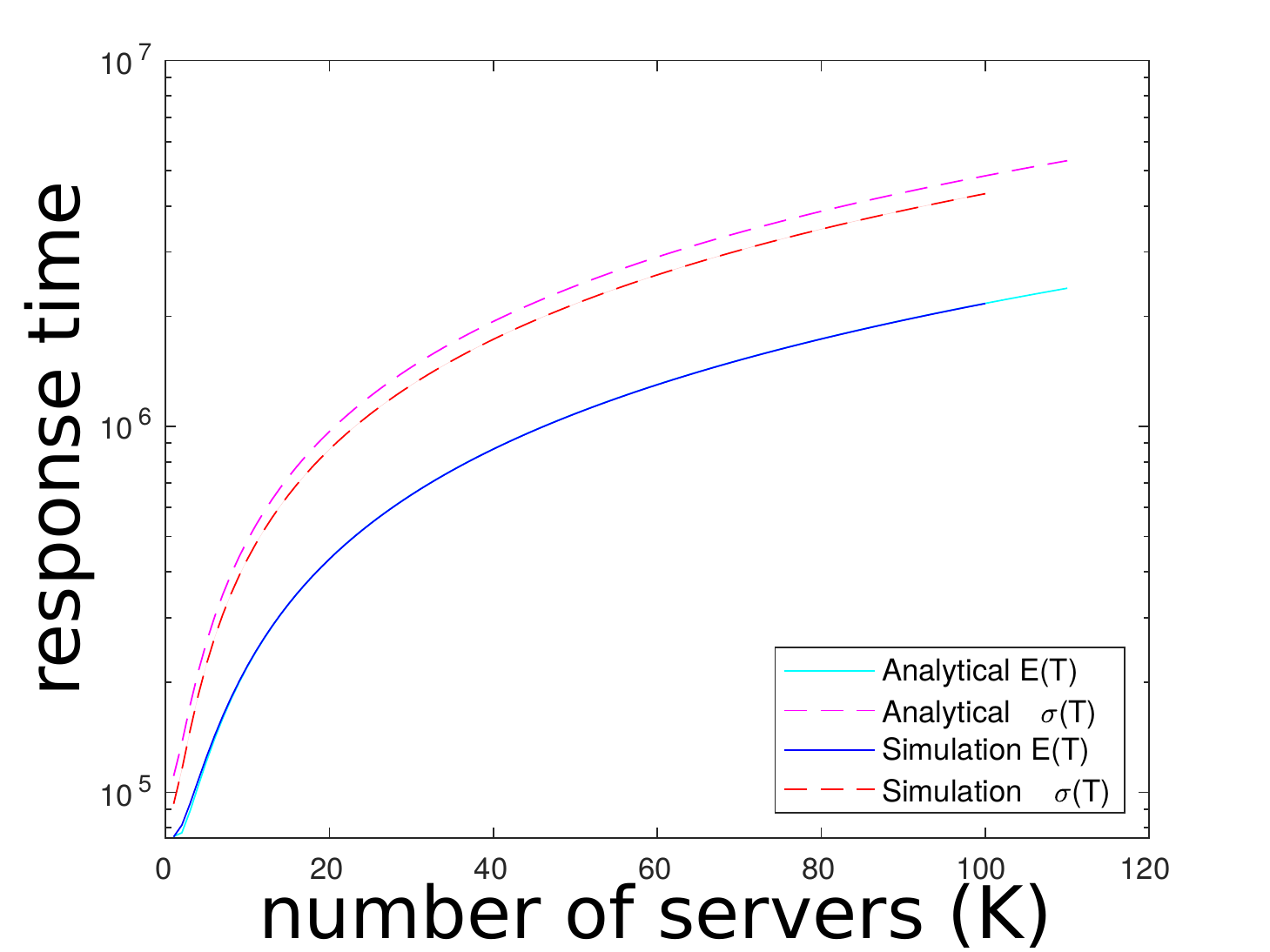}&
		 \includegraphics[width=0.35\textwidth]{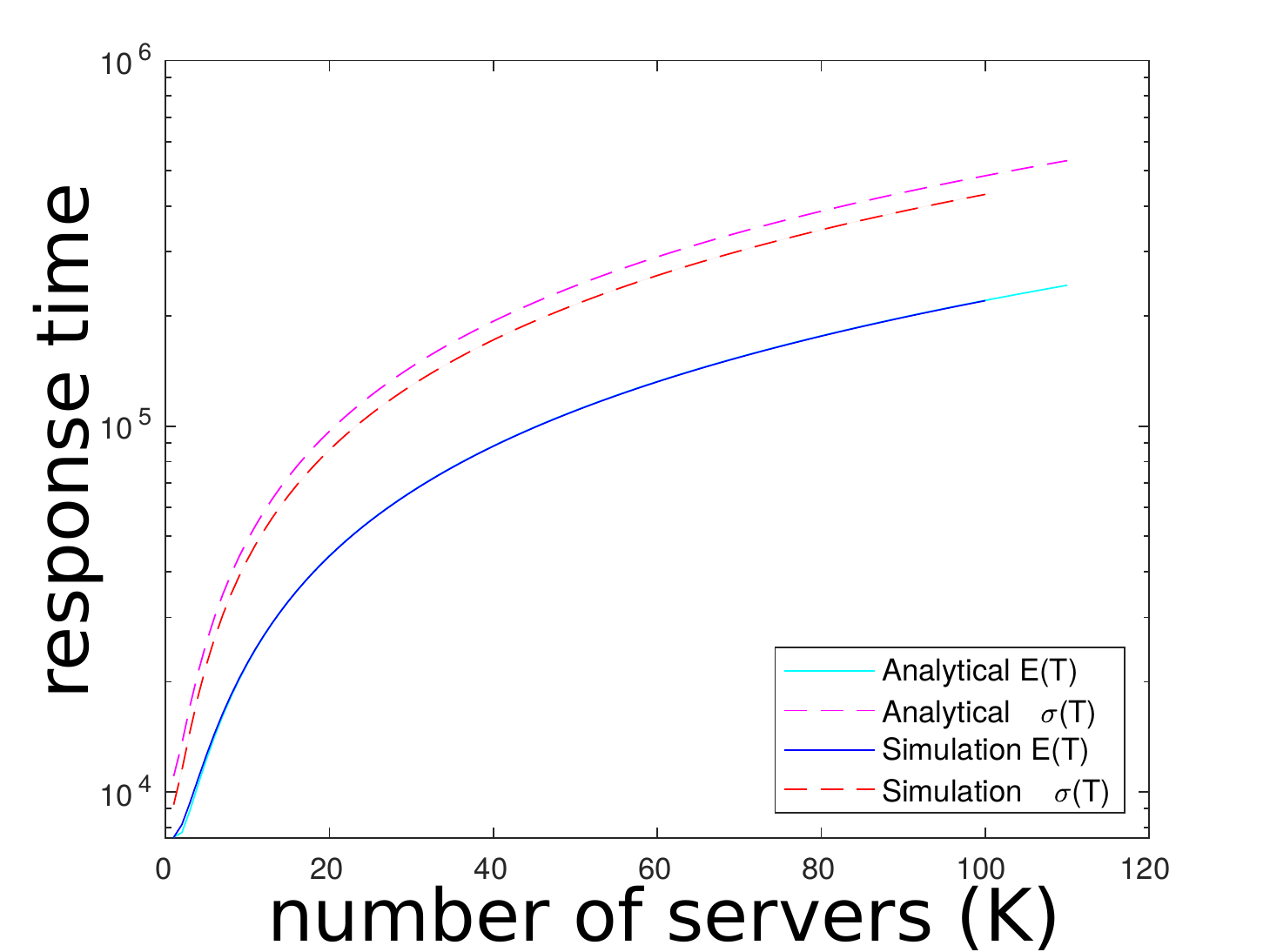}&	\includegraphics[width=0.35\textwidth]{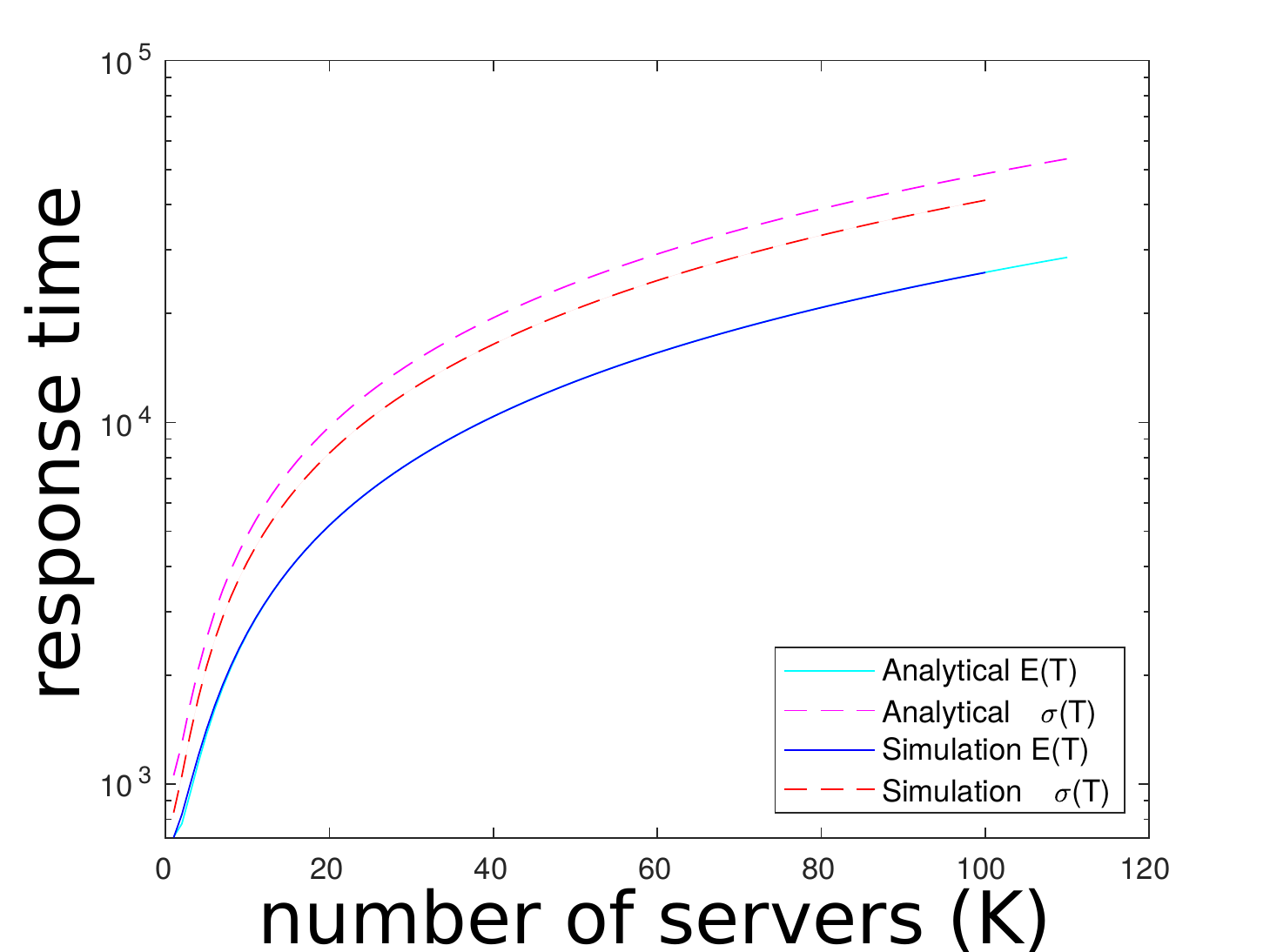}\\
		 (d) $\alpha=0.80$, $\frac{E(X_S)}{E(X_L)}=0.0005$&		 (e) $\alpha=0.80$, $\frac{E(X_S)}{E(X_L)}=0.0050$&		 (f) $\alpha=0.80$, $\frac{E(X_S)}{E(X_L)}=0.0500$
\end{tabular}
\end{center}
\caption{Model validation: response times as a function of  system load (analytical and simulation results), $\rho=0.5$. 
}
\label{fig:response_time_simulation_analitical_rho05}
\end{figure*}

\subsection{Model validation through simulations} \label{sec:2_validation}

To check   the agreement between the analytical model approximate solution presented in Section~\ref{sec:2_responsetime} against the $M/G/K$ exact solution, which is not amenable to a simple closed-form expression~\cite{gupta2010inapproximability}, we conducted a  simulation campaign. 
Our simulations are executed using an efficient simulator which leverages an extension of recursive Lindley equations~\cite{kin2010generalized}, allowing us to produce  numerical results  more efficiently than traditional event-driven simulations.   The simulations were performed in 5 rounds with 1 million samples per round.

Figure~\ref{fig:response_time_simulation_analitical_rho05} 
and Table~\ref{tab:optimalnumbig} report  our simulation results. 
 They show that the proposed model approximations are more accurate for smaller values of  $\rho$.  
 When $\rho= 0.80$ and $\rho= 0.50$,  model approximations  typically have predictive power to determine the optimal values of $K$ that minimize $E(T)$ and $V(T)$ (columns $K_{\mu}^{\star}$ and $K_{\sigma}^{\star}$ in Table~\ref{tab:optimalnumbig}). In addition, for all the considered scenarios, the qualitative behavior of the model and simulations is similar    (Figure~\ref{fig:response_time_simulation_analitical_rho05}).
  Note that scenarios  wherein  $\rho$ is large, e.g., $\rho=0.95$, are arguably of less relevance, as system administrators will typically provision the system to avoid  heavy traffic.

When $K=1$ the corresponding M/G/1 system admits solutions in closed form (last two columns of Table~\ref{tab:optimalnumbig}).  It is interesting to observe that by deviating from the single server setup both the mean and the variance of the residence time can often be significantly reduced.  Nonetheless, the single server system is still the optimal choice in multiple scenarios, e.g., when the utilization is small  and attacks are not frequent ($\rho=0.5$   and $\alpha=0.6$, as shown in the last line of Table~\ref{tab:optimalnumbig}).

\section{Anomaly Detection}
\label{sec:2_anomaly}

Next, our goal is to assess the accuracy of machine learning classifiers to detect anomalies.  To that aim, we vary the number of servers $K$, and report results on the performance of  classifiers as a function  of $K$.  The classifiers, in turn,   classify jobs based on whether they  had their response times impacted by anomalies as detailed below.

\subsection{Accuracy of Anomaly Detection }

We begin by considering an SVM classifier. 

    
    \textbf{Input feature}: A classifier is developed to detect  anomalies. Each job that concludes its service produces a new sample to the classifier.  Each sample, in turn, comprises a  single  feature, namely the corresponding  job response time.  Response times $(i)$ can be shared among multiple stakeholders, $(ii)$ are readily available, as they can be measured by end users under low overhead, and $(iii)$ have minimal privacy implications.  Therefore, we posit that using  response time as    feature for anomaly detection  constitutes one of the simplest privacy-preserving approaches for anomaly detection. 
    
    
      \textbf{Target classes}:  The classifier classifies each job into one of three classes: 
        $(a)$~\emph{short jobs not impaired by anomalies}, i.e., short jobs not affected by long jobs, whose sojourn in the system did not overlap with that of a  long job;
        $(b)$~\emph{short jobs impaired by anomalies}, i.e., short jobs affected by  long jobs, whose sojourn in the system  overlapped with that of a long job; and
        $(c)$~\emph{long jobs}.   The three target classes are illustrated in Figure~\ref{fig:anomalygeneral}. 
        
        \textbf{Challenges: } Classifying among those three classes solely based on response times is non trivial, as the impairment suffered by short jobs may initially be subtle when anomalies are initiated. As we consider deterministic service times in our simplified model,  it is trivial to detect the presence of anomalies after the fact, in retrospect.  Nonetheless, it is challenging to detect anomalies in hindsight, after a small set of short jobs have been slightly impaired by the anomaly.  \emph{The goal of the classifier is to enable an early detection of anomalies leveraging those subtle but possibly statistically significant response time impairments. }

            \textbf{Training and test sets}:   We collected  50,000 samples from our $M/G/K$ simulator to run the experiments. A fraction of 20\% of the samples was used  for evaluating the SVM classifier (test set) and 80\% for training (training set).
               The whole process was repeated with the number of servers varying from 1 to 100.

 \textbf{Accuracy:}  To assess the classifier performance, we estimate its accuracy under the test set. The accuracy is the fraction of jobs correctly classified in one of the three target classes, divided by the  number of jobs classified.

 

        \begin{figure*}[!htb]
            \setlength\tabcolsep{1pt}
            \center
            \begin{tabular}{ccc}
               \includegraphics[width=0.37\textwidth]{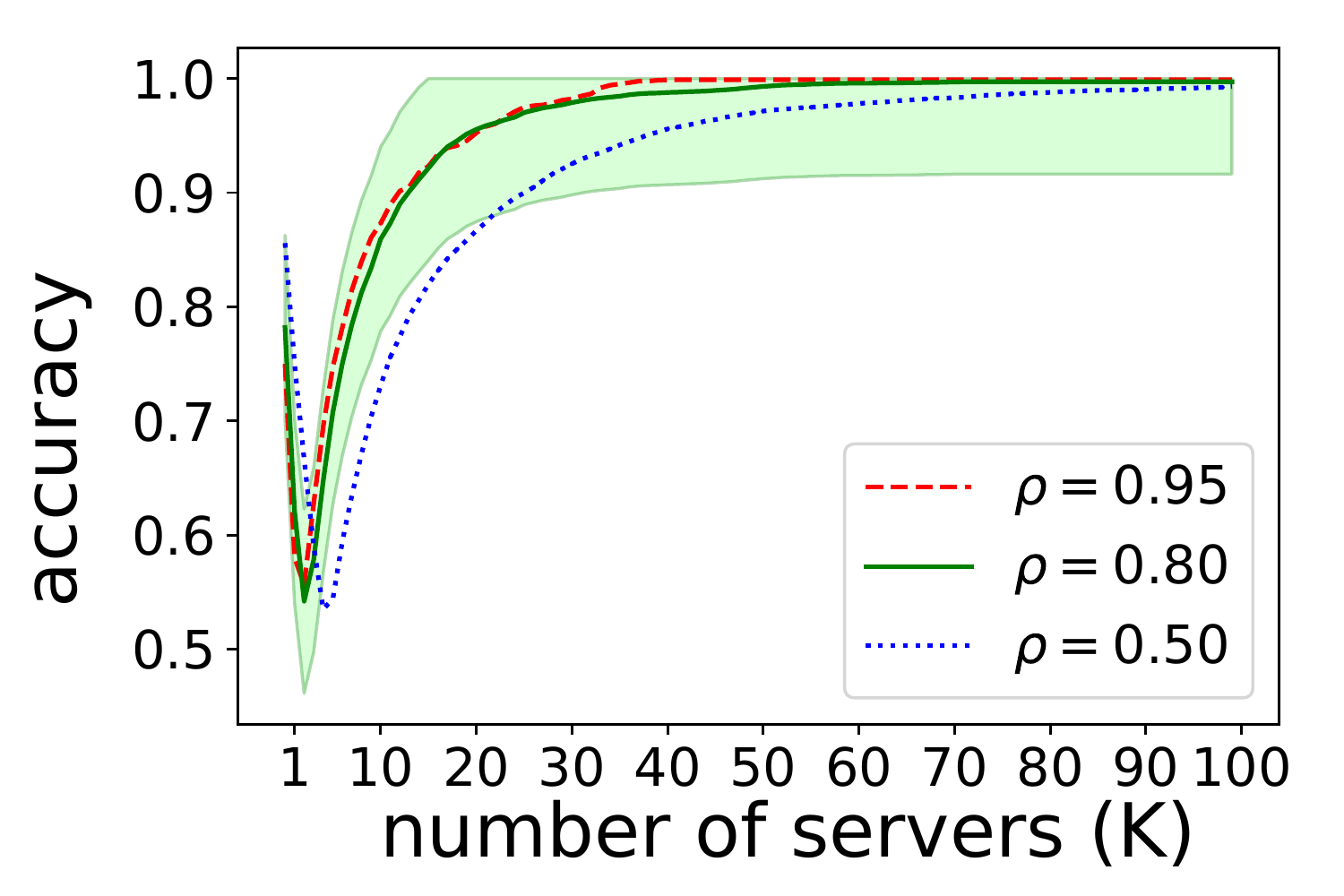} & 
               \includegraphics[width=0.37\textwidth]{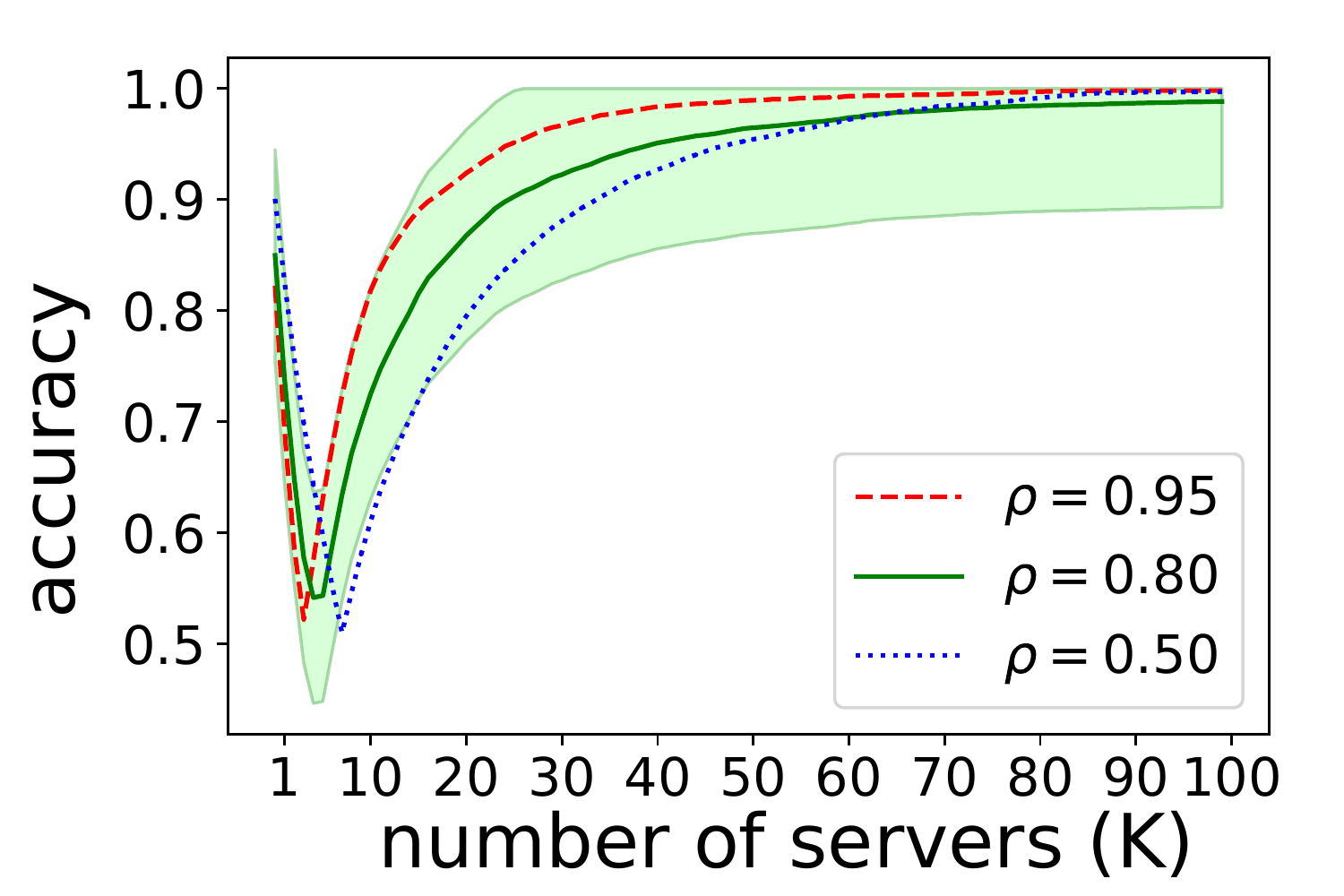} &
               \includegraphics[width=0.37\textwidth]{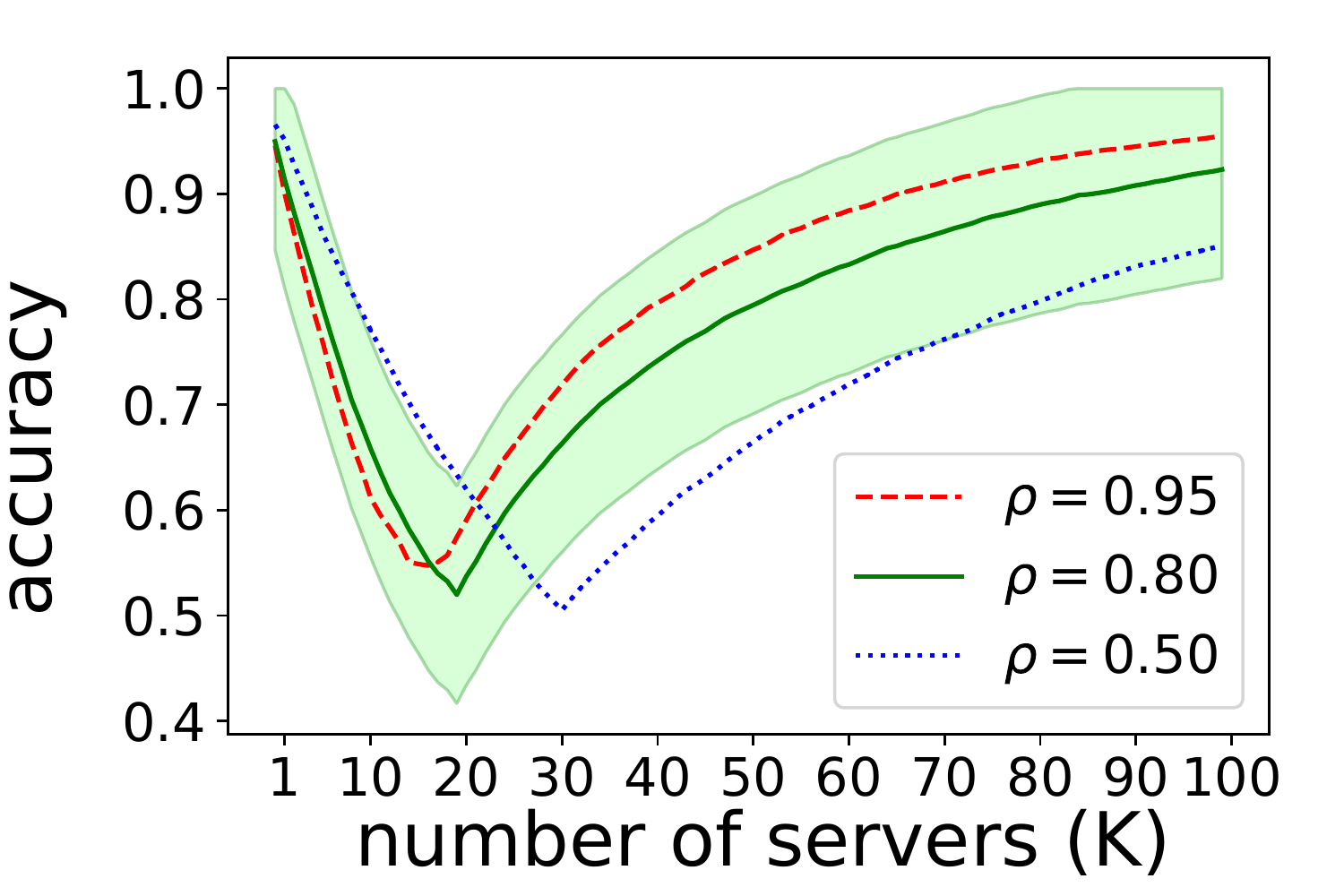} \\
                (a) $E(X_S)/E(X_L)=0.0005$ & 
                (b) $E(X_S)/E(X_L)=0.0050$ & 
                (c) $E(X_S)/E(X_L)=0.0500$ \\
            \end{tabular}
            \caption{
                Classifier's accuracy as a  function of the number of servers, with $\alpha=0.99$.
                }
            \label{fig:accuracies} \vspace{-0.1in}
        \end{figure*}
        
        \begin{figure*}[!htb]
            \setlength\tabcolsep{1pt}
            \center
            \begin{tabular}{ccc}
                \includegraphics[width=0.37\textwidth]{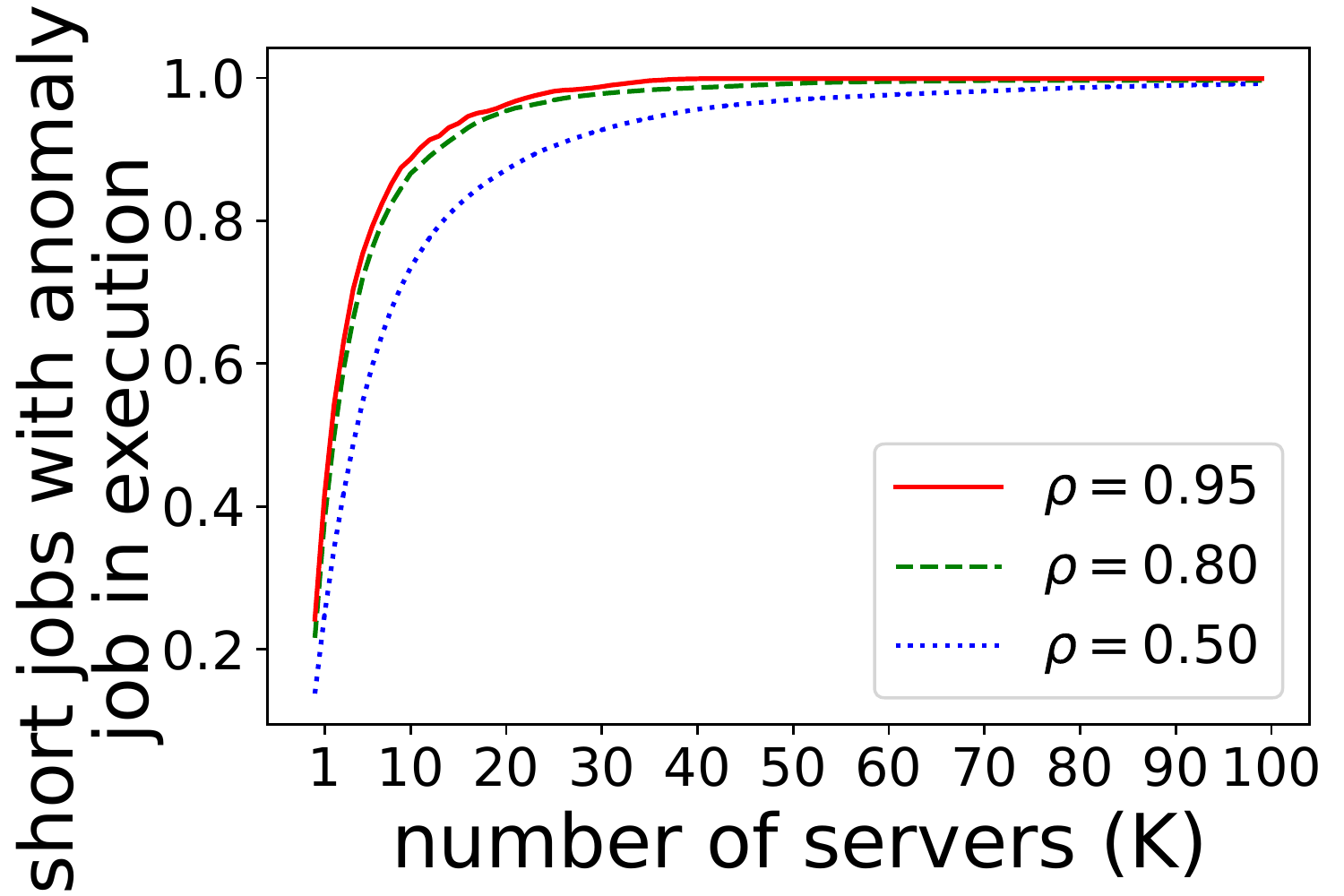} &
                \includegraphics[width=0.37\textwidth]{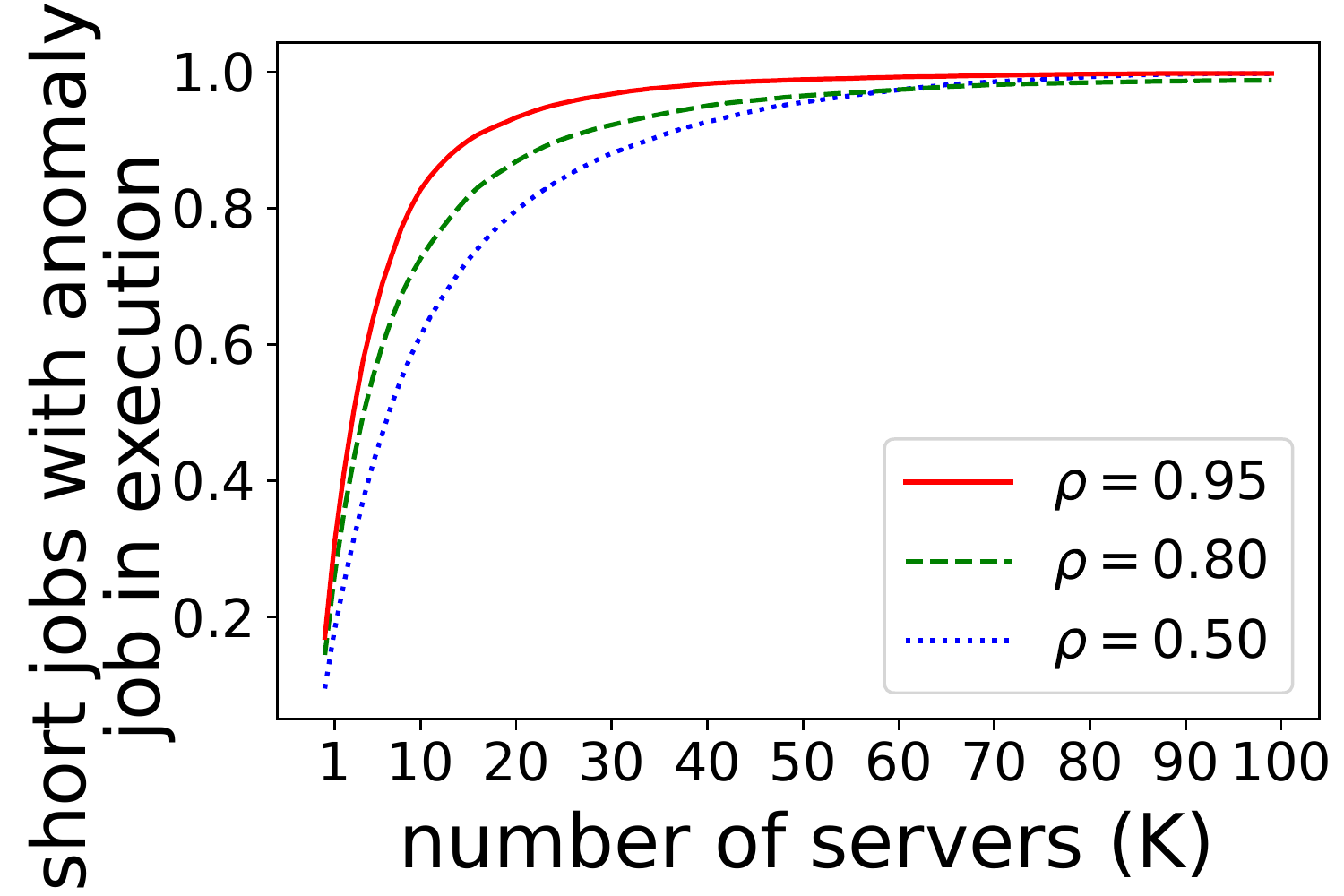} &    
                \includegraphics[width=0.37\textwidth]{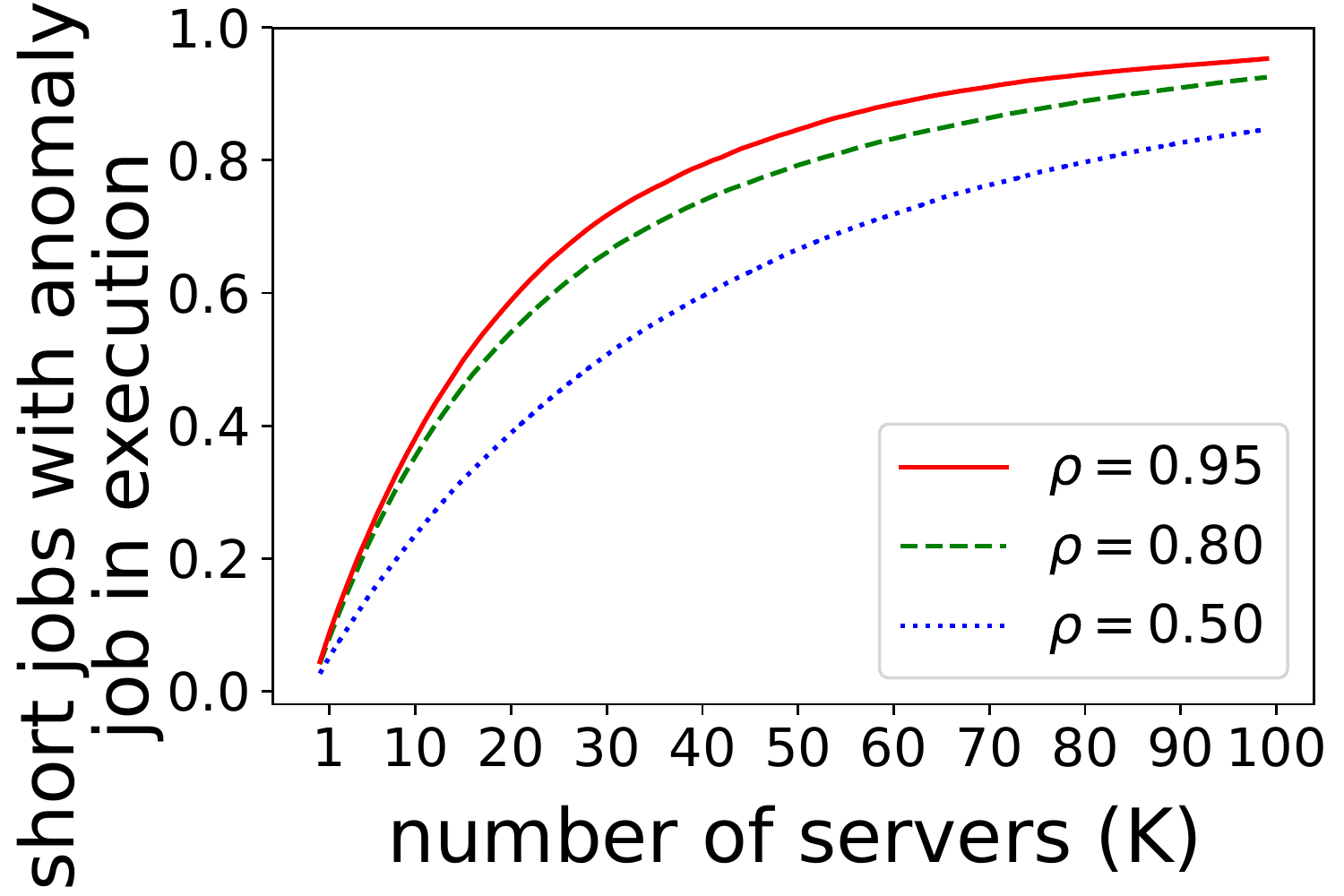} \\
                (a) $E(X_S)/E(X_L)=0.0005$ & 
                (b) $E(X_S)/E(X_L)=0.0050$ & 
                (c) $E(X_S)/E(X_L)=0.0500$ \\
            \end{tabular}
            \caption{
                Proportion of the number of short jobs that find the system with anomalies, with $\alpha=0.99$.
                }
            \label{fig:number_sj} \vspace{-0.1in}
        \end{figure*} 
        
    \textbf{Results}:~Figure~\ref{fig:accuracies} shows how accuracy varies as a function of the number of servers $K$. We let the server utilization $\rho$ vary between 0.95, 0.80 and 0.50, and 
    $\alpha=0.99$ (other values of $\alpha$ were also considered, with similar results). 

\emph{The classifier  accuracy initially decreases as $K$ grows:}  starting from $K= 1$, in most of the considered scenarios  both $E(T)$ and $V(T)$ decrease as $K$ grows. Indeed, an increase in $K$ reduces  blocking, which favors better performance (see discussion in Section~\ref{sec:2_key}).  Nonetheless, this increase in performance comes at the cost of a more challenging classification of jobs.   {When $K=1$, the blocking of shorts jobs   is an early  signal that the system is facing an anomaly.  As $K$ increases, such blocking decreases, and the early detection of anomalies through the assessment of response times becomes more challenging.}

\emph{Further increasing $K$ favors accuracy, but hurts performance: } as $K$ is further increased, 
the impact of the long jobs on the service times of short jobs grows. Then,  the accuracy of the classifier increases.  Such increase in classification accuracy comes   at the cost of performance degradation. 

        \emph{ For  large values of $K$, the classification task is trivial: } when $K$ is large, most short jobs will be affected by long jobs.
        Indeed, Figure~\ref{fig:number_sj} reports the fraction of short jobs whose sojourn in the system overlapped with at least one anomalous job.   As shown in Figure~\ref{fig:number_sj}, such fraction is close to 1 when $K \ge 80$ and $E(X_S)/E(X_L)$ is either $0.0005$ or $0.005$.   
        In a system with many small servers,  most short  jobs will suffer from the impairments of anomalous jobs and the class of short jobs not affected by long jobs vanishes. 

    \textbf{Take away message: }
    Figures~\ref{fig:accuracies} and~\ref{fig:number_sj} show that the detection of anomalies is simplified when the number of servers is either small or large.  Otherwise, one needs to cope with a tradeoff between  response time mean, response time  variance, and  the accuracy of mechanisms for anomaly detection. Those metrics, in turn, are related to fundamental aspects of the system such as its  performance, predictability  and security.

\subsection{Threshold Strategies}

Next, we report illustrative results on the anomaly detection thresholds derived from our numerical evaluation.  To that aim, we consider  decision trees (DT) and naive Bayes (NB)  classifiers to distinguish between the waiting time of jobs that find the system with anomalies from those that did not find anomalies.  Note that whereas in the previous section we considered an SVM classifier, in this section we illustrate our results using DT and NB to indicate that our results are not bound to a specific classifier.   The evaluation methodology is the same as the one described in the previous section, except that in this section we consider the whole dataset for training purposes, i.e., for learning the thresholds.   In addition, we focus solely on short jobs.  If the waiting time of a job is smaller than the learned   threshold, the job is classified as being executed in a system without anomalies. Otherwise, the job is tagged as an execution facing an anomaly.

%
%

\begin{figure*}
    \setlength\tabcolsep{1pt}
    \centering
    \begin{tabular}{@{}c@{}c@{}c@{}}
    \includegraphics[width=0.37\textwidth]{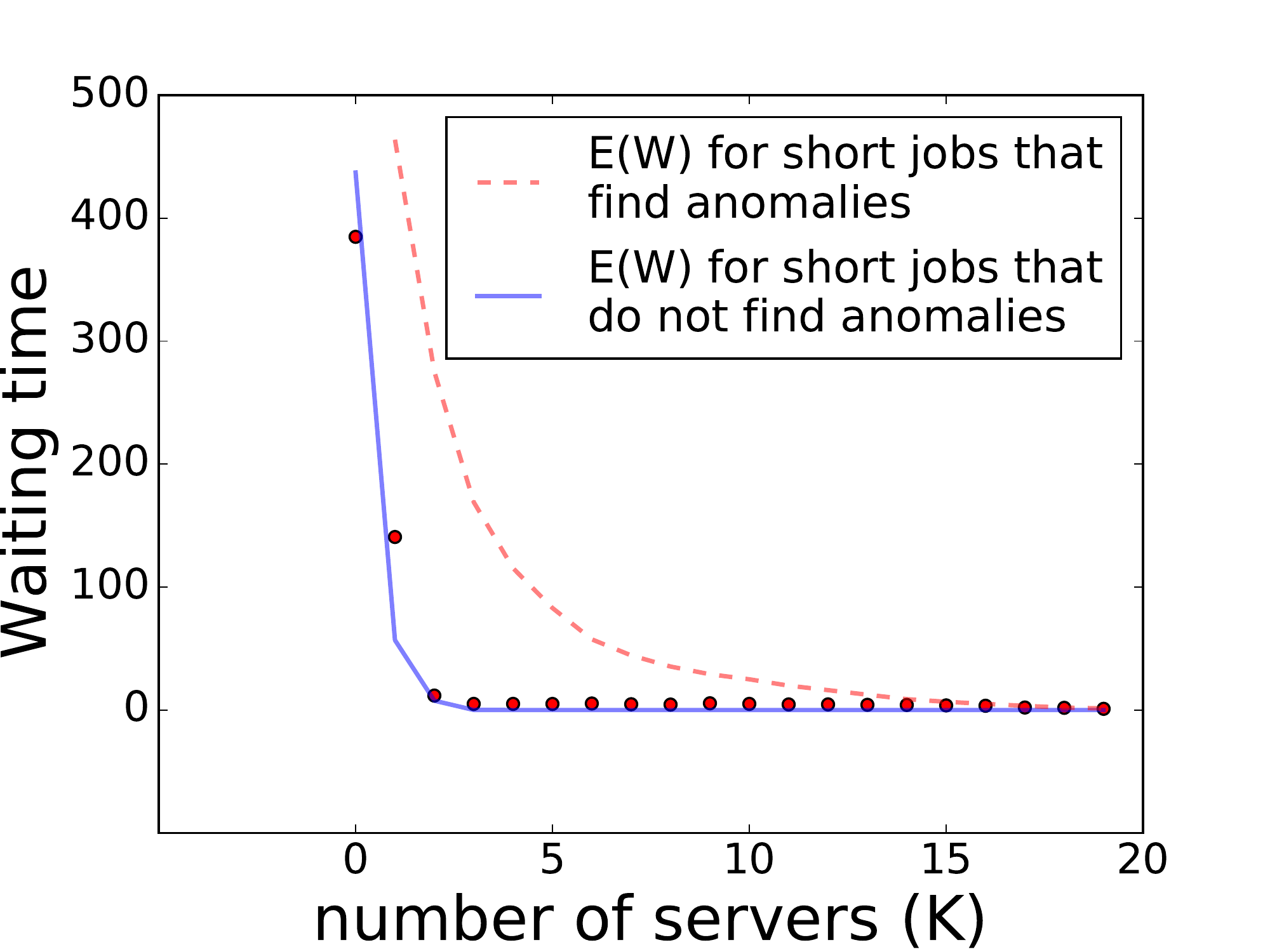} &   \includegraphics[width=0.37\textwidth]{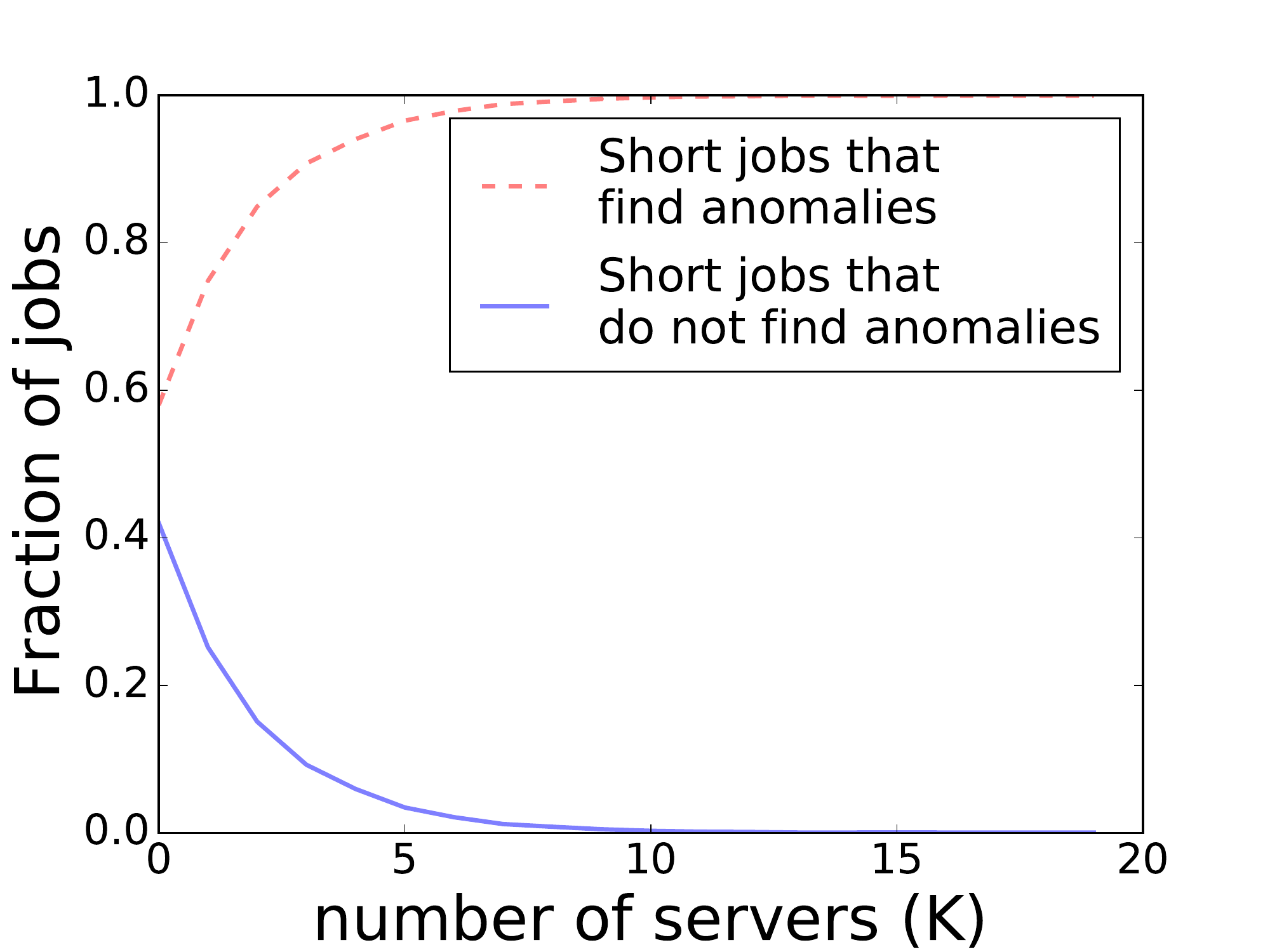} & \includegraphics[width=0.37\textwidth]{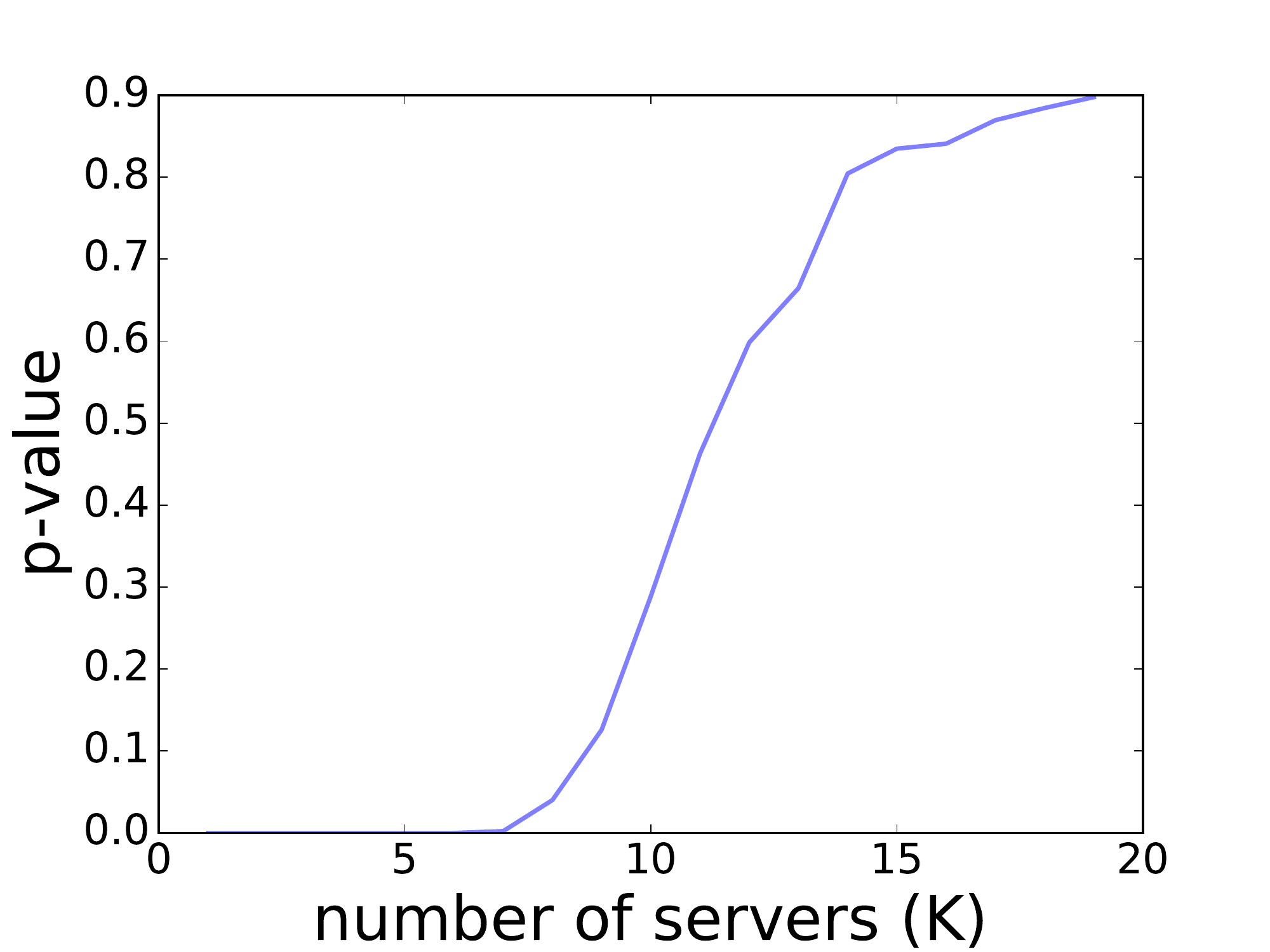}  \\
    (a)  & (b) & (c)
    \end{tabular}
    \caption{Evaluating threshold strategies and feasibility of anomaly detection based on waiting times: (a) threshold strategy; (b) \emph{a priori} class probabilities and (c) p-values and  statistical hypothesis test to verify the feasibility of anomaly detection based on waiting times: as $K$ grows,  waiting times  decrease and it becomes more challenging to detect anomalies based on those. Recall that response times are waiting times plus service times, and that service times are assumed to be deterministic per class. We  let $\rho=0.5, \alpha=0.8$ and  $E(X_S)/E(X_L)=0.05.$}
    \label{fig:thresholds}
\end{figure*}

We let $\rho=0.5$,   $\alpha=0.8$ and $E(X_S)/E(X_L)=0.05$, and   vary $K$ between 1 and 20.  This scenario corresponds to Figures~\ref{fig:accuracies}(c) and~\ref{fig:number_sj}(c). The thresholds learned through NB  and DT classifiers agreed with each other. As shown in Figure~\ref{fig:thresholds}(a),  under the considered scenarios the anomaly detection thresholds closely followed the mean waiting time of jobs that did not find anomalies in the system.

The dependency of the learned  threshold on $K$ is non-trivial, as the threshold is affected by a number of factors, including the fraction     of short jobs that find an anomalous job in execution as well as the variance of the waiting times. The former impacts the \emph{a priori} probability of classifying a job as a short job impaired by an anomaly (Figure~\ref{fig:thresholds}(b)), and the latter impacts the likelihood of observed waiting times given the target classes.  In particular, the imbalance of the number of samples across classes partially explains why the thresholds do not always reside in between the curves corresponding to the average time of short jobs impaired by anomalies and short jobs not impaired by those anomalies. Indeed, the fact that in the considered setup the fraction of short jobs that do not find an anomalous job in execution is below 0.4  (see Figure~\ref{fig:thresholds}(b)) favors a reduction in the threshold values.

\subsection{Feasibility of anomaly detection}

Next, we consider the feasibility of detecting anomalies  solely based on waiting time or response time measurements, from the standpoint of statistical hypothesis tests.   In particular, when the number of servers grows large, the waiting times are negligible, rendering the detection of anomalies based on waiting times unfeasible.  Indeed, we conducted statistical hypothesis tests to check if the waiting times of jobs that found an anomaly in the system have the same mean as the waiting times of jobs that did not find anomalies.   The null hypothesis $H_0$ corresponds to the two means being equal, and the alternative hypothesis $H_1$ corresponds to the two means being different.

We illustrate our results under the same setup as considered in the previous section. As shown in Figure~\ref{fig:thresholds}(c), for $K \leq 6$ the p-values obtained from t-tests are negligible (much smaller than 0.05), meaning that the null hypothesis can be rejected and it is feasible to run statistical hypothesis tests to detect anomalies based on waiting times.  However, for larger values of $K$ the p-values grow. In the limit when $K=\infty$ we have an $M/G/\infty$, for which the residence times of all jobs are decoupled, i.e., short jobs are not affected by long jobs, waiting times are zero, and it is infeasible to detect the presence of anomalies based on the response time of regular jobs, under the assumptions considered in this work.   In those cases, one needs to resort to  additional side information for anomaly detection.  

\label{sec:honeypots}

\section{Related Work} 
\label{sec:2_related}

There is a vast literature on analytical and experimental aspects related to security and performance~\cite{alomari2012autonomic, zhao2016optimizing,alomari2014efficient}.  Nonetheless, to the best of our knowledge this work is the first to analytically  study  the interplay between host-based anomaly detection leveraging customer affecting metrics and system scalability in multi-server systems. 

\subsection{Multi-server queues and optimal number of servers} 


The use of the $M/G/K$ queue to model cloud multi-server systems has been considered in~\cite{psounis2005systems,khazaei2012performance}.  
In this paper, we apply the model introduced in~\cite{psounis2005systems} to the performance analysis of denial of service attacks in multi-server systems.  Security implications of~\cite{psounis2005systems} have been briefly pointed out by \cite{rohr2015workload}. In this work, we take a step further, studying how the mean and standard deviation of response times may impact decisions related to the setup of service infrastructures.

The problem of determining the optimal number of servers in a computer system is one of the most classical problems in queuing theory and autonomic computing~\cite{stidham1970optimality, sharma2012provisioning}. The optimal number of servers in $M/G/K$ queues and its variants has also received some attention in the past~\cite{wierman2006many,qin2010new, scheller2003necessary}. However, none of those previous works accounted for the response time predictability, which is intrinsically related to system security, when tuning the number of servers.  In this paper,  in contrast, we identify a fundamental tradeoff between performance and security when  evaluating  the mean and variance of response times.

In the monitoring literature, mean response time is typically considered a limited metric as  response times are  heavy-tailed~\cite{Mielke2006,Rohr2010WTB}. Therefore, researchers usually suggest the use of quartile-based metrics, e.g., for the specification of service-level objectives or the configuration of performance anomaly detection systems. Nonetheless, most of the previous work on analytical models focused primarily on mean response times \cite{harchol2013performance, menasce2002capacity}. In this paper,
we contribute to bridging that gap between theory and practice, by further investigating the practical implications of the response time variance as predicted by the analytical model.






\subsection{Anomaly detection}

In~\cite{Chandola2009} a survey of anomaly detection techniques  identifies some of the most important application domains for anomaly detection application as intrusion detection, fraud detection, fault-detection, and medical diagnosis. The authors have enumerated the most important  challenges in anomaly detection research  as follows: $(i)$~it is very difficult to comprehensively define normal behavior, $(ii)$~malicious attackers may adapt their behavior to fit the domain definition of ``normal behavior'', $(iii)$ 
anomaly definition varies per domain, $(iv)$~data availability for anomaly training is not easy to obtain, $(v)$ training data is usually noisy. As a consequence of these challenges, researchers usually develop heuristics for anomaly detection that take advantage of specific characteristics of the application domain.

In this paper, we have attempted to address these concerns by: $(i)$ using an analytic model to capture normal behavior in terms of mean and variance of response time, $(ii)$ focus on denial of service attacks, which have impact on mean and variance of the response time, $(iii)$ focus on a given performance engineering domain, $(iv)$  as our anomaly detection algorithm uses mean and variance of response time, our approach does not require extensive training data, $(v)$ use a controlled environment to avoid training data noise.

In~\cite{Gurbani17,Kumar2005}  and~\cite{Denning87} the authors propose threshold-based and more general rule-based models for intrusion detection systems.  The derivation of optimal thresholds and meaningful rules is typically obtained through classical statistical methods or machine learning. 
%
%
%
%
%
In this paper,  we take advantage of an analytic model and an existing  performance testing infrastructure to study some fundamental tradeoffs between anomaly detection and scalability. Our approach {relies on}  experiments that are run in a controlled environment to derive the parameters required to calibrate the performance model and the anomaly detection algorithm.

\subsection{Mirai} 
\label{sec:2_mirai}

In this paper we launched  Mirai DDoS attacks in a controlled lab environment  using the Mirai publicly available source code.    In~\cite{antonakakis2017understanding}  the authors {describe}  the Mirai botnet  used to create a significant DDoS attack in 2016.
This attack  harnessed the power of insecure IoT devices. The authors provided  an analysis of the Mirai timeline covering seven months, which included up to 
600k intrusions.  In~\cite{kambourakis2017mirai},
the authors present a review of Mirai and its mutations and alert to the risks posed by large botnets formed by using compromised IoT devices.
The key lesson from Mirai's attack is that IoT devices are currently a prime target of attacks and can be easily harnessed in large numbers to create a  massive bot attack.

\subsection{Security-scalability tradeoff}

Nylander et al.~\cite{icac2018} propose  
an architecture using a single queue load-balancing front-end combined with
admission control  to support predictable performance in cloud
applications. The authors have shown that 
their approach was able to produce smaller response time variability than other evaluated
strategies. More predictable response time, in turn, translates into higher security levels, as it is easier to tune anomaly detection algorithms when the target system is more predictable.  In general, the set-up of scalable cloud computing services impacts both  performance and  system security aspects.

In~\cite{manjhi2006simultaneous, manjhiincreasing} the authors present a framework for the simultaneous scalability and security confidentiality assessment for data-intensive web applications. The authors presented strategies for database view invalidation to help select which data shall be encrypted without impacting scalability when using a database  service provider. They have proposed a new scalability aware security design using: $(i)$  mandatory encryption of sensitive information, and, $(ii)$  restricting data encryption to  encrypt only data that is not scalability affecting. Our work is related to~\cite{manjhi2006simultaneous}, as we also propose  a security-aware scalability approach. However, in this paper we analyze how to account for tradeoffs between  anomaly detection  and scalability  when computing the optimal number of servers in multi-server central-queueing systems.

In~\cite{gillman2015protecting} the authors also discuss the security-performance tradeoff: increasing security typically degrades performance, as screening users may slow down the site.  As pointed above, this may ultimately  translate into a denial of service,  achieving the goals of the attacker. In this paper, we point that system administrators may control the  server infrastructure (e.g., by adjusting the number of servers accounting for the standard deviation of response times) in order  to circumvent the challenge of inadvertently slowing down users in face of potential attacks.

\subsection{IDS tools}

The literature on IDS tools is vast~\cite{milenkoski2015evaluating,GrottkeAMAA16,Mitchell2014,Zarpela2017,Felemban2018, Zarpela2017}. 
A comprehensive survey of IDS tools is presented {in~\cite{milenkoski2015evaluating}}, where  
three properties of IDS tools are used to classify those systems: $(i)$ monitored platform (host based, network based, or hybrid), $(ii)$ attack detection method (misuse based, anomaly based, hybrid), and, $(iii)$ deployment architecture (non-distributed and distributed)~\cite{milenkoski2015evaluating}. In the anomaly based IDSs, a baseline profile of normal operations is developed and  deviations from the baseline profile are identified as intrusions using performance signatures.  Milenkoski et al.~\cite{milenkoski2015evaluating} report that one of the most important open research questions in this domain is the development of IDSs for detecting zero-day attacks and advanced persistent threats (APTs)~\cite{GrottkeAMAA16}. The anomaly detection approaches based on performance signatures as considered in this paper do not require detailed knowledge of attack history, and for this reason they are suitable to address the   challenges  {associated to zero-day attacks and APTs detections}.

Avritzer et al.~\cite{avritzer2010monitoring} proposed an architecture for intrusion detection systems using off-the-shelf IDSs complemented by performance signatures. The authors have shown that the performance signature of well-behaved systems  and of  several types of security attacks could be identified in terms of certain performance metrics, such as CPU and memory percentage,  number of active threads, etc. Specifically, the following security attacks were evaluated positively for performance signature detection, when compared against off-the-shelf IDS detection: ``man in the middle'', denial of service, buffer overflow, stack overflow, and SQL injection.


The performance of signature-based intrusion detection systems rely on  intrusion detection algorithms that account for workload variability  to avoid a high rate of false positive alerts.  An example of such workload-sensitive algorithms is the bucket algorithm~\cite{Avritzer2005}.

Anomaly detection using the bucket algorithm inspects the last sample of response time and compare it against  calibrated mean and standard deviation of response time. Then, it relies on the central limit theorem~\cite{AvritzerCW07} to assess if the last samples of response time comply with normal behavior or represent an anomaly.  
We envision that the models, measurements and insights  in this paper can be instrumental to set the baselines of those algorithms.

\section{Practical Implications, Assumptions and limitations} \label{sec:2_limitations}

\subsection{Practical engineering implications}

    \label{sec:2_discussion}
    Next, we discuss some of  the practical implications of our work.
    In this work we proposed first steps towards models and measurements to capture the tradeoff between scalability and efficient anomaly detection.  The implications of that tradeoff to architecture design include the tuning of IDS systems and the setup of cloud computing infrastructures.   

    A more sensitive anomaly detection system may filter more malicious jobs, at the expense of also filtering workload from real users due to false positives.   The same holds for an admission control which, by filtering more jobs, may increase the performance of the system, at the expense of indirectly causing the blocking of real users, which translates into a denial of service,  ultimately achieving the goals of the attacker.  
    
    From the infrastructure setup perspective, we analytically observed that  the number of servers that minimizes mean response time is typically larger than or equal to  the number of servers that minimizes response time variance.  
    We experimentally investigated the impact of the workload and of the infrastructure on domain metrics, and discovered that those metrics can be helpful (and in some cases sufficient) to detect attacks and to capture  tradeoffs between scalability and anomaly detection. 

\subsection{Assumptions and limitations}

{Next,} we discuss some of the simplifying assumptions considered in the paper to yield a tractable model, as well as some limitations of the current work.

\textbf{Model accuracy: }   the proposed approximations to solve the analytical model allow us to predict the optimal number of servers that minimizes  the mean or the variance of response times (Section~\ref{sec:2_validation}).  Although there is space for improvement of the approximations, e.g., if the goal is to assess the minimum mean or variance of response times,  such refinements are  possible at the expense of a more complex solution, which is out of the scope of this work.  In this paper, we focus on the simplest closed form expressions  (eqs.~\eqref{eq:et1}-\eqref{eq:et21}) that were able to capture the trends and insights of interest (Section~\ref{sec:2_key}). For ideas of further refinements of the proposed approximations, we refer the reader to~\cite{psounis2005systems}.

\textbf{Optimal number of servers: }  we have separately considered the problem of minimizing the mean and the variance of response times.   To combine the two problems, one may either minimize a weighted sum of the two metrics, or minimize mean response time with constraints on its variance.  The target application will determine how to account for  the two metrics.  Our results indicate that  there is a tension between the two competing metrics, and that decreasing variance may come at the cost of an increase in the mean response time.

\textbf{Practical experiments: }  the performance of anomaly detection algorithms based on response times relies on  reference values for the mean and variance of response times~\cite{AvritzerCW07}.   Section~\ref{sec:2_modelparam} {shows} how real experiments can be used to parameterize the proposed model, and in Section~\ref{sec:2_anomaly} we report  results on how the number of servers  impacts anomaly detection algorithms. {The} model and insights introduced in this paper can be leveraged by anomaly detection algorithms, {noting that} a further integration of the proposed model and anomaly detection algorithms is left as topic for future research.

\section{Conclusions}
\label{sec:2_conclusions}
Anomaly detection  has been widely recognized as an important production tool in several application domains, such as medical systems, banking networks and failure diagnosis in mission-critical domains~\cite{Chandola2009}. 
In this paper, we have  presented an anomaly  detection framework that is supported by standard performance engineering   {processes}, such as workload modeling, performance modeling, performance testing, and performance monitoring.
Our focus  is on  one of the key elements of the proposed framework: an analytic model to estimate the average and standard deviation of response time of a central server system subjected to security attacks.

We have  used the performance testing environment to derive parameters for the calibration of the proposed analytic model. 
Our initial experiments are encouraging,  showing that the customer-affecting metric is sensitive to the security anomalies evaluated. 


{Our results  show}
that the number of servers that minimizes response time variance   is   typically   smaller   than   or   equal   to   the   one that minimizes the response time mean.  Therefore, for security purposes it may be worth slightly sacrificing mean response time to reduce its variance. The  variability  in  response  times  is    a  key  component that impacts the performance of  anomaly detection algorithms, and we believe that this paper is a first step towards bridging the gap between analytical models for performance evaluation and their security implications.

As topics for future research, we envision  the application of  the proposed  framework   to large mission-critical systems. Specifically, additional experiments are required to evaluate  how to  integrate admission control and load balancing~\cite{icac2018} with the cloud computing infrastructure configuration approach  to ensure mean response time predictability. In addition, we plan to compare the performance of anomaly detection algorithms based on different approaches, such as   bucket algorithms~\cite{AvritzerCW07}  which can naturally leverage the statistical characterization of response times  proposed in this paper against other  machine learning algorithms~\cite{jung2004fast}.

\bibliographystyle{ACM-Reference-Format}
\bibliography{200_bibliography}


\begin{thebibliography}{53}


\ifx \showCODEN    \undefined \def \showCODEN     #1{\unskip}     \fi
\ifx \showDOI      \undefined \def \showDOI       #1{#1}\fi
\ifx \showISBNx    \undefined \def \showISBNx     #1{\unskip}     \fi
\ifx \showISBNxiii \undefined \def \showISBNxiii  #1{\unskip}     \fi
\ifx \showISSN     \undefined \def \showISSN      #1{\unskip}     \fi
\ifx \showLCCN     \undefined \def \showLCCN      #1{\unskip}     \fi
\ifx \shownote     \undefined \def \shownote      #1{#1}          \fi
\ifx \showarticletitle \undefined \def \showarticletitle #1{#1}   \fi
\ifx \showURL      \undefined \def \showURL       {\relax}        \fi
\providecommand\bibfield[2]{#2}
\providecommand\bibinfo[2]{#2}
\providecommand\natexlab[1]{#1}
\providecommand\showeprint[2][]{arXiv:#2}

\bibitem[\protect\citeauthoryear{Alomari and Menasc{\'e}}{Alomari and
  Menasc{\'e}}{2012}]%
        {alomari2012autonomic}
\bibfield{author}{\bibinfo{person}{Firas Alomari} {and}
  \bibinfo{person}{Daniel~A Menasc{\'e}}.} \bibinfo{year}{2012}\natexlab{}.
\newblock \showarticletitle{An autonomic framework for integrating security and
  quality of service support in databases}. In \bibinfo{booktitle}{\emph{IEEE
  Sixth Int. Conf. on Software Security and Reliability}}.
  \bibinfo{publisher}{IEEE}, \bibinfo{address}{Gaithersburg, MD, USA},
  \bibinfo{pages}{51--60}.
\newblock
\newblock
\shownote{{doi}: 10.1109/SERE.2012.15.}


\bibitem[\protect\citeauthoryear{Alomari and Menasc{\'e}}{Alomari and
  Menasc{\'e}}{2014}]%
        {alomari2014efficient}
\bibfield{author}{\bibinfo{person}{Firas Alomari} {and}
  \bibinfo{person}{Daniel~A Menasc{\'e}}.} \bibinfo{year}{2014}\natexlab{}.
\newblock \showarticletitle{Efficient response time approximations for
  multiclass fork and join queues in open and closed queuing networks}.
\newblock \bibinfo{journal}{\emph{IEEE Transactions on Parallel and Distributed
  Systems}} \bibinfo{volume}{25}, \bibinfo{number}{6} (\bibinfo{date}{Jun.}
  \bibinfo{year}{2014}), \bibinfo{pages}{1437--1446}.
\newblock
\newblock
\shownote{{doi}: 10.1109/TPDS.2013.70.}


\bibitem[\protect\citeauthoryear{Antonakakis, April, Bailey, Bernhard,
  Bursztein, Cochran, Durumeric, Halderman, Invernizzi, Kallitsis, Kumar,
  Lever, Ma, Mason, Menscher, Seaman, Sullivan, Thomas, and Zhou}{Antonakakis
  et~al\mbox{.}}{2017}]%
        {antonakakis2017understanding}
\bibfield{author}{\bibinfo{person}{Manos Antonakakis}, \bibinfo{person}{Tim
  April}, \bibinfo{person}{Michael Bailey}, \bibinfo{person}{Matt Bernhard},
  \bibinfo{person}{Elie Bursztein}, \bibinfo{person}{Jaime Cochran},
  \bibinfo{person}{Zakir Durumeric}, \bibinfo{person}{J~Alex Halderman},
  \bibinfo{person}{Luca Invernizzi}, \bibinfo{person}{Michalis Kallitsis},
  \bibinfo{person}{Deepak Kumar}, \bibinfo{person}{Chaz Lever},
  \bibinfo{person}{Zane Ma}, \bibinfo{person}{Joshua Mason},
  \bibinfo{person}{Damian Menscher}, \bibinfo{person}{Chad Seaman},
  \bibinfo{person}{Nick Sullivan}, \bibinfo{person}{Kurt Thomas}, {and}
  \bibinfo{person}{Yi Zhou}.} \bibinfo{year}{2017}\natexlab{}.
\newblock \showarticletitle{Understanding the {M}irai botnet}. In
  \bibinfo{booktitle}{\emph{26th {USENIX} Security Symp.}}
  \bibinfo{publisher}{USENIX}, \bibinfo{address}{Vancouver, BC, Canada},
  \bibinfo{pages}{1092--1110}.
\newblock


\bibitem[\protect\citeauthoryear{Avritzer, Bondi, and Weyuker}{Avritzer
  et~al\mbox{.}}{2005}]%
        {Avritzer2005}
\bibfield{author}{\bibinfo{person}{Alberto Avritzer}, \bibinfo{person}{Andre
  Bondi}, {and} \bibinfo{person}{Elaine~J. Weyuker}.}
  \bibinfo{year}{2005}\natexlab{}.
\newblock \showarticletitle{Ensuring Stable Performance for Systems That
  Degrade}. In \bibinfo{booktitle}{\emph{5th Int. Workshop on Software and
  Performance (WOSP)}}. \bibinfo{publisher}{ACM}, \bibinfo{address}{Palma de
  Mallorca, Spain}, \bibinfo{pages}{43--51}.
\newblock
\newblock
\shownote{{doi}: 10.1145/1071021.1071026.}


\bibitem[\protect\citeauthoryear{Avritzer, Cole, and Weyuker}{Avritzer
  et~al\mbox{.}}{2007}]%
        {AvritzerCW07}
\bibfield{author}{\bibinfo{person}{Alberto Avritzer},
  \bibinfo{person}{Robert~G. Cole}, {and} \bibinfo{person}{Elaine~J. Weyuker}.}
  \bibinfo{year}{2007}\natexlab{}.
\newblock \showarticletitle{Using performance signatures and software
  rejuvenation for worm mitigation in tactical {MANETs}}. In
  \bibinfo{booktitle}{\emph{6th Int. Workshop on Software and Performance
  (WOSP)}}. \bibinfo{publisher}{ACM}, \bibinfo{address}{Buenes Aires,
  Argentina}, \bibinfo{pages}{172--180}.
\newblock
\newblock
\shownote{{doi}: 10.1145/1216993.1217023.}


\bibitem[\protect\citeauthoryear{Avritzer, Ferme, Janes, Russo, Schulz, and van
  Hoorn}{Avritzer et~al\mbox{.}}{2018}]%
        {avritzer2018}
\bibfield{author}{\bibinfo{person}{Alberto Avritzer}, \bibinfo{person}{Vincenzo
  Ferme}, \bibinfo{person}{Andrea Janes}, \bibinfo{person}{Barbara Russo},
  \bibinfo{person}{Henning Schulz}, {and} \bibinfo{person}{Andr\'e van Hoorn}.}
  \bibinfo{year}{2018}\natexlab{}.
\newblock \showarticletitle{A quantitative approach for the assessment of
  microservice architecture deployment alternatives using automated performance
  testing}. In \bibinfo{booktitle}{\emph{European Conf. on Software
  Architecture (ECSA)}} \emph{(\bibinfo{series}{LNCS})},
  Vol.~\bibinfo{volume}{11048}. \bibinfo{publisher}{Springer},
  \bibinfo{address}{Madrid, Spain}, \bibinfo{pages}{159--174}.
\newblock
\newblock
\shownote{{doi}: 10.1007/978-3-030-00761-4\_11.}


\bibitem[\protect\citeauthoryear{Avritzer, Tanikella, James, Cole, and
  Weyuker}{Avritzer et~al\mbox{.}}{2010}]%
        {avritzer2010monitoring}
\bibfield{author}{\bibinfo{person}{Alberto Avritzer},
  \bibinfo{person}{Rajanikanth Tanikella}, \bibinfo{person}{Kiran James},
  \bibinfo{person}{Robert~G Cole}, {and} \bibinfo{person}{Elaine~J Weyuker}.}
  \bibinfo{year}{2010}\natexlab{}.
\newblock \showarticletitle{Monitoring for security intrusion using performance
  signatures}. In \bibinfo{booktitle}{\emph{Proc. 1st Joint WOSP/SIPEW Int.
  Conf. on Performance Engineering}}. \bibinfo{publisher}{ACM},
  \bibinfo{address}{San Jose, CA, USA}, \bibinfo{pages}{93--104}.
\newblock
\urldef\tempurl%
\url{https://doi.org/10.1145/1712605.1712623}
\showDOI{\tempurl}
\newblock
\shownote{{doi}: 10.1145/1712605.1712623.}


\bibitem[\protect\citeauthoryear{Brumelle}{Brumelle}{1971}]%
        {brumelle1971some}
\bibfield{author}{\bibinfo{person}{Shelby~L Brumelle}.}
  \bibinfo{year}{1971}\natexlab{}.
\newblock \showarticletitle{Some inequalities for parallel-server queues}.
\newblock \bibinfo{journal}{\emph{Operations Research}} \bibinfo{volume}{19},
  \bibinfo{number}{2} (\bibinfo{date}{Apr.} \bibinfo{year}{1971}),
  \bibinfo{pages}{402--413}.
\newblock
\newblock
\shownote{{doi}: 10.1287/opre.19.2.402.}


\bibitem[\protect\citeauthoryear{Carter and Hero~III}{Carter and
  Hero~III}{2008}]%
        {carter2008variance}
\bibfield{author}{\bibinfo{person}{Kevin~M Carter} {and}
  \bibinfo{person}{Alfred~O Hero~III}.} \bibinfo{year}{2008}\natexlab{}.
\newblock \showarticletitle{Variance reduction with neighborhood smoothing for
  local intrinsic dimension estimation}. In \bibinfo{booktitle}{\emph{2008 IEEE
  Int. Conf. on Acoustics, Speech and Signal Processing}}.
  \bibinfo{publisher}{IEEE}, \bibinfo{address}{Las Vegas, NV, EUA},
  \bibinfo{pages}{3917--3920}.
\newblock
\newblock
\shownote{{doi}: 10.1109/ICASSP.2008.4518510.}


\bibitem[\protect\citeauthoryear{Chandola, Banerjee, and Kumar}{Chandola
  et~al\mbox{.}}{2009}]%
        {Chandola2009}
\bibfield{author}{\bibinfo{person}{Varun Chandola}, \bibinfo{person}{Arindam
  Banerjee}, {and} \bibinfo{person}{Vipin Kumar}.}
  \bibinfo{year}{2009}\natexlab{}.
\newblock \showarticletitle{Anomaly detection: A survey}.
\newblock \bibinfo{journal}{\emph{Comput. Surveys}} \bibinfo{volume}{41},
  \bibinfo{number}{3}, Article \bibinfo{articleno}{15} (\bibinfo{date}{Jul.}
  \bibinfo{year}{2009}), \bibinfo{numpages}{58}~pages.
\newblock
\newblock
\shownote{{doi}: 10.1145/1541880.1541882.}


\bibitem[\protect\citeauthoryear{Chen, Sathe, Aggarwal, and Turaga}{Chen
  et~al\mbox{.}}{2017}]%
        {chen2017outlier}
\bibfield{author}{\bibinfo{person}{Jinghui Chen}, \bibinfo{person}{Saket
  Sathe}, \bibinfo{person}{Charu Aggarwal}, {and} \bibinfo{person}{Deepak
  Turaga}.} \bibinfo{year}{2017}\natexlab{}.
\newblock \showarticletitle{Outlier detection with autoencoder ensembles}. In
  \bibinfo{booktitle}{\emph{Proc. SIAM Int. Conf. on Data Mining}}.
  \bibinfo{publisher}{SIAM}, \bibinfo{address}{Houston, TX, EUA},
  \bibinfo{pages}{90--98}.
\newblock
\newblock
\shownote{{doi}: 10.1137/1.9781611974973.11.}


\bibitem[\protect\citeauthoryear{Denning}{Denning}{1987}]%
        {Denning87}
\bibfield{author}{\bibinfo{person}{Dorothy~E. Denning}.}
  \bibinfo{year}{1987}\natexlab{}.
\newblock \showarticletitle{An Intrusion-Detection Model}.
\newblock \bibinfo{journal}{\emph{IEEE Transactions on Software Engineering}}
  \bibinfo{volume}{SE-13}, \bibinfo{number}{2} (\bibinfo{date}{Feb}
  \bibinfo{year}{1987}), \bibinfo{pages}{222--232}.
\newblock
\newblock
\shownote{{doi}: 10.1109/TSE.1987.232894.}


\bibitem[\protect\citeauthoryear{Felemban, Javed, Kobes, Qadah, Ghafoor, and
  Aref}{Felemban et~al\mbox{.}}{2018}]%
        {Felemban2018}
\bibfield{author}{\bibinfo{person}{Muhamad Felemban}, \bibinfo{person}{Yahya
  Javed}, \bibinfo{person}{Jason Kobes}, \bibinfo{person}{Thamir Qadah},
  \bibinfo{person}{Arif Ghafoor}, {and} \bibinfo{person}{Walid Aref}.}
  \bibinfo{year}{2018}\natexlab{}.
\newblock \bibinfo{title}{Design and Evaluation of A Data Partitioning-Based
  Intrusion Management Architecture for Database Systems}.
  (\bibinfo{date}{out} \bibinfo{year}{2018}).
\newblock
\urldef\tempurl%
\url{https://arxiv.org/abs/1810.02061v2}
\showURL{%
\tempurl}
\newblock
\shownote{~.}


\bibitem[\protect\citeauthoryear{Gillman, Lin, Maggs, and Sitaraman}{Gillman
  et~al\mbox{.}}{2015}]%
        {gillman2015protecting}
\bibfield{author}{\bibinfo{person}{David Gillman}, \bibinfo{person}{Yin Lin},
  \bibinfo{person}{Bruce Maggs}, {and} \bibinfo{person}{Ramesh~K Sitaraman}.}
  \bibinfo{year}{2015}\natexlab{}.
\newblock \showarticletitle{Protecting websites from attack with secure
  delivery networks}.
\newblock \bibinfo{journal}{\emph{Computer}} \bibinfo{volume}{48},
  \bibinfo{number}{4} (\bibinfo{date}{Apr.} \bibinfo{year}{2015}),
  \bibinfo{pages}{26--34}.
\newblock
\newblock
\shownote{{doi}: 10.1109/MC.2015.116.}


\bibitem[\protect\citeauthoryear{Grosof, Scully, and Harchol-Balter}{Grosof
  et~al\mbox{.}}{2018}]%
        {grosof2018srpt}
\bibfield{author}{\bibinfo{person}{Isaac Grosof}, \bibinfo{person}{Ziv Scully},
  {and} \bibinfo{person}{Mor Harchol-Balter}.} \bibinfo{year}{2018}\natexlab{}.
\newblock \showarticletitle{{SRPT} for Multiserver Systems}.
\newblock \bibinfo{journal}{\emph{Performance Evaluation}}
  \bibinfo{volume}{127--128} (\bibinfo{date}{Nov.} \bibinfo{year}{2018}),
  \bibinfo{pages}{154--175}.
\newblock
\newblock
\shownote{{doi}: 10.1016/j.peva.2018.10.001.}


\bibitem[\protect\citeauthoryear{Grottke, Avritzer, Menasch{\'{e}}, de~Aguiar,
  and Altman}{Grottke et~al\mbox{.}}{2016}]%
        {GrottkeAMAA16}
\bibfield{author}{\bibinfo{person}{Michael Grottke}, \bibinfo{person}{Alberto
  Avritzer}, \bibinfo{person}{Daniel~S. Menasch{\'{e}}},
  \bibinfo{person}{Leandro~Pfleger de Aguiar}, {and} \bibinfo{person}{Eitan
  Altman}.} \bibinfo{year}{2016}\natexlab{}.
\newblock \showarticletitle{On the Efficiency of Sampling and Countermeasures
  to Critical-Infrastructure-Targeted Malware Campaigns}.
\newblock \bibinfo{journal}{\emph{{SIGMETRICS} Performance Evaluation Review}}
  \bibinfo{volume}{43}, \bibinfo{number}{4} (\bibinfo{date}{Feb.}
  \bibinfo{year}{2016}), \bibinfo{pages}{33--42}.
\newblock
\newblock
\shownote{{doi}: 10.1145/2897356.2897361.}


\bibitem[\protect\citeauthoryear{Gupta, Harchol-Balter, Dai, and Zwart}{Gupta
  et~al\mbox{.}}{2010}]%
        {gupta2010inapproximability}
\bibfield{author}{\bibinfo{person}{Varun Gupta}, \bibinfo{person}{Mor
  Harchol-Balter}, \bibinfo{person}{JG Dai}, {and} \bibinfo{person}{Bert
  Zwart}.} \bibinfo{year}{2010}\natexlab{}.
\newblock \showarticletitle{On the inapproximability of {M/G/K}: why two
  moments of job size distribution are not enough}.
\newblock \bibinfo{journal}{\emph{Queueing Systems}} \bibinfo{volume}{64},
  \bibinfo{number}{1} (\bibinfo{date}{Jan.} \bibinfo{year}{2010}),
  \bibinfo{pages}{5--48}.
\newblock
\newblock
\shownote{{doi}: 10.1007/s11134-009-9133-x.}


\bibitem[\protect\citeauthoryear{Gurbani, Kushnir, Mendiratta, Phadke, Falk,
  and State}{Gurbani et~al\mbox{.}}{2017}]%
        {Gurbani17}
\bibfield{author}{\bibinfo{person}{Vijay~K. Gurbani}, \bibinfo{person}{Dan
  Kushnir}, \bibinfo{person}{Veena~B. Mendiratta}, \bibinfo{person}{Chitra
  Phadke}, \bibinfo{person}{Eric Falk}, {and} \bibinfo{person}{Radu State}.}
  \bibinfo{year}{2017}\natexlab{}.
\newblock \showarticletitle{Detecting and predicting outages in mobile networks
  with log data}. In \bibinfo{booktitle}{\emph{Int. Conf. on Communications
  ({ICC})}}. \bibinfo{publisher}{IEEE}, \bibinfo{address}{Paris, France}, 7.
\newblock
\newblock
\shownote{{doi}: 10.1109/ICC.2017.7996706.}


\bibitem[\protect\citeauthoryear{Harchol-Balter}{Harchol-Balter}{2013}]%
        {harchol2013performance}
\bibfield{author}{\bibinfo{person}{Mor Harchol-Balter}.}
  \bibinfo{year}{2013}\natexlab{}.
\newblock \bibinfo{booktitle}{\emph{Performance Modeling and Design of Computer
  Systems}}.
\newblock \bibinfo{publisher}{Cambridge University Press},
  \bibinfo{address}{UK}.
\newblock


\bibitem[\protect\citeauthoryear{Heger, van Hoorn, Mann, and Okanovic}{Heger
  et~al\mbox{.}}{2017}]%
        {HegerHMO17}
\bibfield{author}{\bibinfo{person}{Christoph Heger}, \bibinfo{person}{Andr{\'e}
  van Hoorn}, \bibinfo{person}{Mario Mann}, {and} \bibinfo{person}{Dusan
  Okanovic}.} \bibinfo{year}{2017}\natexlab{}.
\newblock \showarticletitle{Application Performance Management: {S}tate of the
  Art and Challenges for the Future}. In \bibinfo{booktitle}{\emph{8th ACM/SPEC
  on Int. Conf. on Performance Engineering (ICPE)}}. \bibinfo{publisher}{ACM},
  \bibinfo{address}{L'Aquila, Italy}, \bibinfo{pages}{429--432}.
\newblock
\newblock
\shownote{{doi}: 10.1145/3030207.3053674, see also:
  \url{https://openapm.io/landscape}.}


\bibitem[\protect\citeauthoryear{Jiang, Hassan, Hamann, and Flora}{Jiang
  et~al\mbox{.}}{2008}]%
        {jiang2008automatic}
\bibfield{author}{\bibinfo{person}{Zhen~Ming Jiang}, \bibinfo{person}{Ahmed~E
  Hassan}, \bibinfo{person}{Gilbert Hamann}, {and} \bibinfo{person}{Parminder
  Flora}.} \bibinfo{year}{2008}\natexlab{}.
\newblock \showarticletitle{Automatic identification of load testing problems}.
  In \bibinfo{booktitle}{\emph{Int. Conf. on Software Maintenance (ICSM)}}.
  \bibinfo{publisher}{IEEE}, \bibinfo{address}{Beijing, China},
  \bibinfo{pages}{307--316}.
\newblock
\newblock
\shownote{{doi}: 10.1109/ICSM.2008.4658079.}


\bibitem[\protect\citeauthoryear{Jung, Paxson, Berger, and Balakrishnan}{Jung
  et~al\mbox{.}}{2004}]%
        {jung2004fast}
\bibfield{author}{\bibinfo{person}{Jaeyeon Jung}, \bibinfo{person}{Vern
  Paxson}, \bibinfo{person}{Arthur~W Berger}, {and} \bibinfo{person}{Hari
  Balakrishnan}.} \bibinfo{year}{2004}\natexlab{}.
\newblock \showarticletitle{Fast portscan detection using sequential hypothesis
  testing}. In \bibinfo{booktitle}{\emph{IEEE Symp. on Security and Privacy}}.
  \bibinfo{publisher}{IEEE}, \bibinfo{address}{Berkeley, CA, USA},
  \bibinfo{pages}{211--225}.
\newblock
\newblock
\shownote{{doi}: 10.1109/SECPRI.2004.1301325.}


\bibitem[\protect\citeauthoryear{Kambourakis, Kolias, and Stavrou}{Kambourakis
  et~al\mbox{.}}{2017}]%
        {kambourakis2017mirai}
\bibfield{author}{\bibinfo{person}{Georgios Kambourakis},
  \bibinfo{person}{Constantinos Kolias}, {and} \bibinfo{person}{Angelos
  Stavrou}.} \bibinfo{year}{2017}\natexlab{}.
\newblock \showarticletitle{The {M}irai botnet and the {IoT} Zombie Armies}. In
  \bibinfo{booktitle}{\emph{IEEE Military Communications Conf. (MILCOM)}}.
  \bibinfo{publisher}{IEEE}, \bibinfo{address}{Baltimore, MD, USA},
  \bibinfo{pages}{267--272}.
\newblock
\newblock
\shownote{{doi}: 10.1109/MILCOM.2017.8170867.}


\bibitem[\protect\citeauthoryear{Khazaei, Mi{\v s}i{\'c}, and Mi{\v
  s}i{\'c}}{Khazaei et~al\mbox{.}}{2012}]%
        {khazaei2012performance}
\bibfield{author}{\bibinfo{person}{Hamzeh Khazaei}, \bibinfo{person}{Jelena
  Mi{\v s}i{\'c}}, {and} \bibinfo{person}{Vojislav~B Mi{\v s}i{\'c}}.}
  \bibinfo{year}{2012}\natexlab{}.
\newblock \showarticletitle{Performance analysis of cloud computing centers
  using {M}/{G}/m/m+r queuing systems}.
\newblock \bibinfo{journal}{\emph{IEEE Transactions on Parallel and Distributed
  Systems}} \bibinfo{volume}{23}, \bibinfo{number}{5} (\bibinfo{date}{May}
  \bibinfo{year}{2012}), \bibinfo{pages}{936--943}.
\newblock
\newblock
\shownote{{doi}: 10.1109/TPDS.2011.199.}


\bibitem[\protect\citeauthoryear{Kin and Chan}{Kin and Chan}{2010}]%
        {kin2010generalized}
\bibfield{author}{\bibinfo{person}{Wai Kin} {and} \bibinfo{person}{Victor
  Chan}.} \bibinfo{year}{2010}\natexlab{}.
\newblock \showarticletitle{Generalized {L}indley-type recursive
  representations for multiserver tandem queues with blocking}.
\newblock \bibinfo{journal}{\emph{ACM Transactions on Modeling and Computer
  Simulation}} \bibinfo{volume}{20}, \bibinfo{number}{4}, Article
  \bibinfo{articleno}{21} (\bibinfo{date}{Nov.} \bibinfo{year}{2010}),
  \bibinfo{numpages}{19}~pages.
\newblock
\newblock
\shownote{{doi}: 10.1145/1842722.1842726.}


\bibitem[\protect\citeauthoryear{Kleinrock}{Kleinrock}{1976}]%
        {kleinrock1976queueing}
\bibfield{author}{\bibinfo{person}{Leonard Kleinrock}.}
  \bibinfo{year}{1976}\natexlab{}.
\newblock \bibinfo{booktitle}{\emph{Queueing Systems: Computer Applications}}.
  Vol.~\bibinfo{volume}{2}.
\newblock \bibinfo{publisher}{Wiley}, \bibinfo{address}{New York, USA}.
\newblock


\bibitem[\protect\citeauthoryear{Kumar}{Kumar}{2005}]%
        {Kumar2005}
\bibfield{author}{\bibinfo{person}{Vipin Kumar}.}
  \bibinfo{year}{2005}\natexlab{}.
\newblock \showarticletitle{Parallel and Distributed Computing for
  Cybersecurity}.
\newblock \bibinfo{journal}{\emph{IEEE Distributed Systems Online}}
  \bibinfo{volume}{6}, \bibinfo{number}{10} (\bibinfo{date}{Nov.}
  \bibinfo{year}{2005}), 9.
\newblock
\newblock
\shownote{{doi}: 10.1109/MDSO.2005.53.}


\bibitem[\protect\citeauthoryear{Laszka, Felegyhazi, and Butty{\'a}n}{Laszka
  et~al\mbox{.}}{2014}]%
        {laszka2012survey}
\bibfield{author}{\bibinfo{person}{Aron Laszka}, \bibinfo{person}{Mark
  Felegyhazi}, {and} \bibinfo{person}{Levente Butty{\'a}n}.}
  \bibinfo{year}{2014}\natexlab{}.
\newblock \showarticletitle{A survey of interdependent security games}.
\newblock \bibinfo{journal}{\emph{Comput. Surveys}} \bibinfo{volume}{47},
  \bibinfo{number}{2}, Article \bibinfo{articleno}{23} (\bibinfo{date}{Aug.}
  \bibinfo{year}{2014}), \bibinfo{numpages}{38}~pages.
\newblock
\newblock
\shownote{{doi}: 10.1145/2635673.}


\bibitem[\protect\citeauthoryear{Li, Sv{\"a}rd, Tordsson, and Elmroth}{Li
  et~al\mbox{.}}{2013}]%
        {li2013cost}
\bibfield{author}{\bibinfo{person}{Wubin Li}, \bibinfo{person}{Petter
  Sv{\"a}rd}, \bibinfo{person}{Johan Tordsson}, {and} \bibinfo{person}{Erik
  Elmroth}.} \bibinfo{year}{2013}\natexlab{}.
\newblock \showarticletitle{Cost-optimal cloud service placement under dynamic
  pricing schemes}. In \bibinfo{booktitle}{\emph{2013 IEEE/ACM 6th Int. Conf.
  on Utility and Cloud Computing}}. \bibinfo{publisher}{IEEE},
  \bibinfo{address}{Dresden, SN, Germany}, \bibinfo{pages}{187--194}.
\newblock
\newblock
\shownote{{doi}: 10.1109/UCC.2013.42.}


\bibitem[\protect\citeauthoryear{Manjhi}{Manjhi}{2006}]%
        {manjhiincreasing}
\bibfield{author}{\bibinfo{person}{Amit Manjhi}.}
  \bibinfo{year}{2006}\natexlab{}.
\newblock \bibinfo{title}{Increasing the Scalability of Data-intensive Web
  Applications}.
\newblock
\newblock
\newblock
\shownote{\url{http://www.cs.cmu.edu/~manjhi/thesis/proposal.pdf}.}


\bibitem[\protect\citeauthoryear{Manjhi, Ailamaki, Maggs, Mowry, Olston, and
  Tomasic}{Manjhi et~al\mbox{.}}{2006}]%
        {manjhi2006simultaneous}
\bibfield{author}{\bibinfo{person}{Amit Manjhi}, \bibinfo{person}{Anastassia
  Ailamaki}, \bibinfo{person}{Bruce~M Maggs}, \bibinfo{person}{Todd~C Mowry},
  \bibinfo{person}{Christopher Olston}, {and} \bibinfo{person}{Anthony
  Tomasic}.} \bibinfo{year}{2006}\natexlab{}.
\newblock \showarticletitle{Simultaneous scalability and security for
  data-intensive web applications}. In \bibinfo{booktitle}{\emph{Int. Conf. on
  Management of Data (SIGMOD)}}. \bibinfo{publisher}{ACM},
  \bibinfo{address}{Chicago, IL, USA}, \bibinfo{pages}{241--252}.
\newblock
\newblock
\shownote{{doi}: 10.1145/1142473.1142501.}


\bibitem[\protect\citeauthoryear{Menascé and Almeida}{Menascé and
  Almeida}{2001}]%
        {menasce2002capacity}
\bibfield{author}{\bibinfo{person}{Daniel~A Menascé} {and}
  \bibinfo{person}{Virgilio A~F Almeida}.} \bibinfo{year}{2001}\natexlab{}.
\newblock \bibinfo{booktitle}{\emph{Capacity Planning for Web Services:
  Metrics, Models, and Methods}}.
\newblock \bibinfo{publisher}{Prentice Hall PTR}, \bibinfo{address}{USA}.
\newblock


\bibitem[\protect\citeauthoryear{Mielke}{Mielke}{2006}]%
        {Mielke2006}
\bibfield{author}{\bibinfo{person}{Andreas Mielke}.}
  \bibinfo{year}{2006}\natexlab{}.
\newblock \showarticletitle{Elements for Response-time Statistics in {E}{R}{P}
  Transaction Systems}.
\newblock \bibinfo{journal}{\emph{Performance Evaluation}}
  \bibinfo{volume}{63}, \bibinfo{number}{7} (\bibinfo{date}{Jul.}
  \bibinfo{year}{2006}), \bibinfo{pages}{635--653}.
\newblock
\newblock
\shownote{{doi}: 10.1016/j.peva.2005.05.006.}


\bibitem[\protect\citeauthoryear{Milenkoski, Vieira, Kounev, Avritzer, and
  Payne}{Milenkoski et~al\mbox{.}}{2015}]%
        {milenkoski2015evaluating}
\bibfield{author}{\bibinfo{person}{Aleksandar Milenkoski},
  \bibinfo{person}{Marco Vieira}, \bibinfo{person}{Samuel Kounev},
  \bibinfo{person}{Alberto Avritzer}, {and} \bibinfo{person}{Bryan~D Payne}.}
  \bibinfo{year}{2015}\natexlab{}.
\newblock \showarticletitle{Evaluating computer intrusion detection systems: A
  survey of common practices}.
\newblock \bibinfo{journal}{\emph{Comput. Surveys}} \bibinfo{volume}{48},
  \bibinfo{number}{1}, Article \bibinfo{articleno}{12} (\bibinfo{date}{Sep.}
  \bibinfo{year}{2015}), \bibinfo{numpages}{41}~pages.
\newblock
\newblock
\shownote{{doi}: 10.1145/2808691.}


\bibitem[\protect\citeauthoryear{Mitchell and Chen}{Mitchell and Chen}{2014}]%
        {Mitchell2014}
\bibfield{author}{\bibinfo{person}{Robert Mitchell} {and}
  \bibinfo{person}{Ing-Ray Chen}.} \bibinfo{year}{2014}\natexlab{}.
\newblock \showarticletitle{A Survey of Intrusion Detection Techniques for
  Cyber-physical Systems}.
\newblock \bibinfo{journal}{\emph{Comput. Surveys}} \bibinfo{volume}{46},
  \bibinfo{number}{4}, Article \bibinfo{articleno}{55} (\bibinfo{date}{Mar.}
  \bibinfo{year}{2014}), \bibinfo{numpages}{29}~pages.
\newblock
\newblock
\shownote{{doi}: 10.1145/2542049.}


\bibitem[\protect\citeauthoryear{Nylander, Andrén, {\AA}rzén, and
  Maggio}{Nylander et~al\mbox{.}}{2018}]%
        {icac2018}
\bibfield{author}{\bibinfo{person}{Tommi Nylander},
  \bibinfo{person}{Marcus~Thelander Andrén}, \bibinfo{person}{Karl-Erik
  {\AA}rzén}, {and} \bibinfo{person}{Martina Maggio}.}
  \bibinfo{year}{2018}\natexlab{}.
\newblock \showarticletitle{Cloud Application Predictability through Integrated
  Load-Balancing and Service Time Control}. In \bibinfo{booktitle}{\emph{IEEE
  Int. Conf. on Autonomic Computing (ICAC)}}. \bibinfo{publisher}{IEEE},
  \bibinfo{address}{Trento, Italy}, \bibinfo{pages}{51--60}.
\newblock
\newblock
\shownote{{doi}: 10.1109/ICAC.2018.00015.}


\bibitem[\protect\citeauthoryear{Psounis, Molinero-Fern{\'a}ndez, Prabhakar,
  and Papadopoulos}{Psounis et~al\mbox{.}}{2005}]%
        {psounis2005systems}
\bibfield{author}{\bibinfo{person}{Konstantinos Psounis},
  \bibinfo{person}{Pablo Molinero-Fern{\'a}ndez}, \bibinfo{person}{Balaji
  Prabhakar}, {and} \bibinfo{person}{Fragkiskos Papadopoulos}.}
  \bibinfo{year}{2005}\natexlab{}.
\newblock \showarticletitle{Systems with multiple servers under heavy-tailed
  workloads}.
\newblock \bibinfo{journal}{\emph{Performance Evaluation}}
  \bibinfo{volume}{62}, \bibinfo{number}{1-4} (\bibinfo{date}{Oct.}
  \bibinfo{year}{2005}), \bibinfo{pages}{456--474}.
\newblock
\newblock
\shownote{{doi}: 10.1016/j.peva.2005.07.030.}


\bibitem[\protect\citeauthoryear{Qin}{Qin}{2010}]%
        {qin2010new}
\bibfield{author}{\bibinfo{person}{Sen Qin}.} \bibinfo{year}{2010}\natexlab{}.
\newblock \showarticletitle{A new optimal cost model of queue systems with
  Heavy-Tailed distribution}. In \bibinfo{booktitle}{\emph{2nd Int. Conf. on
  Information Science and Engineering}}. \bibinfo{publisher}{IEEE},
  \bibinfo{address}{Hangzhou, China}, \bibinfo{pages}{2530--2533}.
\newblock
\newblock
\shownote{{doi}: 10.1109/ICISE.2010.5691743.}


\bibitem[\protect\citeauthoryear{Rayana and Akoglu}{Rayana and Akoglu}{2016}]%
        {rayana2016less}
\bibfield{author}{\bibinfo{person}{Shebuti Rayana} {and} \bibinfo{person}{Leman
  Akoglu}.} \bibinfo{year}{2016}\natexlab{}.
\newblock \showarticletitle{Less is more: Building selective anomaly
  ensembles}.
\newblock \bibinfo{journal}{\emph{Transactions on Knowledge Discovery from Data
  (TKDD)}} \bibinfo{volume}{10}, \bibinfo{number}{4}, Article
  \bibinfo{articleno}{42} (\bibinfo{date}{May} \bibinfo{year}{2016}),
  \bibinfo{numpages}{33}~pages.
\newblock
\newblock
\shownote{{doi}: 10.1145/2890508.}


\bibitem[\protect\citeauthoryear{Rohr}{Rohr}{2015}]%
        {rohr2015workload}
\bibfield{author}{\bibinfo{person}{Matthias Rohr}.}
  \bibinfo{year}{2015}\natexlab{}.
\newblock \emph{\bibinfo{title}{Workload-sensitive Timing Behavior Analysis for
  Fault Localization in Software Systems}}.
\newblock Dissertation ({PhD} thesis). \bibinfo{school}{Faculty of Engineering,
  Kiel University}, \bibinfo{address}{Kiel, Germany}.
\newblock
\showISBNx{978-3-7347-4516-4}
\newblock
\shownote{available: \url{http://eprints.uni-kiel.de/27337/}.}


\bibitem[\protect\citeauthoryear{Rohr, van Hoorn, Hasselbring, L\"{u}bcke, and
  Alekseev}{Rohr et~al\mbox{.}}{2010}]%
        {Rohr2010WTB}
\bibfield{author}{\bibinfo{person}{Matthias Rohr}, \bibinfo{person}{Andr{\'e}
  van Hoorn}, \bibinfo{person}{Wilhelm Hasselbring}, \bibinfo{person}{Marco
  L\"{u}bcke}, {and} \bibinfo{person}{Sergej Alekseev}.}
  \bibinfo{year}{2010}\natexlab{}.
\newblock \showarticletitle{Workload-intensity-sensitive Timing Behavior
  Analysis for Distributed Multi-user Software Systems}. In
  \bibinfo{booktitle}{\emph{1st joint WOSP/SIPEW Int. conference on Performance
  Engineering}}. \bibinfo{publisher}{ACM}, \bibinfo{address}{San Jose, CA,
  USA}, \bibinfo{pages}{87--92}.
\newblock
\newblock
\shownote{{doi}: 10.1145/1712605.1712621.}


\bibitem[\protect\citeauthoryear{Rufino, Nogueira, Avritzer, Menasché, Russo,
  Janes, Ferme, van Hoorn, Schulz, and Lima}{Rufino et~al\mbox{.}}{2019}]%
        {techrep}
\bibfield{author}{\bibinfo{person}{Vilc Rufino}, \bibinfo{person}{Mateus
  Nogueira}, \bibinfo{person}{Alberto Avritzer}, \bibinfo{person}{Daniel
  Menasché}, \bibinfo{person}{Barbara Russo}, \bibinfo{person}{Andrea Janes},
  \bibinfo{person}{Vincenzo Ferme}, \bibinfo{person}{Andr{\'e} van Hoorn},
  \bibinfo{person}{Henning Schulz}, {and} \bibinfo{person}{Cabral Lima}.}
  \bibinfo{year}{2019}\natexlab{}.
\newblock \bibinfo{title}{Technical Report: Companion scripts}.
\newblock
\newblock
\newblock
\shownote{\url{https://github.com/queupe/mgk_simulation/}.}


\bibitem[\protect\citeauthoryear{Salehi, Zhang, Bezdek, and Leckie}{Salehi
  et~al\mbox{.}}{2016}]%
        {salehi2016smart}
\bibfield{author}{\bibinfo{person}{Mahsa Salehi}, \bibinfo{person}{Xuyun
  Zhang}, \bibinfo{person}{James~C Bezdek}, {and} \bibinfo{person}{Christopher
  Leckie}.} \bibinfo{year}{2016}\natexlab{}.
\newblock \showarticletitle{Smart sampling: A novel unsupervised boosting
  approach for outlier detection}. In \bibinfo{booktitle}{\emph{Australasian
  Joint Conf. on Artificial Intelligence}}. \bibinfo{publisher}{Springer},
  \bibinfo{address}{Hobart, TAS, Australia}, \bibinfo{pages}{469--481}.
\newblock
\newblock
\shownote{{doi}: 10.1007/978-3-319-50127-7\_40.}


\bibitem[\protect\citeauthoryear{Scheller-Wolf}{Scheller-Wolf}{2003}]%
        {scheller2003necessary}
\bibfield{author}{\bibinfo{person}{Alan Scheller-Wolf}.}
  \bibinfo{year}{2003}\natexlab{}.
\newblock \showarticletitle{Necessary and sufficient conditions for delay
  moments in FIFO multiserver queues with an application comparing s slow
  servers with one fast one}.
\newblock \bibinfo{journal}{\emph{Operations Research}} \bibinfo{volume}{51},
  \bibinfo{number}{5} (\bibinfo{year}{2003}), \bibinfo{pages}{748--758}.
\newblock
\newblock
\shownote{{doi}: 10.1287/opre.51.5.748.16759.}


\bibitem[\protect\citeauthoryear{Sharma, Shenoy, and Towsley}{Sharma
  et~al\mbox{.}}{2012}]%
        {sharma2012provisioning}
\bibfield{author}{\bibinfo{person}{Upendra Sharma}, \bibinfo{person}{Prashant
  Shenoy}, {and} \bibinfo{person}{Donald~F Towsley}.}
  \bibinfo{year}{2012}\natexlab{}.
\newblock \showarticletitle{Provisioning multi-tier cloud applications using
  statistical bounds on sojourn time}. In \bibinfo{booktitle}{\emph{9th Int.
  Conf. on Autonomic Computing (ICAC)}}. \bibinfo{publisher}{ACM},
  \bibinfo{address}{San Jose, CA, USA}, \bibinfo{pages}{43--52}.
\newblock
\newblock
\shownote{{doi}: 10.1145/2371536.2371545.}


\bibitem[\protect\citeauthoryear{Stidham~Jr}{Stidham~Jr}{1970}]%
        {stidham1970optimality}
\bibfield{author}{\bibinfo{person}{Shaler Stidham~Jr}.}
  \bibinfo{year}{1970}\natexlab{}.
\newblock \showarticletitle{On the optimality of single-server queuing
  systems}.
\newblock \bibinfo{journal}{\emph{Operations Research}} \bibinfo{volume}{18},
  \bibinfo{number}{4} (\bibinfo{date}{Aug.} \bibinfo{year}{1970}),
  \bibinfo{pages}{708--732}.
\newblock
\newblock
\shownote{{doi}: 10.1287/opre.18.4.708.}


\bibitem[\protect\citeauthoryear{Tadakamalla and Menascé}{Tadakamalla and
  Menascé}{2019}]%
        {tadakamalla2020autonomic}
\bibfield{author}{\bibinfo{person}{Venkat Tadakamalla} {and}
  \bibinfo{person}{Daniel~A. Menascé}.} \bibinfo{year}{2019}\natexlab{}.
\newblock \showarticletitle{Autonomic Elasticity Control for Multi-server
  Queues under Generic Workload Surges in Cloud Environments}.
\newblock \bibinfo{journal}{\emph{IEEE Transactions on Cloud Computing}}
  \bibinfo{volume}{14}, \bibinfo{number}{8} (\bibinfo{date}{Aug.}
  \bibinfo{year}{2019}), \bibinfo{pages}{1--12}.
\newblock
\newblock
\shownote{{doi}: 10.1109/TCC.2020.2992949.}


\bibitem[\protect\citeauthoryear{Tim{\v{c}}enko and Gajin}{Tim{\v{c}}enko and
  Gajin}{2017}]%
        {timvcenko2017ensemble}
\bibfield{author}{\bibinfo{person}{Valentina Tim{\v{c}}enko} {and}
  \bibinfo{person}{Slavko Gajin}.} \bibinfo{year}{2017}\natexlab{}.
\newblock \showarticletitle{Ensemble classifiers for supervised anomaly based
  network intrusion detection}. In \bibinfo{booktitle}{\emph{13th IEEE Int.
  Conf. on Intelligent Computer Communication and Processing (ICCP)}}.
  \bibinfo{publisher}{IEEE}, \bibinfo{address}{Cluj-Napoca, Romania},
  \bibinfo{pages}{13--19}.
\newblock
\newblock
\shownote{{doi}: 10.1109/ICCP.2017.8116977.}


\bibitem[\protect\citeauthoryear{Tordsson, Montero, Moreno-Vozmediano, and
  Llorente}{Tordsson et~al\mbox{.}}{2012}]%
        {tordsson2012cloud}
\bibfield{author}{\bibinfo{person}{Johan Tordsson},
  \bibinfo{person}{Rub{\'e}n~S Montero}, \bibinfo{person}{Rafael
  Moreno-Vozmediano}, {and} \bibinfo{person}{Ignacio~M Llorente}.}
  \bibinfo{year}{2012}\natexlab{}.
\newblock \showarticletitle{Cloud brokering mechanisms for optimized placement
  of virtual machines across multiple providers}.
\newblock \bibinfo{journal}{\emph{Future Generation Computer Systems}}
  \bibinfo{volume}{28}, \bibinfo{number}{2} (\bibinfo{year}{2012}),
  \bibinfo{pages}{358--367}.
\newblock
\urldef\tempurl%
\url{https://doi.org/10.1016/j.future.2011.07.003}
\showDOI{\tempurl}
\newblock
\shownote{{doi}: 10.1016/j.future.2011.07.003.}


\bibitem[\protect\citeauthoryear{Wierman, Osogami, Harchol-Balter, and
  Scheller-Wolf}{Wierman et~al\mbox{.}}{2006}]%
        {wierman2006many}
\bibfield{author}{\bibinfo{person}{Adam Wierman}, \bibinfo{person}{Takayuki
  Osogami}, \bibinfo{person}{Mor Harchol-Balter}, {and} \bibinfo{person}{Alan
  Scheller-Wolf}.} \bibinfo{year}{2006}\natexlab{}.
\newblock \showarticletitle{How many servers are best in a dual-priority M/PH/k
  system?}
\newblock \bibinfo{journal}{\emph{Performance Evaluation}}
  \bibinfo{volume}{63}, \bibinfo{number}{12} (\bibinfo{date}{Dec.}
  \bibinfo{year}{2006}), \bibinfo{pages}{1253--1272}.
\newblock
\newblock
\shownote{{doi}: 10.1016/j.peva.2005.12.004.}


\bibitem[\protect\citeauthoryear{Wolter and Reinecke}{Wolter and
  Reinecke}{2010}]%
        {Wolter2010}
\bibfield{author}{\bibinfo{person}{Katinka Wolter} {and}
  \bibinfo{person}{Philipp Reinecke}.} \bibinfo{year}{2010}\natexlab{}.
\newblock \showarticletitle{Performance and Security Tradeoff}. In
  \bibinfo{booktitle}{\emph{10th Int. School on Formal Methods for the Design
  of Computer, Communication and Software Systems}}
  \emph{(\bibinfo{series}{LNCS})}, Vol.~\bibinfo{volume}{6154}.
  \bibinfo{publisher}{Springer-Verlag}, \bibinfo{address}{Bertinoro, Italy},
  \bibinfo{pages}{135--167}.
\newblock
\newblock
\shownote{{doi}: 10.1007/978-3-642-13678-8\_4.}


\bibitem[\protect\citeauthoryear{Zarpelão, Miani, Kawakani, and
  de~Alvarenga}{Zarpelão et~al\mbox{.}}{2017}]%
        {Zarpela2017}
\bibfield{author}{\bibinfo{person}{Bruno~Bogaz Zarpelão},
  \bibinfo{person}{Rodrigo~Sanches Miani}, \bibinfo{person}{Cláudio~Toshio
  Kawakani}, {and} \bibinfo{person}{Sean~Carlisto de Alvarenga}.}
  \bibinfo{year}{2017}\natexlab{}.
\newblock \showarticletitle{A survey of intrusion detection in Internet of
  Things}.
\newblock \bibinfo{journal}{\emph{Journal of Network and Computer
  Applications}}  \bibinfo{volume}{84} (\bibinfo{date}{Apr.}
  \bibinfo{year}{2017}), \bibinfo{pages}{25--37}.
\newblock
\newblock
\shownote{{doi}: 10.1016/j.jnca.2017.02.009.}


\bibitem[\protect\citeauthoryear{Zhao, Lin, Chen, Wang, Yu, and Ming}{Zhao
  et~al\mbox{.}}{2016}]%
        {zhao2016optimizing}
\bibfield{author}{\bibinfo{person}{Xuancai Zhao}, \bibinfo{person}{Qiuzhen
  Lin}, \bibinfo{person}{Jianyong Chen}, \bibinfo{person}{Xiaomin Wang},
  \bibinfo{person}{Jianping Yu}, {and} \bibinfo{person}{Zhong Ming}.}
  \bibinfo{year}{2016}\natexlab{}.
\newblock \showarticletitle{Optimizing security and quality of service in a
  {R}eal-time database system using {M}ulti-objective genetic algorithm}.
\newblock \bibinfo{journal}{\emph{Expert Systems with Applications}}
  \bibinfo{volume}{64} (\bibinfo{date}{Dec.} \bibinfo{year}{2016}),
  \bibinfo{pages}{11--23}.
\newblock
\newblock
\shownote{{doi}: 10.1016/j.eswa.2016.07.023.}


\end{thebibliography}


\end{document}